\documentclass[a4paper, amsfonts, amssymb, amsmath, reprint, showkeys, longbibliography, nofootinbib, superscriptaddress, floatfix, twoside]{revtex4-2}

\usepackage{amsthm}
\usepackage{mathtools}
\usepackage{physics}
\usepackage{xcolor}
\usepackage{graphicx}
\usepackage[left=23mm,right=13mm,top=35mm,columnsep=15pt]{geometry} 
\usepackage{adjustbox}
\usepackage{placeins}
\usepackage[T1]{fontenc}
\usepackage{lipsum}
\usepackage{csquotes}
\usepackage{float}
\usepackage{caption}
\usepackage{comment}

\usepackage{hyperref} 
\hypersetup{pdftex, pdftitle={Article}, pdfauthor={Author}}

\begin{document}

\title{Soft-margin classification of object manifolds}
\author{Uri Cohen}
\affiliation{Edmond and Lily Safra Center for Brain Sciences, Hebrew University of Jerusalem, Israel}
\author{Haim Sompolinsky}
\thanks{Corresponding author: haim@fiz.huji.ac.il}
\affiliation{Edmond and Lily Safra Center for Brain Sciences, Hebrew University of Jerusalem, Israel}
\affiliation{Center for Brain Science, Harvard University, Cambridge, MA, USA}

\date{\today} 

\begin{abstract}
A neural population responding to multiple appearances of a single object defines a manifold in the neural response space. The ability to classify such manifolds is of interest, as object recognition and other computational tasks require a response that is insensitive to variability within a manifold. Linear classification of object manifolds was previously studied for max-margin classifiers. Soft-margin classifiers are a larger class of algorithms and provide an additional regularization parameter used in applications to optimize performance outside the training set by balancing between making fewer training errors and learning more robust classifiers. Here we develop a mean-field theory describing the behavior of soft-margin classifiers applied to object manifolds. Analyzing manifolds with increasing complexity, from points through spheres to general manifolds, a mean-field theory describes the expected value of the linear classifier's norm, as well as the distribution of fields and slack variables. By analyzing the robustness of the learned classification to noise, we can predict the probability of classification errors and their dependence on regularization, demonstrating a finite optimal choice. The theory describes a previously unknown phase transition, corresponding to the disappearance of a non-trivial solution, thus providing a soft version of the well-known classification capacity of max-margin classifiers.
\end{abstract}


\maketitle

\section{Introduction} \label{sec:introduction}

\paragraph*{Max-margin and soft-margin classification }

When performing linear classification, the naive approach would aim for classifying all the training samples correctly with the largest possible margin, an approach known as max-margin classification \cite{vapnik1963recognition,boser1992training}. An alternative approach, known as soft-margin classification \cite{cortes1995support,scholkopf2000new}, is to allow for misclassification of some of the samples, in order to increase the classification margin of most samples. Soft-margin classification is common in applications, where the data is not necessarily linearly separable. Furthermore, it allows for minimizing generalization error by optimizing a regularization parameter that balances between classification errors on the training set and achieving a larger margin. Both max-margin and soft-margin classification problems are solved by Support Vector Machine algorithms (hereafter, SVM).

\paragraph*{Previous works on manifold classification}

The problem of manifold classification arises in neuroscience and machine learning when a population of biological or artificial neurons represents an object, and variability in object appearance would define a manifold in the neural response space. In invariant object recognition tasks, the response of output neurons is determined by object identity alone, which is naturally defined as performing manifold classification, i.e., using target labels that are constant within manifolds. The ability to perform max-margin classification on manifolds of increased complexity was analyzed in recent years. Building on the seminal work of Gardner \cite{gardner1988space} which considered the classification of points, recent works have extended theory to describe manifolds of any shape \cite{chung2016linear,chung2018classification} and to allow for certain correlations between manifolds \cite{cohen2020separability}. Those theoretical advances described only max-margin classifiers, which are not common in applications. Here we close this gap by analyzing soft classification of manifolds of increasing complexity, going from points, through spheres, to general manifolds.

\paragraph*{Previous works on soft classification theory}

Previous theoretical works on soft-margin classifiers have analyzed the classification of points. Statistical learning tools were used to provide bounds on generalization error when using different kernels and different regularization schemes \cite{shawe2002generalization,chen2004support}. A statistical physics analysis of soft-margin classification in a teacher-student setup described the learning curve, i.e., the dependence of training and generalization error on the number of samples \cite{risau2001statistical}, thus extending the max-margin analysis \cite{dietrich1999statistical}. Here we avoid making specific assumptions on the teacher and instead consider soft classification performance when averaging over random choice of labels. 

\paragraph*{The role of noise}
When a soft-margin classifier is learned on a training set and then evaluated on a held-out test set, the classification errors achieved are called the training error and the test error, respectively. In general we expect the training error to be minimized for the max-margin classifier while the test error may be minimized at a finite value of the soft classification regularization parameter, which needs to be found empirically. Here we aim to analyze this setting by considering a test set that is a noisy version of the training set. This corresponds to noise-resistance of the classifier, and not to the notion of generalization error in machine-learning where it is assumed that the training and test set are sampled from the same distribution.
\section{Results}  \label{sec:results}

\subsection{Soft classification of points}

Max-margin classification of points is discussed by \cite{gardner1988space}; here we extend this seminal work to soft classification.
Given $P$ pairs $\left\{ \left(\boldsymbol{x}^{\mu},y^{\mu}\right)\right\} _{\mu=1}^{P}$
of points $\boldsymbol{x}^{\mu}\in\mathbb{R}^{N}$ and labels $y^{\mu}\in\left\{ \pm1\right\} $,
soft classification is defined by a set of weights $\boldsymbol{w}\in\mathbb{R}^{N}$
and slack variables $\vec{s}\in\mathbb{R}^{P}$ such that the fields
at the solution obey for all $\mu\in\left[1..P\right]$
\begin{equation}
h^{\mu}=y^{\mu}\boldsymbol{w}\cdot\boldsymbol{x}^{\mu}\ge1-s^{\mu}\label{eq:points-fields-def}
\end{equation}
The bold notation for $\boldsymbol{x}^{\mu}$ and $\boldsymbol{w}$ indicates
that they are vectors in $\mathbb{R}^{N}$, whereas the arrow notation
is used for other vectors, such as $\vec{s}$.
Given a regularization parameter $c\ge0$ the optimal classifier and slack variables are defined $\boldsymbol{w}^{*},\vec{s}^{*} =\arg\min_{\boldsymbol{w},\vec{s}}L\left(\boldsymbol{w},\vec{s}\right)$ for a Lagrangian
\begin{align}
L & =\|\boldsymbol{w}\|^{2}/N+c\|\vec{s}\|^{2}/N
\ s.t.\ \forall\mu\  h^{\mu}\ge1-s^{\mu}\label{eq:points-optimization-target}
\end{align}
and $L^{*}$ denotes the minimal value of $L$.

\paragraph*{Replica theory}

From the Lagrangian the volume of solutions $V\left(L,c\right)$ for a given value of the loss $L$ and a choice of regularization $c$ is given by:
\begin{align}
V\left(L,c\right) & =\int d^{N}\boldsymbol{w}\int d^{P}\vec{s}\delta\left(\|\boldsymbol{w}\|^{2}+c\|\vec{s}\|^{2}-NL\right) \\ & \cdots\prod_{\mu}^{P}\delta\left(y^{\mu}\boldsymbol{w}\cdot\boldsymbol{x}^{\mu}-h^{\mu}\right)\Theta\left(h^{\mu}-1+s^{\mu}\right)
\end{align}
The volume is defined for any positive $L,c$ but we are interested
in the problem parameters where it vanishes, which is expected
to happen only at the minimal value $L^{*}$. Thus by analyzing
the conditions where $V\to0$ we characterize the optimal solution
achieved by the optimization procedure, without introducing an additional
temperature variable as is usually done \cite{dietrich1999statistical,mezard2002analytic,zdeborova2007phase,ganguli2010statistical,advani2016statistical}.
This allows us to describe not only $L^{*}$ but also the expected norms of the weights $\|\boldsymbol{w}\|$ and slack variables $\|\vec{s}\|$, and the relation between $N$ and $P$ where the solution is achieved. 
For random labels $\vec{y}\in\left\{ \pm1\right\} ^{P}$ and points
$\boldsymbol{x}_{i}^{\mu}\sim{\cal N}\left(0,1/N\right)$ we calculate
the volume through replica identity:
\begin{equation}
\left[\log V\right]_{x,y}=\lim_{n\to0}\left[\frac{V^{n}-1}{n}\right]_{x,y}
\end{equation}
We solve this problem using a (replica symmetric) mean-field theory,
which is expected to be exact in the thermodynamic limit $N,P\to\infty$
with a finite ratio $\alpha=P/N$. Analyzing the case where $V\rightarrow0$ we obtain an expression for the loss $L$ in terms
of two order parameters $q$ and $k$ (see details in
section \ref{subsec:appendix-points-replica-theory}):
\begin{align}
L/q & =\frac{k-1}{k}+\frac{c}{1+ck}\alpha\alpha_{0}^{-1}\left(1/\sqrt{q}\right)\label{eq:points-L-mf}
\end{align}
where $\alpha_{0}^{-1}(\kappa)=\int_{-\infty}^{\kappa}Dt\left(\kappa-t\right)^{2}$
is Gardner's points capacity \cite{gardner1988space}, $q=\|\boldsymbol{w}\|^{2}/N$ is the norm of the weight vector, and the interpretation of $k$ is discussed below. 
Note we assumed here $\|\boldsymbol{x}^{\mu}\|=1$; if instead $\|\boldsymbol{x}^{\mu}\|=a$, then $q,c$ need to be scaled by $1/a^{2}$.

\paragraph*{Self-consistent equations}

We expect the solution to satisfy saddle-point equations $0=\frac{\partial L}{\partial q}=\frac{\partial L}{\partial k}$,
yielding 2 self-consistent equations for $q,k$ (see section \ref{subsec:appendix-points-self-consistent}):
\begin{align}
1 & =\frac{\left(ck\right)^{2}}{\left(1+ck\right)^{2}}\alpha\alpha_{0}^{-1}\left(1/\sqrt{q}\right)\label{eq:points-mf1}\\
1-k & =\frac{ck}{\left(1+ck\right)}\alpha H\left(-1/\sqrt{q}\right)\label{eq:points-mf2}
\end{align}
for $H(x)=\int_{x}^{\infty}\frac{dt}{\sqrt{2\pi}}e^{-x^{2}/2}$ the Gaussian tail function.

\begin{figure}
\includegraphics[width=8.5cm]{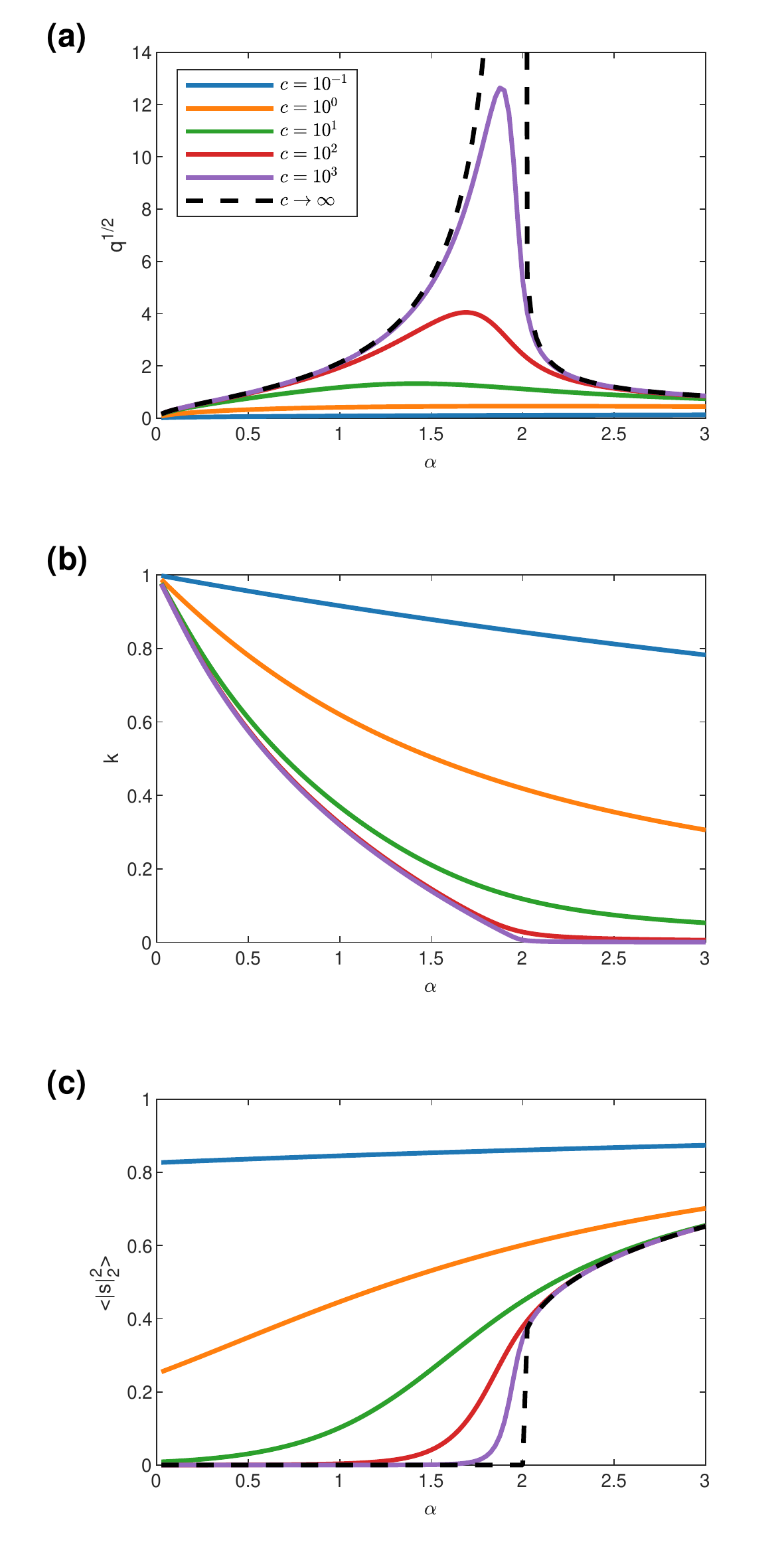}
\caption{\textbf{Order parameters in soft classification of points.}
(a) The optimal
weights' norm $q^{1/2}$ (y-axis) for different values of $\alpha$
(x-axis) and choices of the regularization variable $c$ (color coded), including the $c\to\infty$ limit (dashed line).
(b) The order parameter
$k$ (y-axis) for different values of $\alpha$ (x-axis) and choices
of $c$ (color coded). (c) The mean slack norm $\|\vec{s}\|^{2}$
(y-axis) for different values of $\alpha$ (x-axis) and choices of $c$ (color coded), including the $c\to\infty$ limit (dashed line).}
\label{fig:points-classification}
\end{figure}

The mean-field equations can be solved numerically for any load $\alpha$ (see algorithm
at section \ref{subsec:algs-self-consistent-solution-points}); figure
\ref{fig:points-classification}a-b shows the resulting
values of $q$ and $k$, respectively. 
We observe that $k\text{\ensuremath{\left(\alpha\right)}}$ decreases monotonically from $1$ at $0$ (figure \ref{fig:points-classification}b),
and similarly $\|\vec{s}\|$ increases monotonically from 0 to 1 (figure
\ref{fig:points-classification}c). Those are tightly related as from equations \ref{eq:points-optimization-target},\ref{eq:points-L-mf},\ref{eq:points-mf1} we have that $q/(ck)^{2}=\alpha\left\langle s^{2}\right\rangle$,
so that $ck$ describes the ratio between the weights' norm and the
slack norm at the optimization target (equation \ref{eq:points-optimization-target}). Furthermore, the first moment of the slack is related to the optimal cost through $L^{*}=\alpha c\left\langle s\right\rangle$ (see section \ref{subsec:appendix-optimal-loss}).

In contrast, $q\left(\alpha\right)$ is non-monotonic, increasing
from $0$ to a peak at a finite value, then decreasing (figure \ref{fig:points-classification}a).
This is an indication of the trade-off between achieving a larger margin
(small $q$) and making only small errors (small $\|\vec{s}\|$).
Figure \ref{fig:si-points-solution-props}
compares simulation results for $q$ with the results of solving the self-consistent equations.

We now consider some interesting limits (see details in section \ref{subsec:appendix-points-self-consistent}). When $\alpha\to0$ we have $k\to1$ and $q\to0$ so that $\alpha_{0}^{-1}\left(1/\sqrt{q}\right)\approx1/q$ and thus $k\approx1-\alpha c/\left(1+c\right)$ and $q\approx\alpha c^{2}/\left(1+c\right)^{2}$.
When $\alpha\to\infty$ we have $k\to0$ and $q\to0$ with scaling $k\approx1/c\alpha$, $q\approx1/\alpha$.
Both limits are marked in figure \ref{fig:si-points-solution-props}.

\paragraph*{Infinite $c$ limit}

When $c\to\infty$ and $\alpha<2$ there is a solution for $\boldsymbol{w}$ (of unconstrained norm) where $\vec{s}=\vec{0}$, so the Lagrangian becomes that of max-margin classifiers:
\begin{equation}
L=\min\|\boldsymbol{w}\|^{2}\ s.t.\ \forall\mu\ h^{\mu}\ge1
\end{equation}
In this regime $k$ is finite while $ck$ diverges, so equation \ref{eq:points-mf1} recovers the max-margin theory \cite{gardner1988space} and $q$ diverges for $\alpha$ near 2. On the other hand, for $c\to\infty$ and $\alpha>2$ there is no solution with $\vec{s}=\vec{0}$ so this term dominates the loss and the Lagrangian becomes:
\begin{equation}
L=\min\|\vec{s}\|^{2}\ s.t.\ \forall\mu\ h^{\mu}\ge1-s^{\mu}
\end{equation}
A mean-field solution of this Lagrangian involves two order parameters $q=\|\boldsymbol{w}\|^2/N$ and $K=\lim_{c\to\infty} ck$, which follow the self-consistent equations \ref{eq:points-mf1}-\ref{eq:points-mf2} (where $k$ on the left-hand-side of equation \ref{eq:points-mf2} approaches $0$, see section \ref{subsec:appendix-points-self-consistent}). 
Thus in the limit of $c\to\infty$ the mean-field theory reduces to a simple relation between $q$ and $\alpha$ (dashed line in figure \ref{fig:points-classification}a):
\begin{equation}
\alpha  = \begin{cases}
\alpha_{0}\left(1/\sqrt{q}\right) & \alpha<2\\
\alpha_{0}^{-1}\left(1/\sqrt{q}\right)/H^2\left(-1/\sqrt{q}\right)
& \alpha>2\\
\end{cases}
\end{equation}

\paragraph*{Field distribution}

\begin{figure*}
\includegraphics[width=17.5cm]{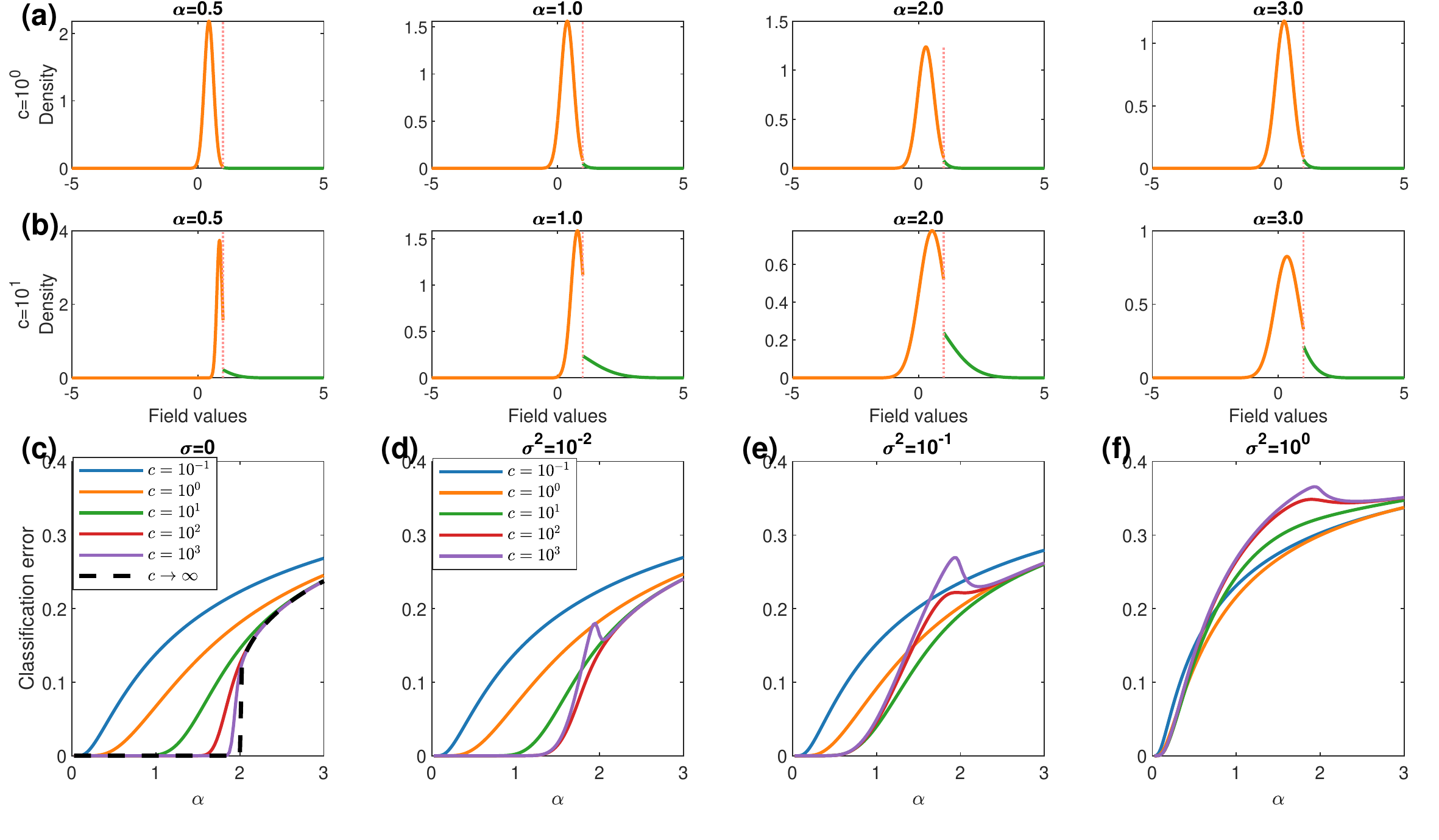}
\caption{\textbf{Field distribution and errors in soft classification of points.}
(a-b) Field distributions at different values of $\alpha$ (panels), with color coded regime (orange: ``touching'' regime; green: ``interior'' regime; a dashed line at $h=1$ indicates regime boundary), using $c=1$ (a) and $c=10$ (b). 
(c-f) Classification error (y-axis) for different values of $\alpha$ (x-axis) and choices of $c$ (color coded), including the $c\to\infty$ limit (dashed line in (c)). Each panel (c-f) shows the error at a different noise level $\sigma^{2}$ (indicated in title).}
\label{fig:points-classification-field-errors}
\end{figure*}

The theory also provides the joint distribution of $h,s$; their variance is due to the quenched variability in the choice of the classification labels and the arrangement of points (see details in section \ref{subsec:appendix-points-fields-slack-distribution}).
The field distribution is a concatenation of two truncated Gaussian variables, each representing a different solution regime:
\begin{equation}
h\sim\begin{cases}
{\cal N}\left(\frac{ck}{1+ck},\frac{q}{\left(1+ck\right)^{2}}\right) & h<1\\
{\cal N}\left(0,q\right) & h\ge1
\end{cases}\label{eq:point-fields}
\end{equation}
Fields $h\ge1$ are the ``interior'' regime (i.e., of points beyond the separating hyper-plane), where $s=0$, while fields $h<1$ are the ``touching'' regime (i.e., of points touching the separating hyper-plane), where $s>0$. This distribution is shown for several choices of $c$ and $\alpha$ in figure \ref{fig:points-classification-field-errors}a-b, and figure \ref{fig:si-points-fields} compares theory to the empirical histogram from simulations. The distribution of slack variables then follows from $s=\max\left\{ 1-h,0\right\}$.

\paragraph*{Classification errors}

We now turn our focus to the classification errors achieved when performing soft classification. 
The classification error on the training set is defined $\varepsilon_{tr}=P\left(h<0\right)=P\left(s>1\right)$ and from the field distribution we have:
\begin{align}
\varepsilon_{tr} & =H\left(ck/\sqrt{q}\right)=H\left(1/\sqrt{\alpha\left\langle s^{2}\right\rangle }\right)\label{eq:points-training-error}
\end{align}
A comparison of the training error observed in simulations with the theoretical predictions is given in figure \ref{fig:si-points-training-error}. As demonstrated in figure \ref{fig:points-classification-field-errors}c, the training error is monotonically increasing with $\alpha$ and monotonically decreasing with $c$ throughout (a proof for this is provided in section \ref{subsec:appendix-points-optimal-c}). For $\alpha<2$ where max-margin classifiers achieve no errors this is to be expected, but surprisingly this is also the case for $\alpha\ge2$ (see classification error for $\alpha\ge2$ and $c\to\infty$ in figure \ref{fig:points-classification-field-errors}c).

Thus we turn to analyze classification error in the presence of noise, where a finite $c$ may be optimal. When Gaussian noise ${\cal N}\left(0,\sigma^{2}/N\right)$ is applied at each component of the input vectors, a noise ${\cal N}\left(0,\sigma^{2}q\right)$ is added to the fields, so test error with respect to such noise is give by $\varepsilon_{g}=P\left(h+\eta\sigma\sqrt{q}<0\right)$, where $\eta$ is a standard Gaussian, or equivalently:
\begin{equation}
\varepsilon_{g}=\left\langle H\left(h/\sigma\sqrt{q}\right)\right\rangle _{h}\label{eq:points-test-error}
\end{equation}
Equation \ref{eq:points-test-error} can be evaluated using the field distribution (equation \ref{eq:point-fields}). The resulting theoretical predictions are shown in figure \ref{fig:points-classification-field-errors}d-f, exhibiting non-monotonic dependence on both $c$ and $\alpha$. A comparison of the theoretical predictions with simulation results for different choices of $c$ and levels of noise is provided in figure \ref{fig:si-points-test-error}.

\paragraph*{Classification errors for small noise}

While an explicit expression for the error is complicated, when the noise is small relative to the margin from the optimal hyper-plane $\sigma\ll1/\sqrt{q}$, we provide a simple approximation for the test error which can be written as a signal-to-noise ratio $\varepsilon_{g} \approx H({\cal S})$ (hereafter: SNR; see details in section \ref{subsec:appendix-points-classfication-error}):
\begin{align}
{\cal S} & =ck/\sqrt{q\left(1+\left(1+ck\right)^{2}\sigma^{2}\right)}\label{eq:points-test-error-SNR}
\end{align}
From the scaling of $q,k$ for large and small $\alpha$-s we have:
\begin{align}
{\cal S} & \approx\begin{cases}
1/\sqrt{\alpha\left(1/\left(1+c\right)^{2}+\sigma^{2}\right)} & \alpha\ll1\\
1/\sqrt{\alpha\left(1+\sigma^{2}\right)} & \alpha\gg1
\end{cases}\label{eq:points-snr-limits}
\end{align}

In this regime the optimal choice of $c$ can be found by maximizing ${\cal S}$ (equation \ref{eq:points-test-error-SNR}) with respect to $c$, that is solving $0=\frac{\partial{\cal S}^{-2}}{\partial c}$ for $c$, which yields (see section \ref{subsec:appendix-points-optimal-c}):
\begin{equation}
c^{*}=\frac{\sigma^{-2}}{1-k}-\frac{1}{k}\label{eq:points-optimal-c}
\end{equation}
which is positive in the regime where the SNR is a valid approximation, and needs to be solved self-consistently as $k$ depends on $c$. Due to the dependence on $k$ we have that $c^{*}$ depends on $\alpha$, but this analysis also suggests a ``canonical choice'' of $c$ which is independent of $\alpha$:
\begin{equation}
c\approx\sigma^{-2}\label{eq:optimal-c-approx}
\end{equation}
This choice is expected to capture the order of magnitude of $c^{*}$, except when $\alpha$ is very small or very large (as equation \ref{eq:points-optimal-c} diverges for both $k\to0$ and $k\to1$). 

\begin{figure}
\includegraphics[width=8.5cm]{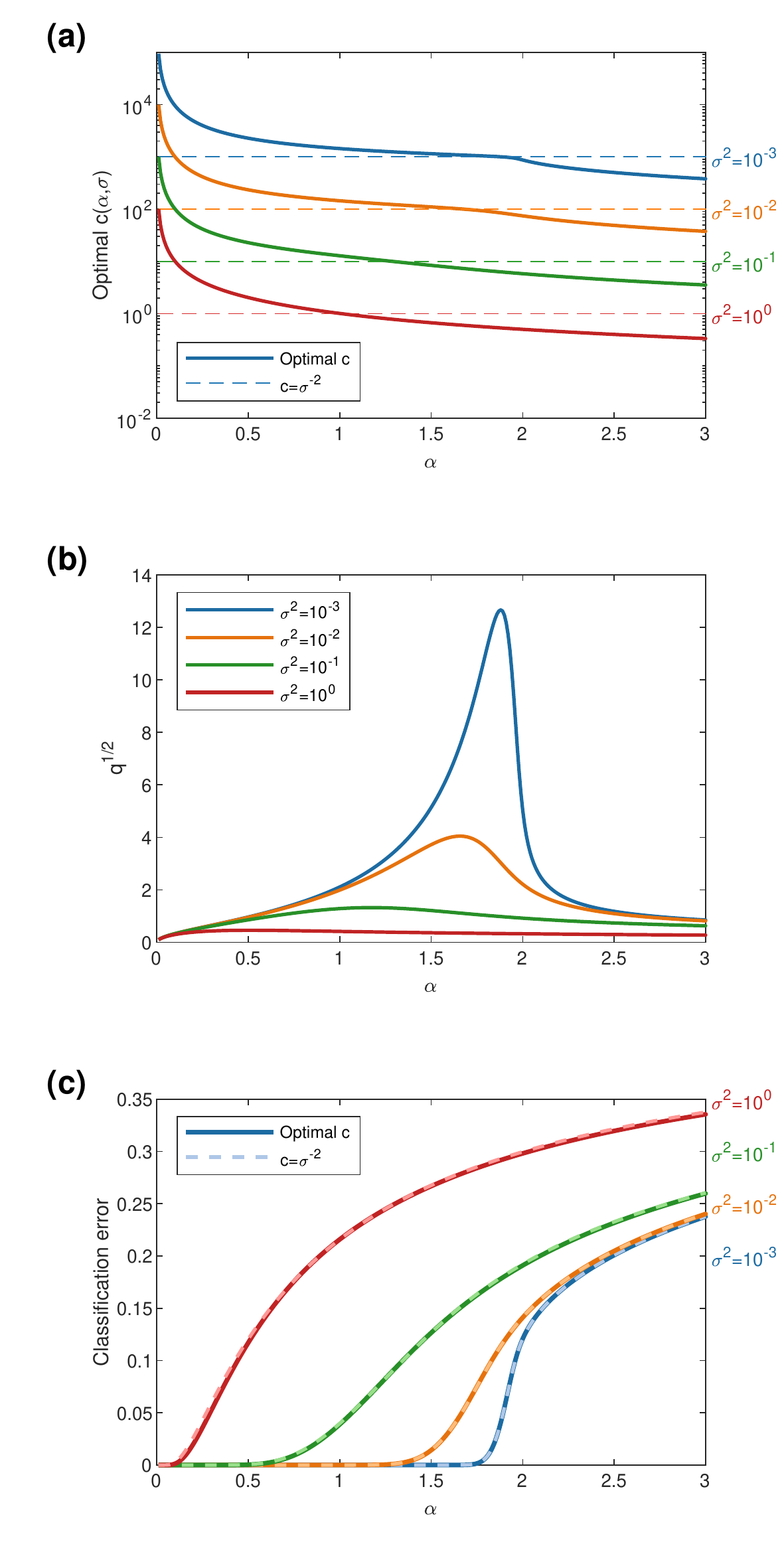}
\caption{\textbf{The optimal choice of $c$ in soft classification of points.}
(a) The optimal choice of $c$ (y-axis, log scale) for different values of $\alpha$ (x-axis) and levels of noise $\sigma^{2}$ (color coded). Compares the optimal choice $c^{*}$ (solid lines) and the canonical choice $c=\sigma^{-2}$ (dashed lines). 
(b) The weights' norm $q^{1/2}$ (y-axis) for different values of $\alpha$ (x-axis) and levels of noise $\sigma^{2}$ (color coded) when using the optimal value of $c$. (c) Classification error (y-axis) for different values of $\alpha$ (x-axis) and levels of noise $\sigma^{2}$ (color coded). Compares the optimal choice $c^{*}$ (solid lines) and the canonical choice $c=\sigma^{-2}$ (dashed lines).}
\label{fig:points-optimal-c}
\end{figure}

Figure \ref{fig:points-optimal-c}a demonstrates the optimal choice of $c$ calculated by solving equation \ref{eq:points-optimal-c} and compares it to equation \ref{eq:optimal-c-approx}, showing this approximation is within the correct scale for a large range of $\alpha$ values.
The resulting norm of the optimal solution changes smoothly with $\alpha$ (figure \ref{fig:points-optimal-c}b) and the canonical choice of $c$ achieves classification error which differs from the optimal one only when the error is much smaller than 1 (figure \ref{fig:points-optimal-c}c), and is superior to other sub-optimal choices of $c$ (figure \ref{fig:si-points-optimal-c}).

\subsection{Methods for soft classification of manifolds}

A manifold $M^{\mu}\subseteq\mathbb{R}^{N}$ for index $\mu\in\left[1..P\right]$ is parameterized by its axes $\left\{ \boldsymbol{u}_{l}^{\mu}\in\mathbb{R}^{N}\right\} _{l=0..D}^{\mu=1..P}$ and the manifold's intrinsic coordinates $\vec{S}\in{\cal M}^{\mu}\subseteq\mathbb{R}^{D+1}$. Each point in the manifold is a vector $\boldsymbol{x}^{\mu}(\vec{S})\in M^{\mu}$ such that:
\begin{equation}
\boldsymbol{x}^{\mu}(\vec{S})=\sum_{l=0}^{D}\boldsymbol{u}_{l}^{\mu}S_{l}\label{eq:def-general-manifold}
\end{equation}
As above, the bold notation for $\boldsymbol{x}^{\mu}$ and $\boldsymbol{u}_{l}^{\mu}$ indicates that they are vectors in $\mathbb{R}^{N}$, whereas the arrow notation is used for other vectors, such as the coordinates $\vec{S}$ (not to be confused with the slack $\vec{s}$). By convention $\boldsymbol{u}_{0}^{\mu}$ is the manifold center and we take $S_{0}=1$, so that distances are measured in units of the center norm. When classifying $P$ manifolds with weights $\boldsymbol{w}\in\mathbb{R}^{N}$, denoting axes projections $v_{l}^{\mu}=y^{\mu}\boldsymbol{u}_{l}^{\mu}\cdot\boldsymbol{w}$ the fields become:
\begin{equation}
h^{\mu}(\vec{S})=y^{\mu}\boldsymbol{w}\cdot\boldsymbol{x}^{\mu}(\vec{S})=v_{0}^{\mu}+\vec{S}\cdot\vec{v}^{\mu}\label{eq:manifold-fields}
\end{equation}

The classic soft classification formalism \cite{cortes1995support}, called here point-slack SVM, uses one slack variable per sample. It is usually inapplicable for manifold classification as the number of samples may be infinite. Thus we consider two simple alternatives which allow for soft classification of manifolds, both require only a single slack variable per manifold. In several specific cases where the point-slack formalism can be used, it will be compared to those formalisms.

\paragraph*{Center-slack method}

A naive approach for the classification of manifolds is to assume the soft classifier is learned using only the manifolds' centers and then evaluated on the entire manifolds.  Formally, soft classification using center-slacks is defined by weights $\boldsymbol{w}\in\mathbb{R}^{N}$ and slack variables $\vec{s}\in\mathbb{R}^{P}$ such that the central fields obey for all $\mu\in\left[1..P\right]$:
\begin{equation}
v_{0}^{\mu}=y^{\mu}\boldsymbol{w}\cdot\boldsymbol{u}_{0}^{\mu}\ge1-s^{\mu}\label{eq:manifolds-center-field}
\end{equation}
Given a regularization parameter $c\ge0$ the optimal classifier is defined by Lagrangian:
\begin{align}
L & =\|\boldsymbol{w}\|^{2}/N+c\|\vec{s}\|^{2}/N\ s.t.\ \forall\mu\  v_{0}^{\mu}\ge1-s^{\mu}\label{eq:center-slack-optimization-target}
\end{align}

Using this method the manifold structure is not used during training, so the weights' norm and field distribution (with respect to the centers) are given by points classification theory from previous section. However, an evaluation of classification errors on the manifold would require additional assumptions on the manifold.

\paragraph*{Manifold-slack method}

The previous method uses a slack variable to constrain the mean of the fields on the manifold. A natural alternative would be to constrain the minimal field on the manifold. Using the fields definition $h^{\mu}(\vec{S})$,
soft classification using manifold-slacks is defined by weights
$\boldsymbol{w}\in\mathbb{R}^{N}$ and slack variables $\vec{s}\in\mathbb{R}^{P}$
where the minimal fields obey for all $\mu\in\left[1..P\right]$:
\begin{equation}
h_{min}^{\mu}\doteq\min_{\vec{S}\in{\cal M}^{\mu}}{h^{\mu}(\vec{S})}\ge1-s^{\mu}
\end{equation}
That is, given a regularization parameter $c\ge0$ the optimal classifier is defined by Lagrangian:
\begin{align}
L & =\|\boldsymbol{w}\|^{2}/N+c\|\vec{s}\|^{2}/N\ s.t.\ \forall\mu\ h_{min}^{\mu}\ge1-s^{\mu}\label{eq:manifold-slack-optimization-target}
\end{align}

\begin{figure*}
\includegraphics[width=17.5cm]{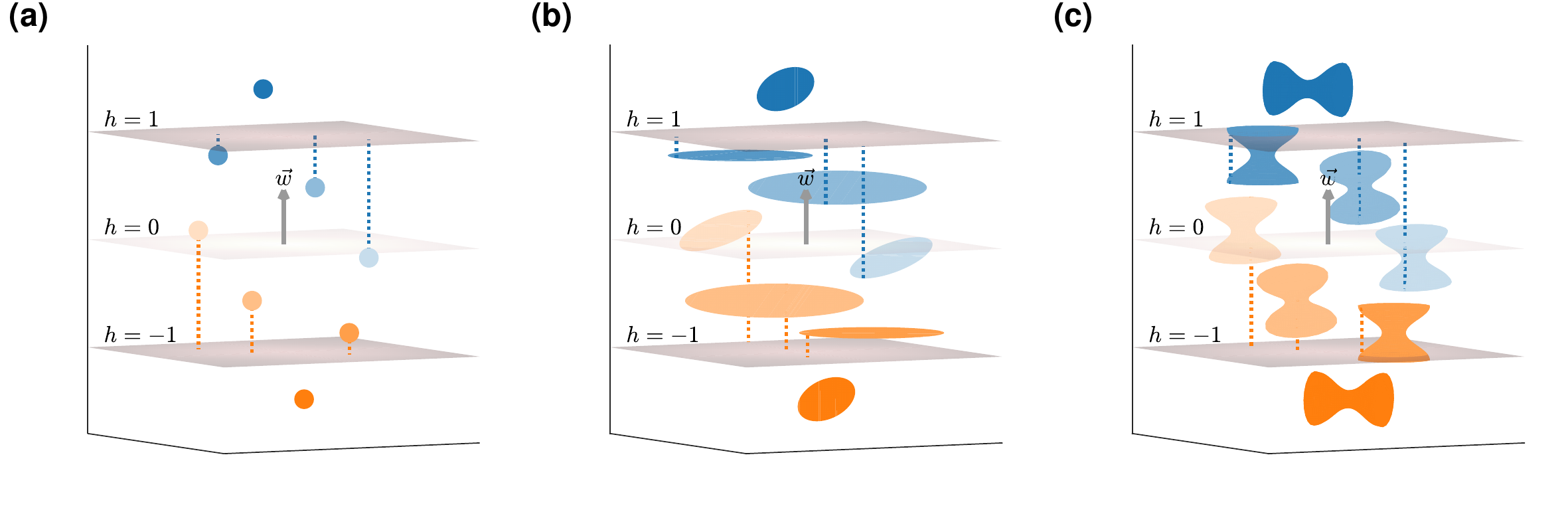}
\caption{\textbf{Illustration of soft classification of points, spheres and general manifolds.}
A weight vector $\boldsymbol{w}$ (gray arrow) defines the signed fields $h=\boldsymbol{w}\cdot\boldsymbol{x}$ on the manifolds being classified, satisfies $y\boldsymbol{w}\cdot\boldsymbol{x}\ge1-s$. The light gray hyper-plane depicts the decision boundary $h=0$; points above it are labeled $+1$ and below it $-1$. The dark gray hyper-planes depicts the boundaries $h=\pm1$. The length of each manifold's slack is indicated by a dashed line from the manifold point with the minimal field $y\boldsymbol{w}\cdot\boldsymbol{x}$ to the hyper-plane $h=y$. Each panel depicts the classification of 4 blue manifolds (target label is $+1$) against 4 orange manifolds (target label is $-1$). The blue and orange manifolds are symmetrically positioned for illustration purposes only. Manifolds are numbered from darkest to lightest. 
(a) Classification of points: the 1st point is in the interior $y\boldsymbol{w}\cdot\boldsymbol{x}>1$, has $s=0$; the 2nd and 3rd points have non-zero slack $0<s<1$, are classified correctly; the 4th point is below the decision boundary $y\boldsymbol{w}\cdot\boldsymbol{x}<0$,  corresponds to an error, has $s>1$. 
(b) Classification of spheres: the 1st sphere is in the interior $y\boldsymbol{w}\cdot\boldsymbol{x}>1$, the 2nd sphere is fully embedded within the hyper-plane $y\boldsymbol{w}\cdot\boldsymbol{x}=1-s$, the 3rd and 4th spheres touching the hyper-plane $y\boldsymbol{w}\cdot\boldsymbol{x}\ge1-s$ with the minimal field above $0$ for the 3rd, below $0$ for the 4th. 
(c) Classification of general manifolds: the 1st manifold is in the interior $y\boldsymbol{w}\cdot\boldsymbol{x}>1$, the 2nd manifold has a face embedded within the hyper-plane $y\boldsymbol{w}\cdot\boldsymbol{x}=1-s$, the 3rd and 4th manifolds touching the hyper-plane $y\boldsymbol{w}\cdot\boldsymbol{x}\ge1-s$ with the minimal field above $0$ for the 3rd, below $0$ for the 4th.}
\label{fig:illustration}
\end{figure*}

Figure \ref{fig:illustration} illustrates soft classification of points (or manifold centers, as noted above), spheres and general manifolds. In what follows we first discuss spheres, then extend the discussion to general manifolds.

\subsection{Soft classification of spheres}

A $D$-dimensional sphere of radius $R$ in $\mathbb{R}^{N}$ is defined:

\begin{equation}
\boldsymbol{x}^{\mu}(\vec{S})=\boldsymbol{u}_{0}^{\mu}+R\sum_{l=1}^{D}S_{l}\boldsymbol{u}_{l}^{\mu}\ s.t.\ \|\vec{S}\|\le1
\end{equation}
As in the case of points we would analyze the classification problem for random labels $\vec{y}\in\left\{ \pm1\right\} ^{P}$ and random axes $\boldsymbol{u}_{li}^{\mu}\sim{\cal N}\left(0,1/N\right)$, i.e., again scaling $\|\boldsymbol{u}_{l}^{\mu}\|\approx 1$.

\subsubsection{Center-slack}

Using center-slacks the classifier properties are given by the theory of soft classification of points, self-consistent equations \ref{eq:points-mf1}-\ref{eq:points-mf2}, and the distribution of the fields on the centers follows equation \ref{eq:point-fields}.

The classification error on the sphere is defined $\varepsilon=P\left(v_{0}+R\sum_{l}v_{l}S_{l}\le0\right)$ but as $v_{l}=y\boldsymbol{w}\cdot\boldsymbol{u}_{l}$ where $\boldsymbol{w}$ is independent of $\boldsymbol{u}_{l}$ in this case, we have that $v_{l}\sim{\cal N}\left(0,q\right)$, and as $\|\vec{S}\|=1$ on the sphere $R\sum_{l}v_{l}S_{l}\sim{\cal N}\left(0,qR^{2}\right)$.
If we assume Gaussian noise ${\cal N}\left(0,\sigma^{2}/N\right)$ is added independently for each sample component, as we have done for points, we have noise of ${\cal N}\left(0,q\left(\sigma^{2}+R^{2}\right)\right)$ at the fields. Thus the error is given by $ \varepsilon=P\left(v_{0}+\sqrt{\left(\sigma^{2}+R^{2}\right)q}\eta\le0\right)$ where $\eta$ is a standard Gaussian, or equivalently:
\begin{align}
\varepsilon & =\left\langle H\left(v_{0}/\sqrt{\left(\sigma^{2}+R^{2}\right)q}\right)\right\rangle _{v_{0}}\label{eq:spheres-center-slack-error}
\end{align}
where surprisingly, the dimensionality $D$ of the spheres plays no role in this setting.

\begin{figure}
\includegraphics[width=8.5cm]{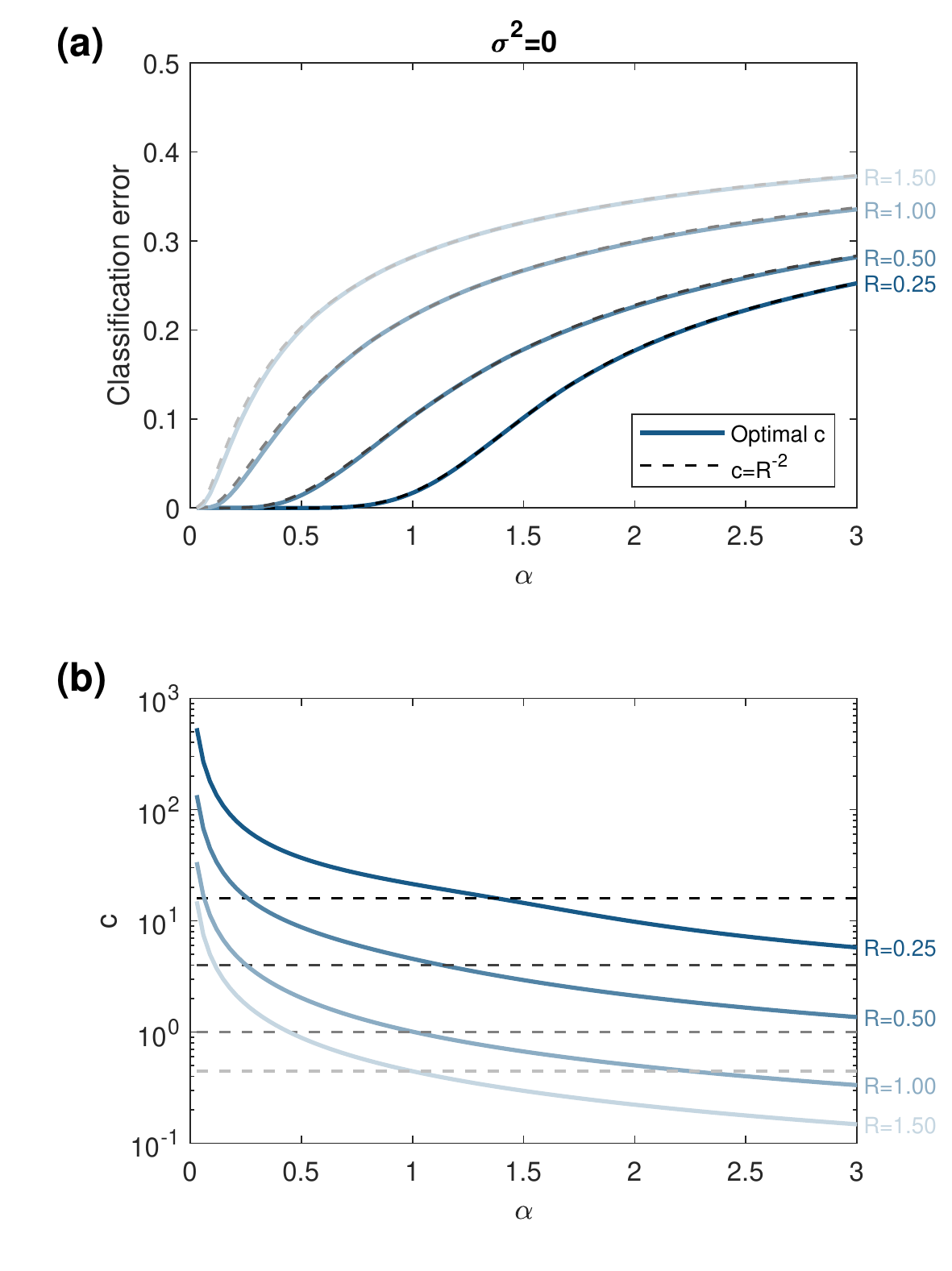}
\caption{\textbf{Soft classification of spheres using center-slacks.}
(a) Classification error (y-axis) for different values of $\alpha$ (x-axis) and $R$ (color coded), without noise $\sigma^{2}=0$, using the optimal choice of $c$ (solid lines) and the canonical choice $c=R^{-2}$ (dashed lines).
(b) The optimal choice of $c$ (y-axis, log scale) for different values of $\alpha$ (x-axis) and $R$ (color coded), without noise $\sigma^{2}=0$. The canonical choice $c=R^{-2}$ is indicated by the dashed horizontal lines.\\
Those results are independent of $D$, see main text.
}
\label{fig:spheres-center-slack}
\end{figure}

We conclude that soft classification of spheres of radius $R$ using center-slacks with noise level of $\sigma^2$ is equivalent to soft classification of points with effective noise  $\sigma_{eff}^2=\sigma^2+R^2$. Several corollaries can be made from the analysis of points, by using the effective noise $\sigma_{eff}^2$ instead of $\sigma^2$.
First, when  $\left(\sigma^{2}+R^{2}\right)q\ll1$ we expect a good SNR approximation $\varepsilon \approx H({\cal S})$ using:
\begin{align}
{\cal S} & =ck/\sqrt{q\left(1+\left(\sigma^{2}+R^{2}\right)\left(1+ck\right)^{2}\right)}
\end{align}
Figure \ref{fig:spheres-center-slack}a show the resulting error when sampling from the sphere (i.e., $\sigma=0$) for different values of $R$, and figure \ref{fig:si-spheres-center-slack-empiric} compares the theory to the error measured empirically. 
Second, the optimal choice of $c$ is then given by equation \ref{eq:points-optimal-c}, as well as the ``canonical choice''
\begin{equation}
c\approx1/\left(\sigma^{2}+R^{2}\right)\end{equation}
Contrary to the result from classification of points, due to the contribution of $R$, here the optimal choice for $c$ is finite even for $\sigma=0$, as illustrated in figure \ref{fig:spheres-center-slack}b.

\subsubsection{Manifold-slack}

We now consider soft classification of the entire manifold, that is $h_{min}^\mu \ge1-s^{\mu}$, thus generalizing the analysis of max-margin classifiers for spheres \cite{chung2016linear}. For spheres the point with the ``worst'' field, or minimal overlap with $\boldsymbol{w}$, is given by $\vec{S}=-\hat{v}$ (where $\hat{v}=\vec{v}/\|\vec{v}\|$), and hence a necessary and sufficient condition for the soft classification of the entire sphere is given by $v_{0}^{\mu}-R\|\vec{v}^{\mu}\|\ge1-s^{\mu}$.

\paragraph*{Replica theory}

This observation allows us to write an expression for the volume $V\left(L,c\right)$ of solutions achieving a target value of the loss $L$:
\begin{align}
V\left(L,c\right) & =\int d^{N}\boldsymbol{w}\int d^{P}\vec{s}\delta\left(\|\boldsymbol{w}\|^{2}+c\|\vec{s}\|^{2}-NL\right) \\
\cdots& \prod_{\mu}^{P}\delta\left(v_{0}^{\mu}-R\|\vec{v}^{\mu}\|-h^{\mu}\right)\Theta\left(h^{\mu}-1+s^{\mu}\right)
\end{align}
A replica analysis yields the following relation between $L,\alpha$ and the two order parameters $q,k$ when the volume of solutions vanishes (see details in section \ref{subsec:appendix-spheres-replica-theory}):
\begin{align}
L/q & =\frac{k-1}{k}+\frac{\alpha}{k}\int D^{D}\vec{t}\int Dt_{0}F\left(\vec{t},t_{0}\right)\label{eq:spheres-mf-equation}\\
F\left(\vec{t},t_{0}\right) & =\min_{v_{0}-R\|\vec{v}\|\ge1/\sqrt{q}}\left\{ \|\vec{v}-\vec{t}\|^{2}+\frac{ck}{1+ck}\left(v_{0}-t_{0}\right)^{2}\right\} \label{eq:spheres-inner-problem}
\end{align}
where $q=\|\boldsymbol{w}\|^{2}/N$ and $Dt_0=dt_0e^{-t_0^2/2}/\sqrt{2\pi}$ so $\vec{t},t_{0}$ are $D+1$ Gaussian variables representing the quenched noise in the solution, due to the variability of the labels $\left\{ y^{\mu}\right\}$ and the manifolds' axes $\left\{ \boldsymbol{u}_{l}^{\mu}\right\}$.

Solving the inner problem (equation \ref{eq:spheres-inner-problem}) using Karush-Kuhn-Tucker conditions \cite{kuhn1951nonlinear} (hereafter: KKT), allows us to describe the joint distribution of $v_{0}$, $v=\|\vec{v}\|$, and $s$ conditioned on $t_{0},t=\|\vec{t}\|$ at different solution regimes (see details in section \ref{subsec:appendix-spheres-KKT}):
\begin{enumerate}
\item ``Interior'' regime: the entire sphere is classified correctly with $h>1$ and a margin larger than $1/\sqrt{q}$ from the hyper-plane $h=0$; in this regime the slack is not utilized $s=0$ and the solution satisfies $v_{0}=t_{0}$, $v_{l}=t_{l}$ so that $F=0$.
This regime is in effect for $1/\sqrt{q}+Rt\le t_{0}\le\infty$.
\item ``Touching'' regime: the tip of the sphere touches the hyper-plane $h=1-s$; in this regime $v_{0},v,s$ have non-trivial values. 
This regime is in effect for $1/\sqrt{q}-\frac{1+ck}{ck}t/R\le t_{0}\le1/\sqrt{q}+Rt$.
\item ``Embedded'' regime: the entire sphere is within the hyper-plane $h=1-s$; in this regime $v=0$ but $v_{0},s$ have non-trivial values. 
This regime is in effect for $-\infty<t_{0}\le1/\sqrt{q}-\frac{1+ck}{ck}t/R$.
\end{enumerate}
The KKT analysis also provides the minimization value $F\left(t_{0},t\right)$ achieved at each regime, so that denoting $f\left(R,D,ck,q\right)=\int D^{D}\vec{t}\int Dt_{0}F\left(\vec{t},t_{0}\right)$ we have (see details in section \ref{subsec:appendix-spheres-KKT}):
\begin{widetext}
\begin{align}
f\left(R,D,ck,q\right) & =\int\chi_{D}\left(t\right)\int_{-\infty}^{1/\sqrt{q}-\frac{1+ck}{ck}t/R}Dt_{0}\left[\frac{ck}{1+ck}\left(1/\sqrt{q}-t_{0}\right)^{2}+t^{2}\right] \label{eq:spheres-capacity}\\
 & +\int\chi_{D}\left(t\right)\int_{1/\sqrt{q}-\frac{1+ck}{ck}t/R}^{1/\sqrt{q}+Rt}Dt_{0}\frac{ck}{1+ck\left(1+R^{2}\right)}\left(1/\sqrt{q}+Rt-t_{0}\right)^{2}
\end{align}
\end{widetext}
where $\chi_{D}\left(t\right)$ is the Chi distribution with $D$ degrees of freedom, $\chi_{D}\left(t\right)=\frac{2^{1-D/2}}{\Gamma\left(D/2\right)}t^{D-1}e^{-t^{2}/2}dt$, and the mean-field equation becomes:
\begin{align}
L/q & =\frac{k-1}{k}+\frac{1}{k}\alpha f\left(R,D,ck,q\right)\label{eq:spheres-L-mf}
\end{align}

\paragraph*{Self-consistent equations}

\begin{figure}
\includegraphics[width=8.5cm]{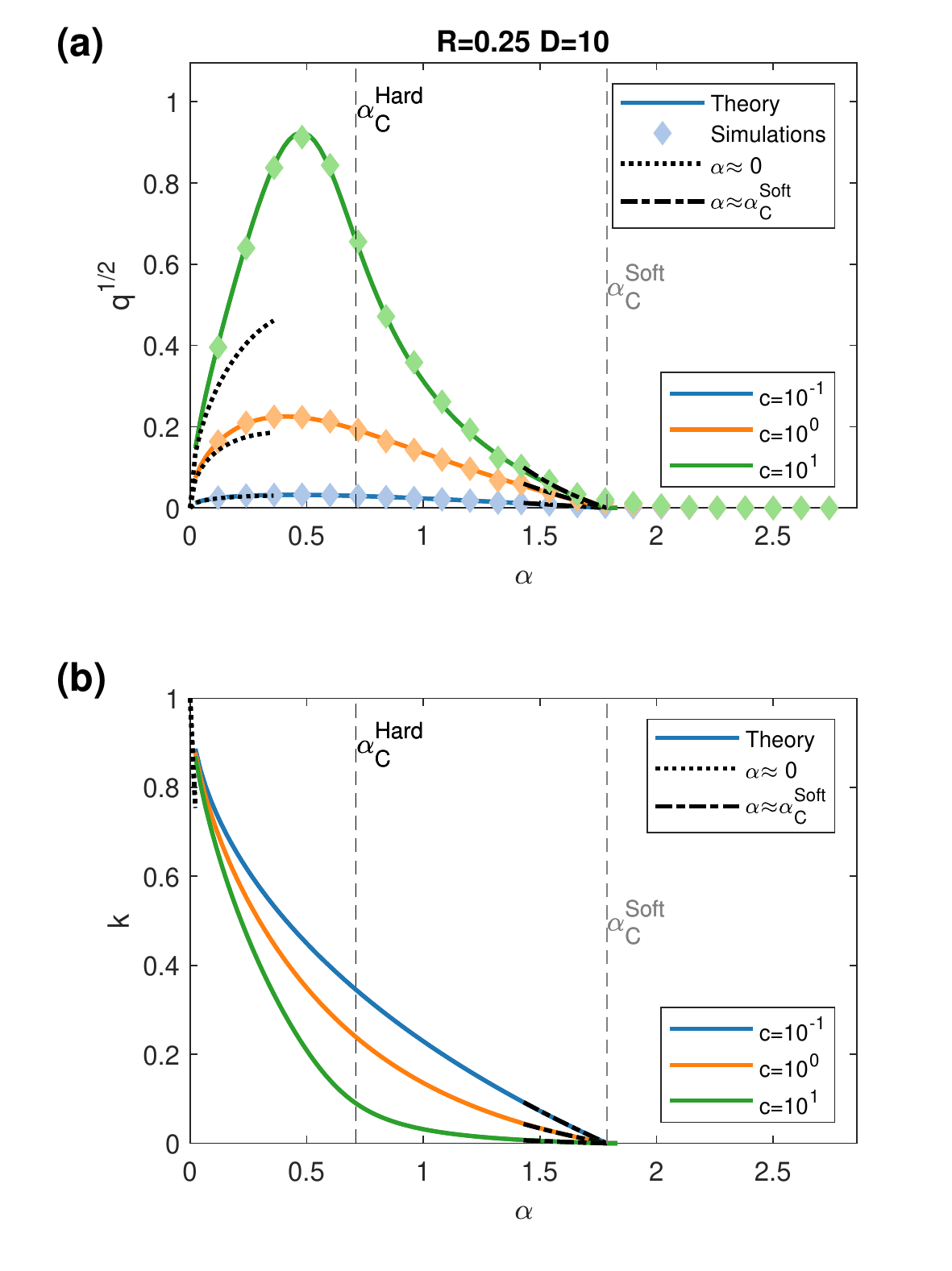}
\caption{\textbf{Order parameters in soft classification of spheres using manifold-slacks.}
(a) The weights' norm $q^{1/2}$ (y-axis) for different values of $\alpha$ (x-axis), and choices of $c$ (color coded), for radius $R=0.25$ and dimension $D=10$. Compares theory results (solid lines) to simulation results (diamonds).\\
(b) The order parameter $k$ (y-axis) for different values of $\alpha$ (x-axis), and choices of $c$ (color coded). \\
(a-b) Theory for the limits of $\alpha\to0$, $\alpha\to\alpha_{C}^{Soft}$ is marked as black dotted, dash-dot lines, respectively.}
\label{fig:spheres-q-k}
\end{figure}

Assuming the optimal loss $L^{*}$ satisfies saddle-point conditions $0=\frac{\partial L}{\partial q}=\frac{\partial L}{\partial k}$, we have 2 self-consistent equations for $k,q$, similar to those found in the case of points:
\begin{align}
1 & =\alpha f-\alpha k\frac{\partial}{\partial k}f\label{eq:spheres-mf1-abstract}\\
1-k & =\alpha f+\alpha q\frac{\partial}{\partial q}f\label{eq:spheres-mf2-abstract}
\end{align}
See the concrete form, equations \ref{eq:si-spheres-mf1},\ref{eq:si-spheres-mf2}, in section \ref{subsec:appendix-spheres-self-consistent-equations}.
Those equations can be solved numerically to predict the weights' norm; the algorithm is formally described in section \ref{subsec:algs-self-consistent-solution-spheres}. 
This prediction is compared to the norm observed in simulations (i.e., by finding the optimal weights for classification of spheres, using an algorithm described in section \ref{subsec:algs-numeric-classification-spheres}).
Figure \ref{fig:spheres-q-k} shows the resulting $q,k$ for specific values of $R,D$ (and additional ones are presented in figure \ref{fig:si-spheres-q-k}); $k\left(\alpha\right)$ decrease monotonically from $1$ to $0$ while $q\left(\alpha\right)$ has a single peak, increasing from $0$ to a finite value at the peak, then decreasing monotonically.

As in the case of points, in the limit $c\to\infty$, we find a different behavior below and above $\alpha_{C}^{Hard}$, the max-margin capacity. For $\alpha<\alpha_{C}^{Hard}$ we have that $k$ is finite while $ck$ diverges, with equation \ref{eq:spheres-mf1-abstract} becoming the mean-field equation from max-margin classification \cite{chung2016linear}, and the underlying Lagrangian is given by 
\begin{equation}
L  = \|\boldsymbol{w}\|^{2}/N\ \ s.t.\ \forall\mu\ h_{min}^{\mu}\ge1
\end{equation}
On the other hand, for $\alpha>\alpha_C^{Hard}$ we have that $k$ approaches $0$ while $q$ and $K=\lim_{c\to\infty} ck$ are finite (see details in section \ref{subsec:appendix-spheres-simplified-mf}), with the underlying Lagrangian
\begin{equation}
L  = \|\vec{s}\|^{2}/N\ \ s.t.\ \forall\mu\ h_{min}^{\mu}\ge1-s^{\mu}
\end{equation}

A second interesting limit is $\alpha\to0$. In this limit we expect the order parameters to behave as in the case of points, $q\to0$ and $k\to1$. We find that for small $\alpha$ the self-consistent equations are simplified and for $\alpha\ll1$ we have the approximations $k\approx1-\alpha\left(1+D\right)$ and $q\approx\alpha\left(ck\right)^{2}/\left(1+ck\right)^{2}$ (see figures \ref{fig:spheres-q-k}, \ref{fig:si-spheres-q-k} where those approximations are marked; see details in section \ref{subsec:appendix-spheres-simplified-mf}).

\paragraph*{Phase-transition}

An analysis of the mean-field equations reveals that for spheres (unlike points) there is a finite value of $\alpha$ where $q\to0$, and above which there is no solution with $q>0$ (visible also in figures \ref{fig:spheres-q-k}, \ref{fig:si-spheres-q-k}). The corresponding simulation results indicate that when the theory equations cannot be solved the optimal classifier is $\boldsymbol{w}=\boldsymbol{0}$, that is $q=0$, with all the slack variable saturating at $\vec{s}\equiv1$. Thus, soft-margin classification problems always have a solution, unlike max-margin classification problem; but when there is no solution with loss below $L=c\alpha$, the optimal choice uses zero weights and unit slack variables.

The critical value for $\alpha$ can be found by assuming that both $k,\sqrt{q}\ll1$; using a scaling of $x=ck/\sqrt{q}$ we get that $\alpha=\alpha_{C}$ would satisfy:
\begin{align}
\alpha_{C}^{-1} & =\int_{0}^{xR}\chi_{D}\left(t\right)t^{2}+xR\int_{xR}^{\infty}\chi_{D}\left(t\right)t \label{eq:spheres-phase-transition}\\
x & =\left(1+R^{2}\int_{xR}^{\infty}\chi_{D}\left(t\right)\right)^{-1}R\int_{xR}^{\infty}\chi_{D}\left(t\right)t \label{eq:spheres-phase-transition-x}
\end{align}
where $x$ is the self-consistent solution of equation \ref{eq:spheres-phase-transition-x}. Above this value of $\alpha$ there is no solution for $k,q$ (see details in section \ref{subsec:appendix-spheres-capacity}).

\begin{figure}
\includegraphics[width=8.5cm]{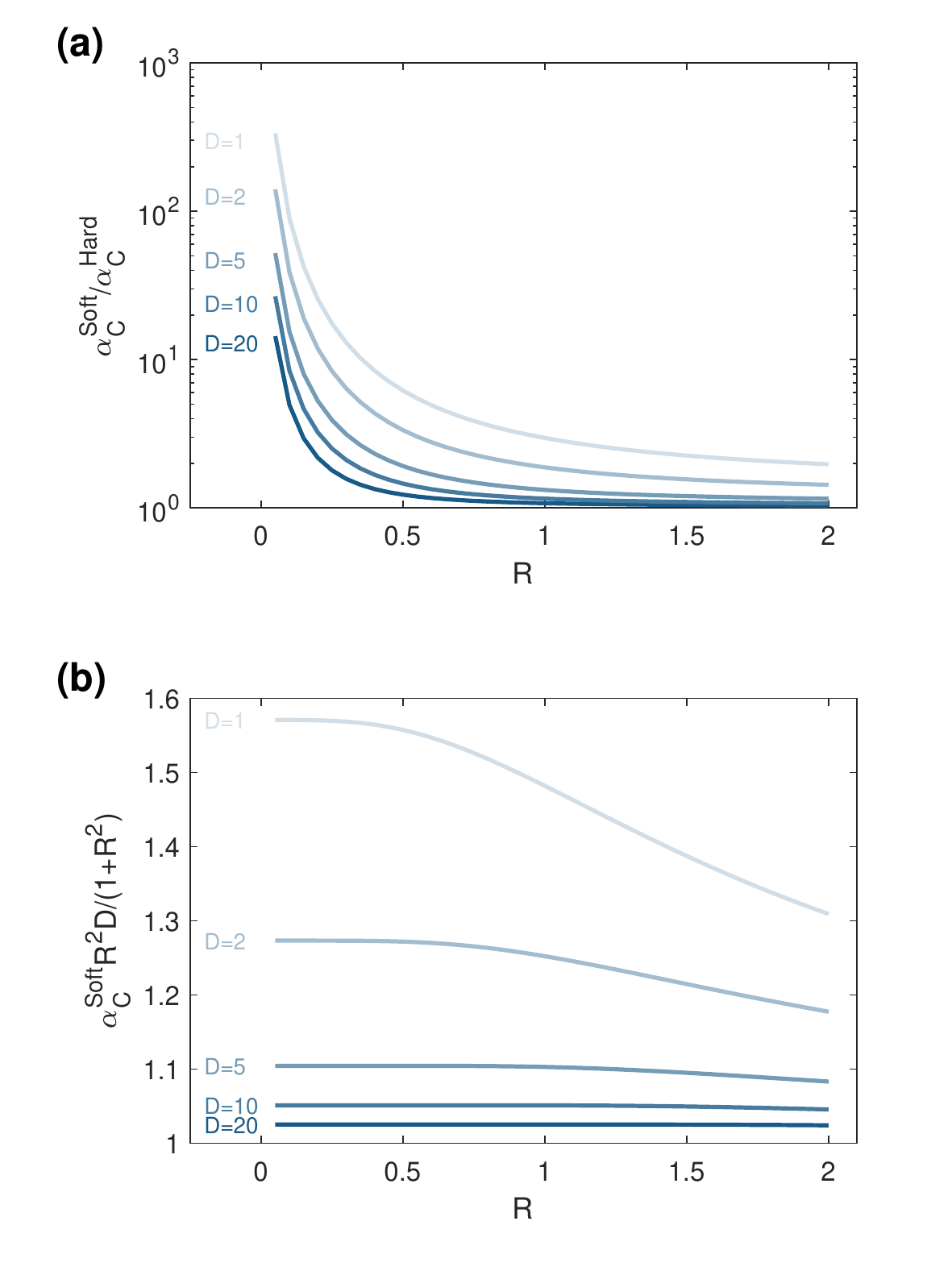}
\caption{\textbf{Capacity in manifold-slack classification of spheres.}
(a) The ratio between $\alpha_{C}^{Soft}$ and $\alpha_{C}^{Hard}$ (y-axis, log scale) for different values of $R$ (x-axis) and $D$ (color coded).
(b) The ratio between $\alpha_{C}^{Soft}$ and equation \ref{eq:spheres-alpha-critical} approximation (y-axis) for different values of $R$ (x-axis) and $D$ (color coded).}
\label{fig:spheres-capacity}
\end{figure}

Surprisingly, the critical value is independent of $c$
and we denote it $\alpha_{C}^{Soft}$, as a soft analog of the max-margin capacity $\alpha_{C}^{Hard}$ \cite{chung2016linear}. Notably, the former is always larger $\alpha_{C}^{Soft}\ge\alpha_{C}^{Hard}$, as shown in figure \ref{fig:spheres-capacity}a. 

For $R\to0$ we have that $x=R\int_{0}^{\infty}\chi_{D}\left(t\right)t=R\sqrt{2}\Gamma\left(\frac{D}{2}+\frac{1}{2}\right)\big/\Gamma\left(\frac{D}{2}\right)$ and $\alpha_{C}^{-1}=x^{2}$. Thus, for small $R$, the critical value $\alpha_{C}^{Soft}$ diverges as $R^{-2}$ (and in the limit of points there is no phase transition). Conversely, for $R\to\infty$ we have $x\approx0$ and $\alpha_{C}^{Soft}=D^{-1}$, whereas in this limit $\alpha_{C}^{Hard}=(D+1/2)^{-1}$ \cite{chung2016linear}. 
Intuitively, in both cases $\boldsymbol{w}$ must be perpendicular to the $PD$ manifold axes; for soft classification this implies just $N>PD$ or $\alpha<D^{-1}$, while for max-margin classification due to the finite capacity when classifying the centers this means $P/\left(N-PD\right)<2$ or $\alpha<\left(D+1/2\right)^{-1}$.

The existence of a sharp transition in the manifold-slack problem is the result of the thermodynamic limit. For small $N$, the existence of a solution at any given $\alpha$ depends on the particular labels realization. As $N$ increases, the probability of having a solution approaches $1$ for $\alpha<\alpha_C$ and zero for $\alpha>\alpha_C$ (figure \ref{fig:si-spheres-phase-transition-finite-size-effect}).

\paragraph*{Phase-transition for large D regime}
When $D\gg1$ the phase-transition equations \ref{eq:spheres-phase-transition}-\ref{eq:spheres-phase-transition-x} implies a simple expression for capacity:
\begin{equation}
\alpha_{C}^{Soft}\approx \left(1+R^{2}\right)\big/R^{2}D\label{eq:spheres-alpha-critical}
\end{equation}
Figure \ref{fig:spheres-capacity}b compares this approximation to the full expression for different values of $R,D$; as observed, this approximation is reasonable for large $D$ independently of the value of $R$ (see details in section \ref{subsec:appendix-spheres-capacity}).
In this regime the max-margin capacity is given by \cite{chung2016linear}:
\begin{equation}
\alpha_{C}^{Hard}\approx \left(1+R^{2}\right)\alpha_{0}(R\sqrt{D})
\end{equation}

\paragraph*{Classification errors}

As for points, the mean-field theory also provides the full distribution of the fields and slack variables (see details in section \ref{subsec:appendix-spheres-fields-and-slack-distribution}). Figure \ref{fig:si-spheres-slack-distribution} compares the theoretical slack distribution to the histogram of the values observed in simulations. In what follows we use these distributions to calculate different kinds of classification errors.

In the framework of manifold-slacks it is natural to consider the probability of error anywhere on the manifold, or equivalently the fraction of manifolds where the worst point is misclassified. This is the fraction of slack variables that are larger than $1$, i.e.,  $\varepsilon_{tr}^{manifold}=P\left(s\ge1\right)$, which can be evaluated from the slack distribution (equation \ref{eq:si-spheres-s} in section \ref{subsec:appendix-spheres-fields-and-slack-distribution}). This entire-manifold classification error is given by $\varepsilon_{tr}^{manifold}=H({\cal S})$ for ${\cal S}$ from
equation \ref{eq:si-training-error-any} in section \ref{subsec:appendix-spheres-classification-errors}.

A different kind of error is the probability of classification error on uniformly sampled points from the sphere, that is $\varepsilon_{tr}^{sample}=P\left(h<0\right)$, similar to the error considered above for center-slacks. These fields can be written as $h=v_{0}+R v z$, where $z=\cos(\theta)$ for $\theta$ the angle between the weight vector and the point on the sphere, $v_0$ and $v=\|\vec{v}\|$ are the projections of the weight vector on the center and the sphere subspace.
Thus, $\varepsilon_{tr}^{sample}=P\left(v_{0}+R v z <0\right)$, where the joint distribution of $v_0,v$ is given by theory (equations \ref{eq:si-spheres-v0},\ref{eq:si-spheres-v} in section \ref{subsec:appendix-spheres-fields-and-slack-distribution}), and for a uniform sampling from a sphere $z \in\left[-1,1\right]$ has a bell-shaped distribution:
\begin{align}
P\left(z\right) & =\frac{1}{\sqrt{\pi}}\left(1-z^{2}\right)^{\frac{D-3}{2}}\Gamma\left(\frac{D}{2}\right)\big/\Gamma\left(\frac{D-1}{2}\right)
\end{align}
with moments $\left\langle z\right\rangle =0$ and $\left\langle \delta z^{2}\right\rangle =1/D$ (see details in section \ref{subsec:appendix-spheres-classification-errors}). 
In this setting classification error monotonically decreases with $c$ so the optimal value of  $\varepsilon_{tr}^{sample}$ is achieved for $c=\infty$. 

We now consider the classification error of points on the sphere in the presence of noise, where the classifier is trained on the entire manifold (i.e., with no noise), and tested on noisy  samples from the manifold. Assuming Gaussian noise ${\cal N}\left(0,\sigma^2/N\right)$ is added to each component of manifold samples, the fields are affected by noise ${\cal N}\left(0,\sigma^{2}q\right)$, so the probability of error in a sample is given by $P\left(h+\sigma\sqrt{q}\eta<0\right)$ where $\eta$ is standard Gaussian, and equivalently:
\begin{equation}
\varepsilon_{g}^{sample}=\left\langle H\left(\frac{v_{0}+R z v}{\sigma\sqrt{q}}\right)\right\rangle _{v_{0},v,z}\label{eq:spheres-test-error}
\end{equation}

\begin{figure}
\includegraphics[width=8.5cm]{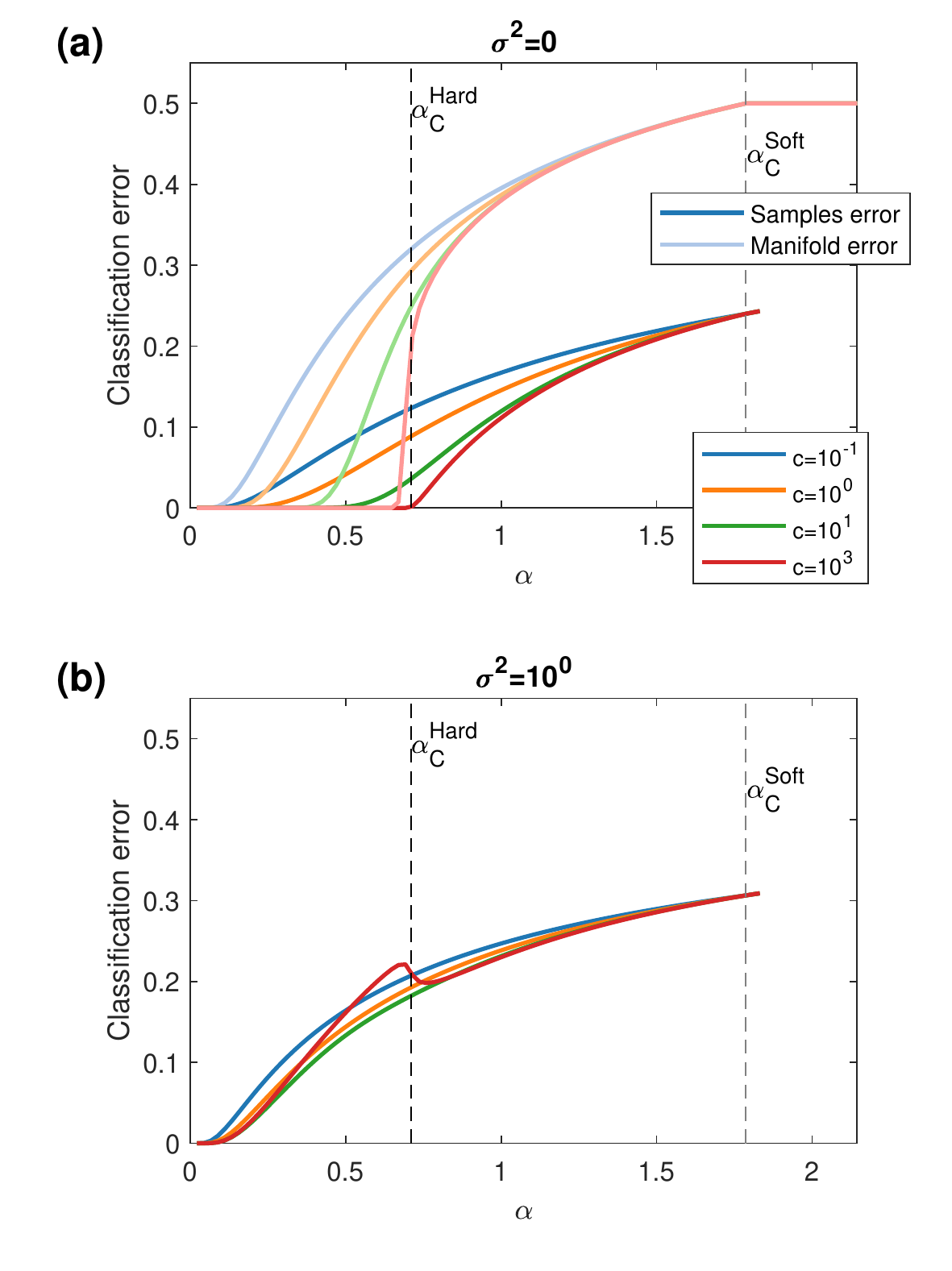}
\caption{\textbf{Errors in soft classification of spheres.}
Results for spheres of radius $R=0.25$ and dimension $D=10$.
(a) Classification error without noise (y-axis) for different values of $\alpha$ (x-axis) and choices of $c$ (color coded). Compares samples' classification error (dark lines) and entire-manifold classification error (light lines). 
(b) Classification error at noise level $\sigma^2=1$ (y-axis) for different values of $\alpha$ (x-axis) and choices of $c$ (color coded). }
\label{fig:spheres-errors}
\end{figure}

\paragraph*{Large D regime}

The regime of spheres with $D\gg1$ is important as real-world manifolds are expected to be high-dimensional, and in this regime it is possible to derive an SNR approximation of equation \ref{eq:spheres-test-error}.




When $R\sim O\left(1\right)$, $\alpha_{C}^{Soft}$ is close to $\alpha_{C}^{Hard}$ (see figure \ref{fig:spheres-capacity}). Thus in this regime the benefit of soft classification, in terms of the range of valid solutions, is small.
On the other hand, when $R\sqrt{D}\sim O\left(1\right)$, $\alpha_{C}^{Soft}$ can be much larger than $\alpha_{C}^{Hard}$ (figure \ref{fig:spheres-capacity}a), and thus we focus on this regime in our analysis of classification errors.

To derive an SNR approximation we assume that in this regime $v_{0}+R z v$ is approximately Gaussian, and that only the ``touching'' regime contributes to the error, thus substituting the values of $v_{0},v$ derived from the mean-field theory in that regime (equations \ref{eq:si-spheres-v0},\ref{eq:si-spheres-v} in section \ref{subsec:appendix-spheres-fields-and-slack-distribution}). The resulting SNR is provided in equations \ref{eq:si-spheres-snr-approx-sigma0}, \ref{eq:si-spheres-snr-approx} in section \ref{subsec:appendix-spheres-classification-errors}. 

Importantly, from this analysis we can calculate the limiting behavior of the SNR. In the $\alpha\to0$ limit the error anywhere on the manifold (equation \ref{eq:si-training-error-any} in section \ref{subsec:appendix-spheres-classification-errors}) scales as $\lim_{\alpha\to0}\varepsilon_{tr}^{manifold}=H\left(ck/\sqrt{q}\right)$, and using the order parameters in this limit leads to:
\begin{align}
\lim_{\alpha\to0}\varepsilon_{tr}^{manifold} & =H\left(\left(1+c\right)/\sqrt{\alpha}\right)\label{eq:sphere-error-alpha-to-0}
\end{align}
which is exactly the scaling for classification of the center points alone ($\varepsilon_{tr}^{centers}$, equation \ref{eq:points-snr-limits} with $\sigma^2=0$). Thus in this regime (i.e., $N\to\infty$) the manifold structure does not affect the classification error and furthermore the error in classification of the entire sphere is the same as the error in classification of samples $\varepsilon_{tr}^{sample}$, as the former is bounded between the two classification errors $\varepsilon_{tr}^{centers}\le\varepsilon_{tr}^{sample}\le\varepsilon_{tr}^{manifold}$.

On the other hand, in the $\alpha\to\alpha_{C}^{Soft}$ limit, from the scaling of $k,q$ in this limit (see section \ref{subsec:appendix-spheres-simplified-mf}) the error in classifying the entire manifold saturates (using equation \ref{eq:si-training-error-any} in section \ref{subsec:appendix-spheres-classification-errors}), but not the error classifying samples (see section \ref{subsec:appendix-spheres-classification-errors}):
\begin{align}
\lim_{\alpha\to\alpha_{C}^{Soft}}\varepsilon_{tr}^{manifold} & =H\left(0\right)=1/2\\
\lim_{\alpha\to\alpha_{C}^{Soft}}\varepsilon_{g}^{sample} & =H\left(\frac{R\sqrt{D}}{1+R^{2}}\frac{1}{\sqrt{1+\sigma^{2}}}\right)\label{eq:spheres-error-as-capacity}
\end{align}

Figure \ref{fig:spheres-errors}a presents both types of training errors and their dependence on $\alpha$ and $c$ at specific values of $R,D$, demonstrating that they are monotonically decreasing with $c$ and monotonically increasing with $\alpha$.
Figure \ref{fig:spheres-errors}b presents the test error at a specific noise level; unlike the training error, the test error is not monotonic in $c$ and thus is minimized for a finite value of $c$. Theory's agreement with empirical simulations is presented for different parameter values and choices of $c$ in figure \ref{fig:si-spheres-training-error} for the training error, and similarly in figure \ref{fig:si-spheres-test-error-per-sigma} for the test error.
Thus the theory predicts that errors at the phase transition are independent of $c$ (as seen in figure \ref{fig:spheres-errors}) and jump from this finite value to $0.5$ (and for a finite $N$ this transition is smoothed, as already discussed above).

\begin{figure}
\includegraphics[width=8.4cm]{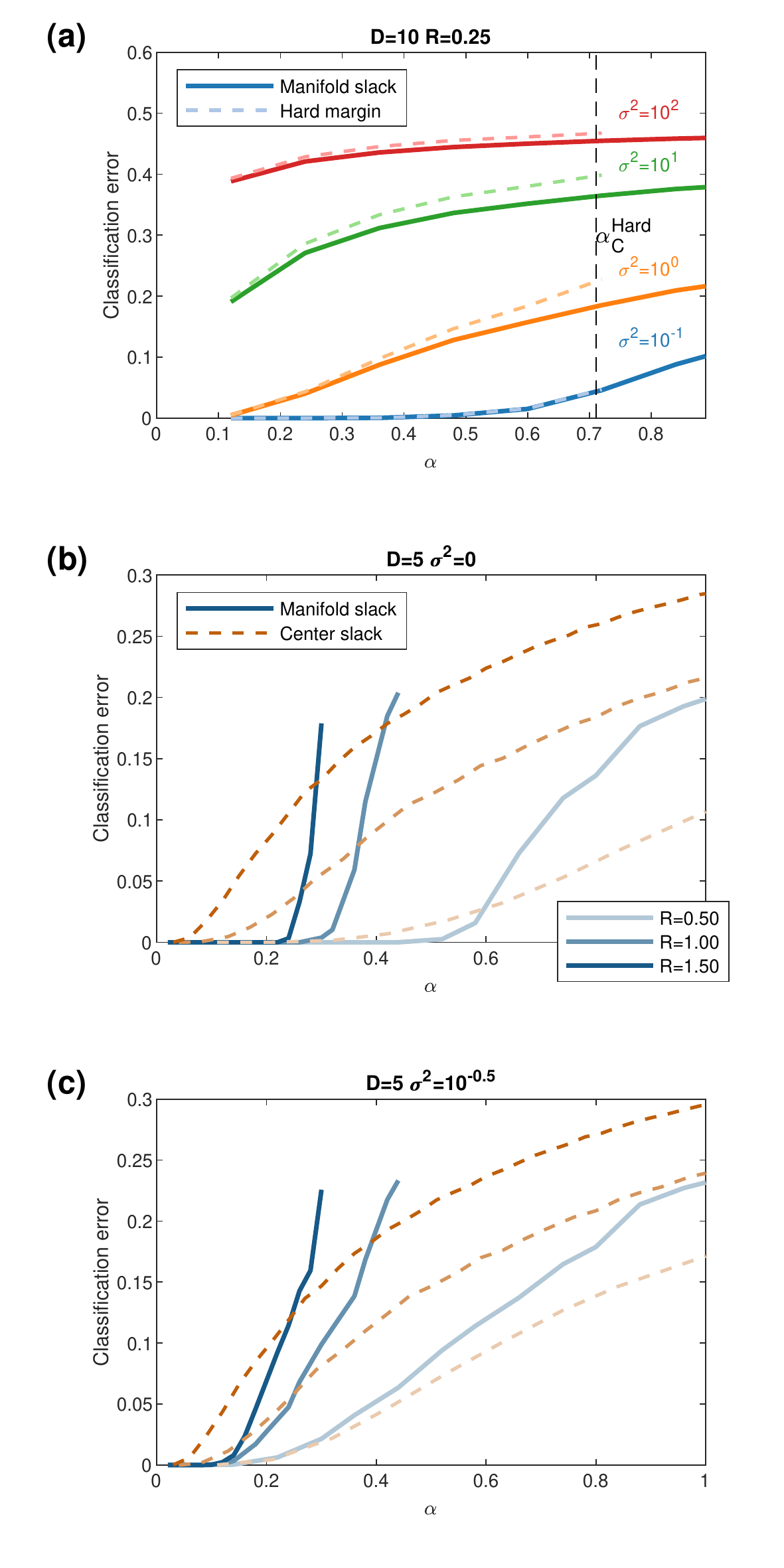}
\caption{\textbf{Comparison of classification errors for spheres using different methods.}
(a) Classification error (y-axis) using manifold-slacks (at the optimal choice of $c$, solid lines) or max-margin classification (dashed lines) at different values of $\alpha$ (x-axis) for radius $R=0.25$ and dimension $D=10$. Compares simulation results at different noise levels (color coded).
(b-c) Classification error using the optimal choice of $c$ (y-axis) for different values of $\alpha$ (x-axis) and values of $R$ (color coded), for dimension $D=5$. Compares simulation results of manifold-slack classifiers (solid lines) and center-slack classifiers (dashed lines), without noise (b) and with noise (c).
}
\label{fig:spheres-methods-comparison}
\end{figure}

The described theory can be used to choose the optimal value of $c$. Figure \ref{fig:si-spheres-test-error-optimal-c} compares, for different values of $R,D$ and noise levels, the optimal error achieved in simulations and by optimizing the theoretical value. Figure \ref{fig:si-spheres-methods-comparison} presents the optimal value of $c$ for different values of $\alpha$ and levels of noise, demonstrating a non-trivial behavior for manifold-slacks, unlike the monotonic behavior predicted by theory for center-slacks.

\paragraph*{Comparison with other methods}

Comparing the performance of the manifold-slack method with other methods requires optimization of the regularization value $c$ independently for each method. When there is no noise, below max-margin capacity $\alpha<\alpha_{C}^{Hard}$, the optimal choice of $c$ is infinite such that manifold-slack classification converges to max-margin classification. However, in the presence of noise the optimal value of $c$ is finite and using manifold-slacks reduces classification error relative to max-margin classification (figure \ref{fig:spheres-methods-comparison}a).
While the manifold-slack method is strictly better than the max-margin method due to choosing from a larger pool of classifiers, the improvement is usually small and is achieved toward $\alpha_C^{Hard}$ (see figure \ref{fig:si-spheres-manifold-max-comparison}). 

A systematic comparison of the manifold-slack and center-slack methods finds that manifold-slacks are better for small $\alpha$ values, with notable benefits at larger $R$ and smaller $\sigma$ values (see figure \ref{fig:spheres-methods-comparison}b-c).
Intuitively, when the noise is small, manifold-slacks may achieve near-zero error at a range of $\alpha$ values, while center-slacks performance depends on $R$ as a noise term and thus may be order $1$ when $R$ is order $1$.
For larger $\alpha$ values the performance of center-slacks surpasses that of manifold-slacks, and finally above $\alpha_{C}^{Soft}$ only the center-slack method is a viable option. Figure \ref{fig:si-spheres-methods-comparison-fields} presents the field distribution (at the manifold center) when using either center-slack or manifold-slack methods. The differences between those distributions provide intuition for the observed difference in the behavior of errors: at $\alpha_{C}^{Hard}$ the central fields using manifold-slacks are much larger than using center-slacks, but at $\alpha_{C}^{Soft}$ the central field distribution using manifold-slacks becomes $\delta\left(x\right)$.

As noted above, the point-slack method cannot in general be used for classification of manifolds with an infinite number of points. However, for classification of line segments (i.e., spheres with $D=1$), a correct classification of the $2P$ end-points is enough to classify the entire line. Figure \ref{fig:si-spheres-manifold-point-comparison-d1} compares manifold-slack with point-slack classification of the $2P$ end-points, both using the optimal choice of c for a given level of noise. The performance of point-slack SVM is usually close to that of the manifold-slack method, but provides a significant improvement toward $\alpha_{C}^{Soft}$.
It is interesting to observe that while using the manifold-slack method (with $P$ slack variables) there is a phase transition where the non-trivial classifier vanishes at a finite $\alpha$, there is no such transition using the point-slack method (with $2P$ slack variables), as expected from the point-slack theory (compare the weights' norms in figure \ref{fig:si-spheres-manifold-point-comparison-d1}a,b).

\subsection{Soft classification of general manifolds}

\subsubsection{Center-slack}

The center-slack method is straightforward to generalize to general manifolds, with the centers defined per our definition of a general manifold ($\boldsymbol{u}_{0}$ in equation \ref{eq:def-general-manifold}). A classifier trained on the centers would have a norm per points theory (equations \ref{eq:points-mf1}-\ref{eq:points-mf2}), and central field distribution per equation \ref{eq:point-fields}.

The probability of classification error for a point on the manifold $\boldsymbol{x}(\vec{S})$ would be $\varepsilon(\vec{S})=P\left(v_{0}+\vec{S}\cdot\vec{v}\le0\right)$ with $\vec{S}\cdot\vec{v}\sim{\cal N}\left(0,q\|\boldsymbol{x}(\vec{S})-\boldsymbol{u}_{0}\|^{2}\right)$. A calculation of classification error on a general manifold requires to make further assumptions on the sampling of $\vec{S}\in{\cal M}$ (see discussion). However, for the simple case of uniform sampling from a point-cloud manifold where $\boldsymbol{x}_{m}=\boldsymbol{u}_{0}+\delta\boldsymbol{x}_{m}$ for $m=1..M$ we have that: 
\begin{align}
\varepsilon & =\frac{1}{M}\sum_{m=1}^{M}\left\langle H\left(v_{0}/\sqrt{\left(\sigma^{2}+\|\delta\boldsymbol{x}_{m}\|^{2}\right)q}\right)\right\rangle _{v_{0}}\label{eq:general-manifolds-point-cloud-center-slack-error}
\end{align}
where $\sigma^{2}/N$ is the variance of Gaussian noise added to each component, which generalize equation \ref{eq:spheres-center-slack-error} from spheres, with the empirical $\|\delta\boldsymbol{x}_{m}\|^{2}$ taking the role of $R^{2}$. Furthermore, when the number of samples is large we expect self-averaging:
\begin{align}
\varepsilon & =\left\langle H\left(v_{0}/\sqrt{\left(\sigma^{2}+\hat{R}^{2}\right)q}\right)\right\rangle _{v_{0}}\label{eq:general-manifolds-point-cloud-center-slack-error-approx}
\end{align}
for $\hat{R}^{2}=\frac{1}{M}\sum_{m=1}^{M}\|\delta\boldsymbol{x}_{m}\|^{2}$ the total variance of the manifold points. Figure \ref{fig:si-general-manifolds-center-slack-empiric} compares the full theory (equation \ref{eq:general-manifolds-point-cloud-center-slack-error}) and the approximation (equation \ref{eq:general-manifolds-point-cloud-center-slack-error-approx}) to empirical measurement of the error using center-slacks.

\subsubsection{Manifold-slack}

\paragraph*{Replica theory}

Generalizing the mean-field equations of spheres (equations \ref{eq:spheres-mf-equation}-\ref{eq:spheres-inner-problem}) to the case of general manifolds, following the approach used by \cite{chung2018classification} for max-margin classifiers, the theory implies:
\begin{align}
L/q & =\frac{k-1}{k}+\frac{\alpha}{k}\int D^{D}\vec{t}\int Dt_{0}F\left(\vec{t},t_{0}\right)\\
F\left(\vec{t},t_{0}\right) & =\min_{\min_{S\in{\cal M} }\vec{v}\cdot\vec{S}\ge1/\sqrt{q}}\left\{ \|\vec{v}-\vec{t}\|^{2}+\frac{ck}{1+ck}\left(v_{0}-t_{0}\right)^{2}\right\}
\end{align}

Generalizing the notion of anchor points from \cite{chung2018classification}, we define them formally as the subgradient $\frac{\partial}{\partial v}$ of the support function $\min_{S\in{\cal M} }\vec{v}\cdot\vec{S}$:
\begin{align}
\tilde{S}\left(\vec{v}\right) & =\frac{\partial}{\partial v}\min_{S\in{\cal M} }\vec{v}\cdot\vec{S}
\end{align}
When the support function is differentiable, the subgradient is unique and is equivalent to the gradient \cite{chung2018classification}:
\begin{align}
\tilde{S}\left(\vec{v}\right) & =\arg\min_{\vec{S}\in {\cal M}}\vec{S}\cdot\vec{v}\label{eq:unique-Stilde}
\end{align}

For a given data manifold ${\cal M}^{\mu}$ and known values of $q,k$, one can sample from the anchor point distribution using the mean-field theory (see details in section \ref{subsec:appendix-general-manifold-iterative}):
\begin{align}
\tilde{S}\left(\vec{t},t_{0}\right) & =\frac{\vec{v}^{*}-\vec{t}}{\frac{ck}{1+ck}\left(v_{0}^{*}-t_{0}\right)}\label{eq:anchor-points}
\end{align}
where $\vec{v}^{*},v_{0}^{*}$ are the values which minimize $F\left(\vec{t},t_{0}\right)$, which can be found using least-squares optimization methods. This algorithm for sampling from the anchor point distribution is formally described in section \ref{subsec:algs-least-square-method-v-s-general-manifolds}.


\paragraph*{Large D regime}

For large $D$ we may define manifold properties $R_{M},D_{M}$ through the statistics of anchor points, as in max-margin classifiers \cite{chung2018classification}:
\begin{align}
R_{M}^{2} & =\left\langle \|\delta\tilde{S}\|^{2}\right\rangle _{\vec{t},t_{0}}\label{eq:Rm}\\
D_{M} & =\left\langle \left(\vec{t}\cdot\delta\tilde{S}\right)^{2}/\|\delta\tilde{S}\|^{2}\right\rangle _{\vec{t},t_{0}}\label{eq:Dm}
\end{align}
Now we may use $R_{M},D_{M}$ to solve for $q,k$ using the self-consistent equations from the theory of spheres. Thus for each value of $\alpha,c$ we can iteratively calculate $R_{M},D_{M}$ by sampling anchor points using the current values of $q,k$, then update the estimation of $q,k$, until convergence. This algorithm is formally described in section \ref{subsec:algs-iterative-general-manifolds}.

As was the case for spheres, when $D$ is large we expect only the ``touching'' regime to contribute, and from KKT condition applied to the minimization problem $F\left(\vec{t},t_{0}\right)$ we get a self-consistent relation:
\begin{align}
\vec{v} & =\vec{t}+\frac{ck}{1+ck}\left(1/\sqrt{q}-\vec{v}\cdot\tilde{S}-t_{0}\right)\tilde{S}\label{eq:touching-regime-iteration}
\end{align}
Thus equations \ref{eq:unique-Stilde},\ref{eq:touching-regime-iteration} can be used to iteratively update $\vec{v}$ and $\tilde{S}$ (see section \ref{subsec:appendix-general-manifold-iterative}). This iterative approach allows for finding the anchor points without solving a least-squares optimization problem for each value of $\vec{t},t_{0}$ (as the least-squares algorithm, section \ref{subsec:algs-least-square-method-v-s-general-manifolds}). This algorithm is formally described in section \ref{subsec:algs-iterative-method-v-s-general-manifolds}.

\begin{figure*}
\includegraphics[width=17.5cm]{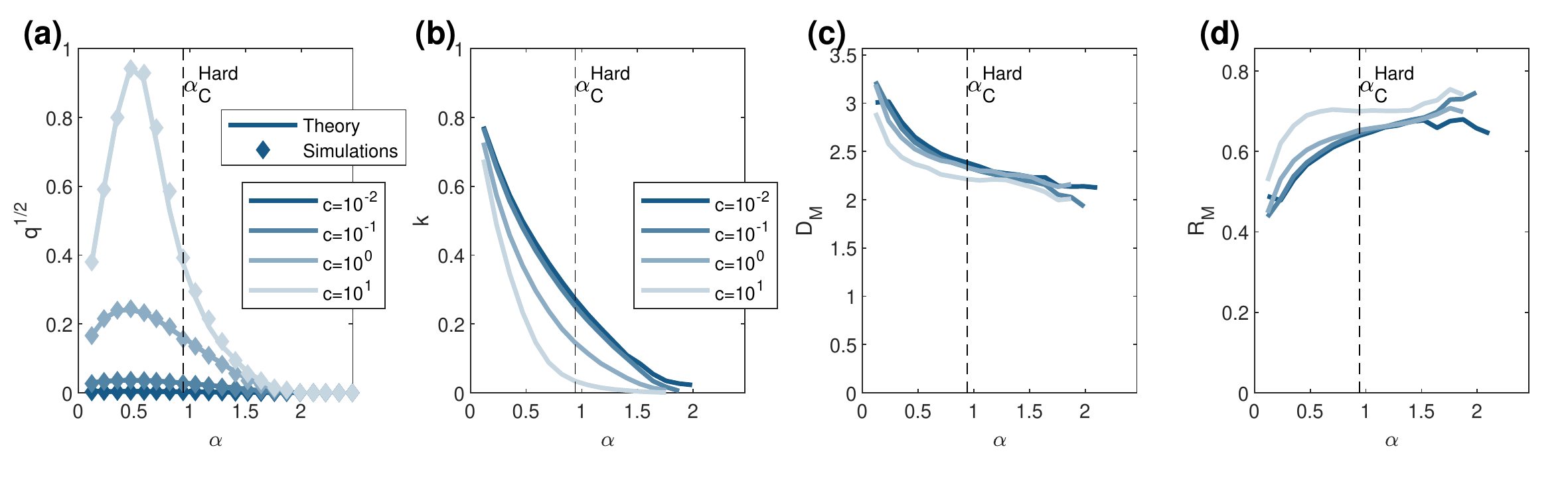}
\caption{\textbf{Order parameters and manifold properties for point-cloud manifolds.} 
Sampling $M=100$ points from an ellipsoid with $\gamma=1.5$, $R=0.25, D=20$. 
(a) The weights' norm $q^{1/2}$ (y-axis) for different values of $\alpha$ (x-axis) and choices of $c$ (color coded). Compares theory (solid lines) and simulation results (diamonds).
(b-d) The corresponding values of the order parameter $k$ (b), manifold dimension $D_{M}$ (c) and manifold radius $R_{M}$ (d) (y-axis) for different values of $\alpha$ (x-axis) and choices of c (color coded).}
\label{fig:general-manifolds-properties}
\end{figure*}

To use a concrete example, for simulations of general manifolds we have used point-cloud manifolds created by sampling $M$ points from a $D$-dimensional ellipsoid with radii $r_{l}\sim l^{-\gamma}$. Denoting $R^{2}=\sum_{l=1}^{D}r_{l}^{2}$ the ellipsoid shape is defined by parameters $R,D,\gamma$. Figures \ref{fig:general-manifolds-properties}a-b, \ref{fig:si-general-manifolds-properties}a-b demonstrate the existence of finite capacity when using manifold-slacks also for those manifolds. The predicted values of $q$ matches the empirically observed values, which vanish at a finite $\alpha$ value (figures \ref{fig:general-manifolds-properties}a, \ref{fig:si-general-manifolds-properties}a). The dependence of the measured $D_{M}$ on $c$ and $\alpha$ is quite small (see figures \ref{fig:general-manifolds-properties}c, \ref{fig:si-general-manifolds-properties}c) and similarly for the measured $R_{M}$ (see figures \ref{fig:general-manifolds-properties}d,  \ref{fig:si-general-manifolds-properties}d). 

Figure \ref{fig:si-general-manifolds-properties2} presents the weights' norm for the classification of point-cloud manifolds and the theoretical values predicted for $q,k,R_{M},D_{M}$, using either the iterative or the least-squares algorithm.
The two algorithms give very similar results, with a notable difference at large $R$ where the assumption that only the ``touching'' regime contributes to the solution no longer holds. 

As it is favorable to have manifold properties $D_{M}$ and $R_{M}$ which do not depend on $\alpha$, figure \ref{fig:si-general-manifolds-q-k-vs-alpha} shows that using a single choice of $D_{M},R_{M}$, calculated for $\alpha$ near $\alpha_{C}^{Soft}$ (i.e., at the largest solvable values) to predict $q$ provides a good approximation for the entire range of $\alpha$ (but not using a single choice of $D_{M},R_{M}$ calculated from a small $\alpha$ value). 

\paragraph*{Classification errors}

For general manifolds, the classification error is defined assuming manifold points are sampled according to some measure on the manifold (see discussion); for the simpler case of point-cloud manifolds, we assume this is a uniform distribution. 

\begin{figure*}
\includegraphics[width=17.5cm]{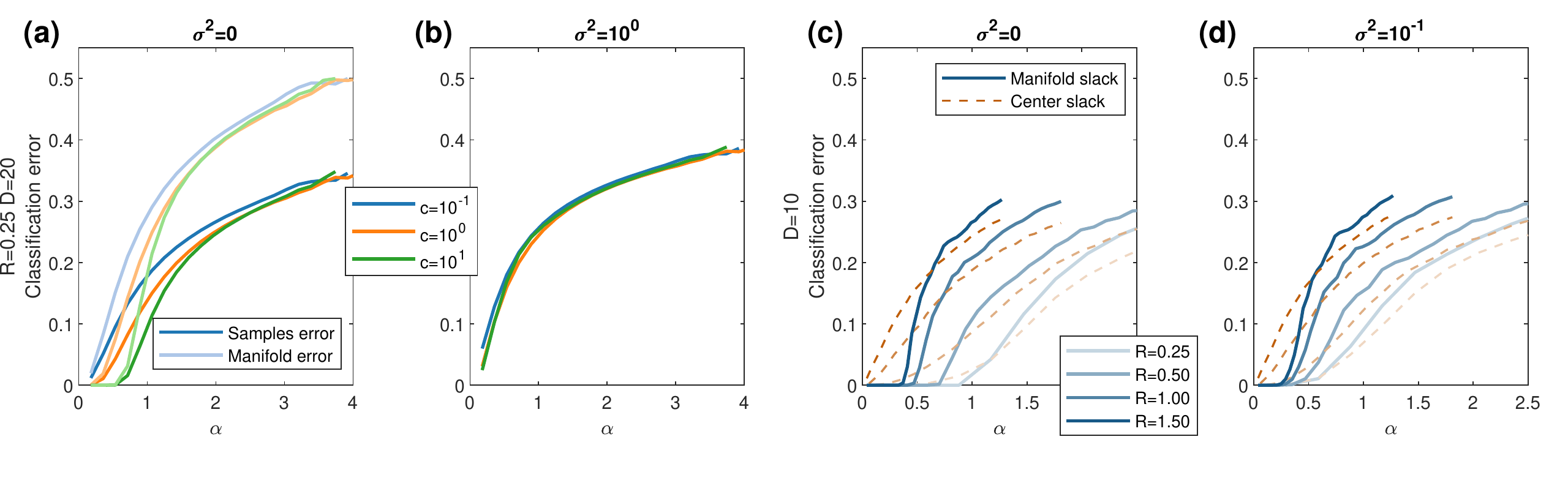}
\caption{\textbf{Errors in soft classification of point-cloud manifolds.} 
(a-b) Results for manifolds of $M=100$ points from an ellipsoid with $\gamma=1.5$, radius $R=0.25$ and dimension $D=20$.
(a) Classification error when using manifold-slacks without noise (y-axis) for different values of $\alpha$ (x-axis) and choices of $c$ (color coded). Compares samples' classification error (dark lines) and entire-manifold classification error (light lines). 
(b) Classification error when using manifold-slacks (y-axis) for different values of $\alpha$ (x-axis) and choices of $c$ (color coded), at noise level of $\sigma^{2}=1$.
(c-d) Results for manifolds of $M=100$ points from an ellipsoid with $\gamma=1.5$, radius $R$ and dimension $D=10$. Classification error using the optimal choice of $c$ (y-axis) for different values of $\alpha$ (x-axis)  and values of $R$ (color coded). Compares simulation results of manifold-slack classifiers (solid lines) and center-slack classifiers (dashed lines), without noise (c) and at noise level $\sigma^2=0.1$ (d).}
\label{fig:general-manifolds-error}
\end{figure*}

Figure \ref{fig:general-manifolds-error}a presents the predictions for the training error using the theory of spheres (equations \ref{eq:si-training-error-any}, \ref{eq:si-spheres-snr-approx-sigma0}, \ref{eq:si-spheres-snr-approx} in section \ref{subsec:appendix-spheres-classification-errors}) where the theoretical values of $q,k,R_{M},D_{M}$, calculated using the least-squares algorithm, are plugged-in. Figure \ref{fig:si-general-manifolds-training-error} compares the training error predicted using the theory of spheres classification with the error measured in simulations, finding good match.
Similarly, figure \ref{fig:general-manifolds-error}b presents the predicted test error for specific noise level and different choices of $c$, and figure \ref{fig:si-general-manifolds-generalization-error} compares the predicted test error for several noise levels and choices of $c$ with simulation results, demonstrating again the applicability of measuring the manifolds' $R_M$, $D_M$ and plugging them into the equations from the theory of spheres classification to make predictions regarding non-spherical manifolds.

\paragraph*{Comparison with other methods}
Comparing the performance of different classification methods on point-cloud manifolds reveals a similar behavior to that observed for spheres. Figure \ref{fig:si-general-manifolds-generalization-error-optimal-c} compares the manifold-slack method with both center-slack and max-margin methods, using the optimal choice of $c$ for each method. Below $\alpha_C^{Hard}$ manifold-slack classification exhibits improved performance compared to max-margin classification, but this improvement is usually small (figure  \ref{fig:si-general-manifolds-generalization-error-optimal-c}a-b). Figures \ref{fig:general-manifolds-error}c-d shows that as in the case of spheres, for small $\alpha$ values manifold-slacks are superior to center-slacks, with large qualitative difference at low noise level when $R$ is order $1$, while for larger $\alpha$ values the performance of the center-slack method is better (see additional noise levels in figure \ref{fig:si-general-manifolds-generalization-error-optimal-c}c).

For point-cloud manifolds, when the number of samples per manifold is not too large, the point-slack method can also be used for manifold classification. 
Figure \ref{fig:si-general-manifolds-points-slack-vs-manifold-slack-error-optimalC}a-d shows that using the point-slack method, there is no phase-transition to zero weights as for the manifold-slack method. Despite this marked difference, the classification error achieved by the point-slack method is only slightly better than that achieved by the manifold-slack method (both using the optimal choice of $c$, figure \ref{fig:si-general-manifolds-points-slack-vs-manifold-slack-error-optimalC}e-f). This improvement is significant only at small levels of noise and towards $\alpha_{C}^{Soft}$. Thus point-slack SVM uses the additional degrees of freedom (and additional computational costs) from assigning a separate slack variable per sample to slightly outperform the manifold-slack method.
\section{Discussion} \label{sec:discussion}

The introduction of slack variables to SVMs allows linear classification of data which is not linearly separable, and for optimizing performance by choosing the right balance between making training errors and increasing classification margin (using the regularization parameter $c$, equations \ref{eq:points-optimization-target},\ref{eq:center-slack-optimization-target},\ref{eq:manifold-slack-optimization-target}). Here we analyze the noise resilience of such classification by considering test performance with respect to input noise (with variance $\sigma^2/N$ applied to each input component).

\paragraph*{Point-slack}
We first study the statistical mechanics of a point-slack model where a set of $P$ random points in $N$ dimensions are independently labeled, and each is assigned a slack variable. We show that the problem has a well defined solution for all load values $\alpha=P/N$ (figure \ref{fig:points-classification}). 
In the absence of input noise, the optimal choice of $c$ is infinite for all $\alpha$; however, in the presence of noise in the test data, the optimal $c$ is finite (figure \ref{fig:points-classification-field-errors}). Furthermore, the optimal choice of $c$ can be calculated from theory (equation \ref{eq:points-optimal-c}), and is roughly given by the ``canonical choice'' $c=\sigma^{-2}$ (figure \ref{fig:points-optimal-c}).

\paragraph*{Manifold classification}
Our main interest is the case of points arranged in $P$ randomly labeled manifolds, such that all points within a manifold have the same target label. Assuming the number of points per manifold is large (and possibly infinite) assigning a slack variable to each point is not feasible.  We introduced and analyzed two schemes of slack algorithms for classification of manifolds, which differ in the manner in which slack variables are attached to manifolds. In the center-slack method, each manifold center is associated with a slack variable, reducing the learning to point-slack SVM of the centers. In the manifold-slack method, a slack variable is associated with the ``worst'' point in each manifold, relative to the separating hyper-plane. 
The relation between slack variables and errors is different in the two methods (figure \ref{fig:illustration}); when using center-slacks, if the center is misclassified, most of the manifold may follow, but using manifold-slacks most of it may be classified correctly even if the ``worst'' point is not.

\paragraph*{Center-slack}
The relatively simple center-slack scheme has several attractive features. First, it has a well defined, non-zero, solution for the weights for all values of $\alpha$. Second, the associated optimal $c$ is provided by theory (figure \ref{fig:spheres-center-slack}) and is approximately given by the simple ``canonical choice'' $c=(R^2+\sigma^2)^{-1}$, where $R$ is the manifold radius, expressing the intuition that the variability of the manifold data relative to the center (quantified by $R^2$) is an intrinsic noise on top of the extrinsic noise $\sigma^2$. Finally, for large $\alpha$ values its performance is superior to the more sophisticated manifold-slack method (figures \ref{fig:spheres-methods-comparison}b-c, \ref{fig:general-manifolds-error}c-d), as discussed below. 
The disadvantages of the center-slack method are its performance for small $\alpha$ values and that it does not generalize max-margin manifold classification.


\paragraph*{Manifold-slack}
The manifold-slack scheme is a natural extension of max-margin manifold classification \cite{chung2016linear,chung2018classification} in which the optimal weight vector is a sum of anchor points, one per manifold, which are the closest points in each manifold to the separating hyper-plane. Here each such point is assigned a slack variable. For $\alpha$ below the error-less classification capacity $\alpha_{C}^{Hard}$, when $c$ approaches $\infty$, manifold-slack classification approaches max-margin classification. However, the optimal $c$ may not be infinite even in this $\alpha$ regime in the presence of noise (figure \ref{fig:spheres-errors}). As for larger values of $\alpha$, a surprising result of our mean-field theory is that the manifold-slack method possesses a solution with non-zero weight vector only below a second critical value, $\alpha_{C}^{Soft}$ (figure \ref{fig:spheres-q-k}). Thus, this method allows for extending the range of linear classification above the error-less capacity, but for a limited range (figure \ref{fig:spheres-capacity}).

The classification-error performance of manifold-slacks is always better than max-margin and may be superior to center-slacks, depending on parameters.
The main improvement over max-margin is the extended range of $\alpha$ values (figure \ref{fig:spheres-capacity}), as the reduction of the classification error (below max-margin capacity) is usually small (figures \ref{fig:spheres-methods-comparison}a, \ref{fig:si-spheres-manifold-max-comparison}, \ref{fig:si-general-manifolds-generalization-error-optimal-c}a-b). 
The improved performance compared to center-slacks is substantial for small $\alpha$ values when the noise is small and $R$ is order $1$, where manifold-slacks achieves near-zero error while center-slacks error is order $1$ (figures \ref{fig:spheres-methods-comparison}b-c, \ref{fig:general-manifolds-error}c-d, \ref{fig:si-general-manifolds-generalization-error-optimal-c}c). 

While many of the results for manifolds were derived in the context of spheres, the theory extends well to general manifolds by recovering their effective radius and dimension (equations \ref{eq:Rm}, \ref{eq:Dm},  figures \ref{fig:general-manifolds-properties}, \ref{fig:si-general-manifolds-properties}). Importantly, their classification performance is well predicted by plugging those values into the theory of spheres (figures \ref{fig:si-general-manifolds-training-error}, \ref{fig:si-general-manifolds-generalization-error}), thus demonstrating they capture the classification-relevant aspects of manifolds' geometry.

\paragraph*{Measure on manifolds}
The use of manifold-slacks benefits from being insensitive to the exact measure assumed on the manifolds (as long as it is non-zero). In the case of center-slacks, the center of mass of the manifolds depends in general on the measure. Nevertheless, in some cases, there is a natural choice for the center, as in spheres or ellipsoids (due to symmetry), or in a points-cloud, where using the points' average corresponds to a uniform measure on the points. Furthermore, one can use the measure-independent Steiner point \cite{shephard1966steiner} as the manifold center. Regardless of the employed classification method, the evaluation of the errors depends in general on the measure.

\paragraph*{Future work}
Extending the theory of max-margin classification of manifolds to soft classification is an important step in connecting the theory to applications, where soft-margin classifiers are more commonly used.
We believe the theory of general manifolds is relevant for the analysis of real-world data. To properly do so, the theory needs to be extended to allow for center correlations, as was done for max-margin classifiers \cite{cohen2020separability}; we expect this to be straightforward as the methods from \cite{cohen2020separability} involve mostly preprocessing of the manifolds, independently of the analysis of the manifolds' geometry.

The issue of robustness to noise would naturally come up when aiming to apply the theory to neural data analysis where noise is a common attribute of the problem, unlike the artificial networks analyzed in \cite{cohen2020separability}. It would be interesting to apply the methods described here to analyze object representations with non-Gaussian noise, such as neural noise with Poisson-like characteristics.

On a broader scope, the discussion of robustness to noise is a limited form of generalization. In general, we would like to be able to discuss generalization with respect to a finite number of samples from a manifold, where the scaling behavior of the classification error with the number of samples is an open question. Recent work on the few-shot learning setup, where the number of samples is very small, has revealed relatively simple behavior of the classification error \cite{sorscher2021geometry}.

\section*{Acknowledgements} \label{sec:acknowledgements}
HS is partially supported by the Gatsby Charitable Foundation, the Swartz Foundation, the National Institutes of Health (Grant No. 1U19NS104653) and the MAFAT Center for Deep Learning.
\bigskip
\bibliography{library}

\begin{thebibliography}{20}%
\makeatletter
\providecommand \@ifxundefined [1]{%
 \@ifx{#1\undefined}
}%
\providecommand \@ifnum [1]{%
 \ifnum #1\expandafter \@firstoftwo
 \else \expandafter \@secondoftwo
 \fi
}%
\providecommand \@ifx [1]{%
 \ifx #1\expandafter \@firstoftwo
 \else \expandafter \@secondoftwo
 \fi
}%
\providecommand \natexlab [1]{#1}%
\providecommand \enquote  [1]{``#1''}%
\providecommand \bibnamefont  [1]{#1}%
\providecommand \bibfnamefont [1]{#1}%
\providecommand \citenamefont [1]{#1}%
\providecommand \href@noop [0]{\@secondoftwo}%
\providecommand \href [0]{\begingroup \@sanitize@url \@href}%
\providecommand \@href[1]{\@@startlink{#1}\@@href}%
\providecommand \@@href[1]{\endgroup#1\@@endlink}%
\providecommand \@sanitize@url [0]{\catcode `\\12\catcode `\$12\catcode
  `\&12\catcode `\#12\catcode `\^12\catcode `\_12\catcode `\%12\relax}%
\providecommand \@@startlink[1]{}%
\providecommand \@@endlink[0]{}%
\providecommand \url  [0]{\begingroup\@sanitize@url \@url }%
\providecommand \@url [1]{\endgroup\@href {#1}{\urlprefix }}%
\providecommand \urlprefix  [0]{URL }%
\providecommand \Eprint [0]{\href }%
\providecommand \doibase [0]{https://doi.org/}%
\providecommand \selectlanguage [0]{\@gobble}%
\providecommand \bibinfo  [0]{\@secondoftwo}%
\providecommand \bibfield  [0]{\@secondoftwo}%
\providecommand \translation [1]{[#1]}%
\providecommand \BibitemOpen [0]{}%
\providecommand \bibitemStop [0]{}%
\providecommand \bibitemNoStop [0]{.\EOS\space}%
\providecommand \EOS [0]{\spacefactor3000\relax}%
\providecommand \BibitemShut  [1]{\csname bibitem#1\endcsname}%
\let\auto@bib@innerbib\@empty
\bibitem [{\citenamefont {Vapnik}\ and\ \citenamefont
  {Lerner}(1963)}]{vapnik1963recognition}%
  \BibitemOpen
  \bibfield  {author} {\bibinfo {author} {\bibfnamefont {V.}~\bibnamefont
  {Vapnik}}\ and\ \bibinfo {author} {\bibfnamefont {A.~Y.}\ \bibnamefont
  {Lerner}},\ }\bibfield  {title} {\bibinfo {title} {Recognition of patterns
  with help of generalized portraits},\ }\href@noop {} {\bibfield  {journal}
  {\bibinfo  {journal} {Avtomat. i Telemekh}\ }\textbf {\bibinfo {volume}
  {24}},\ \bibinfo {pages} {774} (\bibinfo {year} {1963})}\BibitemShut
  {NoStop}%
\bibitem [{\citenamefont {Boser}\ \emph {et~al.}(1992)\citenamefont {Boser},
  \citenamefont {Guyon},\ and\ \citenamefont {Vapnik}}]{boser1992training}%
  \BibitemOpen
  \bibfield  {author} {\bibinfo {author} {\bibfnamefont {B.~E.}\ \bibnamefont
  {Boser}}, \bibinfo {author} {\bibfnamefont {I.~M.}\ \bibnamefont {Guyon}},\
  and\ \bibinfo {author} {\bibfnamefont {V.~N.}\ \bibnamefont {Vapnik}},\
  }\bibfield  {title} {\bibinfo {title} {A training algorithm for optimal
  margin classifiers},\ }in\ \href@noop {} {\emph {\bibinfo {booktitle}
  {Proceedings of the fifth annual workshop on Computational learning
  theory}}}\ (\bibinfo {year} {1992})\ pp.\ \bibinfo {pages}
  {144--152}\BibitemShut {NoStop}%
\bibitem [{\citenamefont {Cortes}\ and\ \citenamefont
  {Vapnik}(1995)}]{cortes1995support}%
  \BibitemOpen
  \bibfield  {author} {\bibinfo {author} {\bibfnamefont {C.}~\bibnamefont
  {Cortes}}\ and\ \bibinfo {author} {\bibfnamefont {V.}~\bibnamefont
  {Vapnik}},\ }\bibfield  {title} {\bibinfo {title} {Support-vector networks},\
  }\href@noop {} {\bibfield  {journal} {\bibinfo  {journal} {Machine learning}\
  }\textbf {\bibinfo {volume} {20}},\ \bibinfo {pages} {273} (\bibinfo {year}
  {1995})}\BibitemShut {NoStop}%
\bibitem [{\citenamefont {Sch{\"o}lkopf}\ \emph {et~al.}(2000)\citenamefont
  {Sch{\"o}lkopf}, \citenamefont {Smola}, \citenamefont {Williamson},\ and\
  \citenamefont {Bartlett}}]{scholkopf2000new}%
  \BibitemOpen
  \bibfield  {author} {\bibinfo {author} {\bibfnamefont {B.}~\bibnamefont
  {Sch{\"o}lkopf}}, \bibinfo {author} {\bibfnamefont {A.~J.}\ \bibnamefont
  {Smola}}, \bibinfo {author} {\bibfnamefont {R.~C.}\ \bibnamefont
  {Williamson}},\ and\ \bibinfo {author} {\bibfnamefont {P.~L.}\ \bibnamefont
  {Bartlett}},\ }\bibfield  {title} {\bibinfo {title} {New support vector
  algorithms},\ }\href@noop {} {\bibfield  {journal} {\bibinfo  {journal}
  {Neural computation}\ }\textbf {\bibinfo {volume} {12}},\ \bibinfo {pages}
  {1207} (\bibinfo {year} {2000})}\BibitemShut {NoStop}%
\bibitem [{\citenamefont {Gardner}(1988)}]{gardner1988space}%
  \BibitemOpen
  \bibfield  {author} {\bibinfo {author} {\bibfnamefont {E.}~\bibnamefont
  {Gardner}},\ }\bibfield  {title} {\bibinfo {title} {The space of interactions
  in neural network models},\ }\href@noop {} {\bibfield  {journal} {\bibinfo
  {journal} {Journal of physics A: Mathematical and general}\ }\textbf
  {\bibinfo {volume} {21}},\ \bibinfo {pages} {257} (\bibinfo {year}
  {1988})}\BibitemShut {NoStop}%
\bibitem [{\citenamefont {Chung}\ \emph {et~al.}(2016)\citenamefont {Chung},
  \citenamefont {Lee},\ and\ \citenamefont {Sompolinsky}}]{chung2016linear}%
  \BibitemOpen
  \bibfield  {author} {\bibinfo {author} {\bibfnamefont {S.}~\bibnamefont
  {Chung}}, \bibinfo {author} {\bibfnamefont {D.~D.}\ \bibnamefont {Lee}},\
  and\ \bibinfo {author} {\bibfnamefont {H.}~\bibnamefont {Sompolinsky}},\
  }\bibfield  {title} {\bibinfo {title} {Linear readout of object manifolds},\
  }\href@noop {} {\bibfield  {journal} {\bibinfo  {journal} {Physical Review
  E}\ }\textbf {\bibinfo {volume} {93}},\ \bibinfo {pages} {060301} (\bibinfo
  {year} {2016})}\BibitemShut {NoStop}%
\bibitem [{\citenamefont {Chung}\ \emph {et~al.}(2018)\citenamefont {Chung},
  \citenamefont {Lee},\ and\ \citenamefont
  {Sompolinsky}}]{chung2018classification}%
  \BibitemOpen
  \bibfield  {author} {\bibinfo {author} {\bibfnamefont {S.}~\bibnamefont
  {Chung}}, \bibinfo {author} {\bibfnamefont {D.~D.}\ \bibnamefont {Lee}},\
  and\ \bibinfo {author} {\bibfnamefont {H.}~\bibnamefont {Sompolinsky}},\
  }\bibfield  {title} {\bibinfo {title} {Classification and geometry of general
  perceptual manifolds},\ }\href@noop {} {\bibfield  {journal} {\bibinfo
  {journal} {Physical Review X}\ }\textbf {\bibinfo {volume} {8}},\ \bibinfo
  {pages} {031003} (\bibinfo {year} {2018})}\BibitemShut {NoStop}%
\bibitem [{\citenamefont {Cohen}\ \emph {et~al.}(2020)\citenamefont {Cohen},
  \citenamefont {Chung}, \citenamefont {Lee},\ and\ \citenamefont
  {Sompolinsky}}]{cohen2020separability}%
  \BibitemOpen
  \bibfield  {author} {\bibinfo {author} {\bibfnamefont {U.}~\bibnamefont
  {Cohen}}, \bibinfo {author} {\bibfnamefont {S.}~\bibnamefont {Chung}},
  \bibinfo {author} {\bibfnamefont {D.~D.}\ \bibnamefont {Lee}},\ and\ \bibinfo
  {author} {\bibfnamefont {H.}~\bibnamefont {Sompolinsky}},\ }\bibfield
  {title} {\bibinfo {title} {Separability and geometry of object manifolds in
  deep neural networks},\ }\href@noop {} {\bibfield  {journal} {\bibinfo
  {journal} {Nature communications}\ }\textbf {\bibinfo {volume} {11}},\
  \bibinfo {pages} {1} (\bibinfo {year} {2020})}\BibitemShut {NoStop}%
\bibitem [{\citenamefont {Shawe-Taylor}\ and\ \citenamefont
  {Cristianini}(2002)}]{shawe2002generalization}%
  \BibitemOpen
  \bibfield  {author} {\bibinfo {author} {\bibfnamefont {J.}~\bibnamefont
  {Shawe-Taylor}}\ and\ \bibinfo {author} {\bibfnamefont {N.}~\bibnamefont
  {Cristianini}},\ }\bibfield  {title} {\bibinfo {title} {On the generalization
  of soft margin algorithms},\ }\href@noop {} {\bibfield  {journal} {\bibinfo
  {journal} {IEEE Transactions on Information Theory}\ }\textbf {\bibinfo
  {volume} {48}},\ \bibinfo {pages} {2721} (\bibinfo {year}
  {2002})}\BibitemShut {NoStop}%
\bibitem [{\citenamefont {Chen}\ \emph {et~al.}(2004)\citenamefont {Chen},
  \citenamefont {Wu}, \citenamefont {Ying},\ and\ \citenamefont
  {Zhou}}]{chen2004support}%
  \BibitemOpen
  \bibfield  {author} {\bibinfo {author} {\bibfnamefont {D.-R.}\ \bibnamefont
  {Chen}}, \bibinfo {author} {\bibfnamefont {Q.}~\bibnamefont {Wu}}, \bibinfo
  {author} {\bibfnamefont {Y.}~\bibnamefont {Ying}},\ and\ \bibinfo {author}
  {\bibfnamefont {D.-X.}\ \bibnamefont {Zhou}},\ }\bibfield  {title} {\bibinfo
  {title} {Support vector machine soft margin classifiers: error analysis},\
  }\href@noop {} {\bibfield  {journal} {\bibinfo  {journal} {The Journal of
  Machine Learning Research}\ }\textbf {\bibinfo {volume} {5}},\ \bibinfo
  {pages} {1143} (\bibinfo {year} {2004})}\BibitemShut {NoStop}%
\bibitem [{\citenamefont {Risau-Gusman}\ and\ \citenamefont
  {Gordon}(2001)}]{risau2001statistical}%
  \BibitemOpen
  \bibfield  {author} {\bibinfo {author} {\bibfnamefont {S.}~\bibnamefont
  {Risau-Gusman}}\ and\ \bibinfo {author} {\bibfnamefont {M.~B.}\ \bibnamefont
  {Gordon}},\ }\bibfield  {title} {\bibinfo {title} {Statistical mechanics of
  learning with soft margin classifiers},\ }\href@noop {} {\bibfield  {journal}
  {\bibinfo  {journal} {Physical Review E}\ }\textbf {\bibinfo {volume} {64}},\
  \bibinfo {pages} {031907} (\bibinfo {year} {2001})}\BibitemShut {NoStop}%
\bibitem [{\citenamefont {Dietrich}\ \emph {et~al.}(1999)\citenamefont
  {Dietrich}, \citenamefont {Opper},\ and\ \citenamefont
  {Sompolinsky}}]{dietrich1999statistical}%
  \BibitemOpen
  \bibfield  {author} {\bibinfo {author} {\bibfnamefont {R.}~\bibnamefont
  {Dietrich}}, \bibinfo {author} {\bibfnamefont {M.}~\bibnamefont {Opper}},\
  and\ \bibinfo {author} {\bibfnamefont {H.}~\bibnamefont {Sompolinsky}},\
  }\bibfield  {title} {\bibinfo {title} {Statistical mechanics of support
  vector networks},\ }\href@noop {} {\bibfield  {journal} {\bibinfo  {journal}
  {Physical review letters}\ }\textbf {\bibinfo {volume} {82}},\ \bibinfo
  {pages} {2975} (\bibinfo {year} {1999})}\BibitemShut {NoStop}%
\bibitem [{\citenamefont {M{\'e}zard}\ \emph {et~al.}(2002)\citenamefont
  {M{\'e}zard}, \citenamefont {Parisi},\ and\ \citenamefont
  {Zecchina}}]{mezard2002analytic}%
  \BibitemOpen
  \bibfield  {author} {\bibinfo {author} {\bibfnamefont {M.}~\bibnamefont
  {M{\'e}zard}}, \bibinfo {author} {\bibfnamefont {G.}~\bibnamefont {Parisi}},\
  and\ \bibinfo {author} {\bibfnamefont {R.}~\bibnamefont {Zecchina}},\
  }\bibfield  {title} {\bibinfo {title} {Analytic and algorithmic solution of
  random satisfiability problems},\ }\href@noop {} {\bibfield  {journal}
  {\bibinfo  {journal} {Science}\ }\textbf {\bibinfo {volume} {297}},\ \bibinfo
  {pages} {812} (\bibinfo {year} {2002})}\BibitemShut {NoStop}%
\bibitem [{\citenamefont {Zdeborov{\'a}}\ and\ \citenamefont
  {Krz{\k{a}}ka{\l}a}(2007)}]{zdeborova2007phase}%
  \BibitemOpen
  \bibfield  {author} {\bibinfo {author} {\bibfnamefont {L.}~\bibnamefont
  {Zdeborov{\'a}}}\ and\ \bibinfo {author} {\bibfnamefont {F.}~\bibnamefont
  {Krz{\k{a}}ka{\l}a}},\ }\bibfield  {title} {\bibinfo {title} {Phase
  transitions in the coloring of random graphs},\ }\href@noop {} {\bibfield
  {journal} {\bibinfo  {journal} {Physical Review E}\ }\textbf {\bibinfo
  {volume} {76}},\ \bibinfo {pages} {031131} (\bibinfo {year}
  {2007})}\BibitemShut {NoStop}%
\bibitem [{\citenamefont {Ganguli}\ and\ \citenamefont
  {Sompolinsky}(2010)}]{ganguli2010statistical}%
  \BibitemOpen
  \bibfield  {author} {\bibinfo {author} {\bibfnamefont {S.}~\bibnamefont
  {Ganguli}}\ and\ \bibinfo {author} {\bibfnamefont {H.}~\bibnamefont
  {Sompolinsky}},\ }\bibfield  {title} {\bibinfo {title} {Statistical mechanics
  of compressed sensing},\ }\href@noop {} {\bibfield  {journal} {\bibinfo
  {journal} {Physical review letters}\ }\textbf {\bibinfo {volume} {104}},\
  \bibinfo {pages} {188701} (\bibinfo {year} {2010})}\BibitemShut {NoStop}%
\bibitem [{\citenamefont {Advani}\ and\ \citenamefont
  {Ganguli}(2016)}]{advani2016statistical}%
  \BibitemOpen
  \bibfield  {author} {\bibinfo {author} {\bibfnamefont {M.}~\bibnamefont
  {Advani}}\ and\ \bibinfo {author} {\bibfnamefont {S.}~\bibnamefont
  {Ganguli}},\ }\bibfield  {title} {\bibinfo {title} {Statistical mechanics of
  optimal convex inference in high dimensions},\ }\href@noop {} {\bibfield
  {journal} {\bibinfo  {journal} {Physical Review X}\ }\textbf {\bibinfo
  {volume} {6}},\ \bibinfo {pages} {031034} (\bibinfo {year}
  {2016})}\BibitemShut {NoStop}%
\bibitem [{\citenamefont {Kuhn}\ and\ \citenamefont
  {Tucker}(1951)}]{kuhn1951nonlinear}%
  \BibitemOpen
  \bibfield  {author} {\bibinfo {author} {\bibfnamefont {H.}~\bibnamefont
  {Kuhn}}\ and\ \bibinfo {author} {\bibfnamefont {A.}~\bibnamefont {Tucker}},\
  }\bibfield  {title} {\bibinfo {title} {Nonlinear programming. berkeley},\
  }\href@noop {} {\bibfield  {journal} {\bibinfo  {journal} {University of
  California Press}\ }\textbf {\bibinfo {volume} {13}},\ \bibinfo {pages} {54}
  (\bibinfo {year} {1951})}\BibitemShut {NoStop}%
\bibitem [{\citenamefont {Shephard}(1966)}]{shephard1966steiner}%
  \BibitemOpen
  \bibfield  {author} {\bibinfo {author} {\bibfnamefont {G.~C.}\ \bibnamefont
  {Shephard}},\ }\bibfield  {title} {\bibinfo {title} {The steiner point of a
  convex polytope},\ }\href@noop {} {\bibfield  {journal} {\bibinfo  {journal}
  {Canadian Journal of Mathematics}\ }\textbf {\bibinfo {volume} {18}},\
  \bibinfo {pages} {1294} (\bibinfo {year} {1966})}\BibitemShut {NoStop}%
\bibitem [{\citenamefont {Sorscher}\ \emph {et~al.}(2021)\citenamefont
  {Sorscher}, \citenamefont {Ganguli},\ and\ \citenamefont
  {Sompolinsky}}]{sorscher2021geometry}%
  \BibitemOpen
  \bibfield  {author} {\bibinfo {author} {\bibfnamefont {B.}~\bibnamefont
  {Sorscher}}, \bibinfo {author} {\bibfnamefont {S.}~\bibnamefont {Ganguli}},\
  and\ \bibinfo {author} {\bibfnamefont {H.}~\bibnamefont {Sompolinsky}},\
  }\bibfield  {title} {\bibinfo {title} {The geometry of concept learning},\
  }\href@noop {} {\bibfield  {journal} {\bibinfo  {journal} {bioRxiv}\ }
  (\bibinfo {year} {2021})}\BibitemShut {NoStop}%
\bibitem [{\citenamefont {Owen}(1980)}]{owen1980table}%
  \BibitemOpen
  \bibfield  {author} {\bibinfo {author} {\bibfnamefont {D.~B.}\ \bibnamefont
  {Owen}},\ }\bibfield  {title} {\bibinfo {title} {A table of normal integrals:
  A table},\ }\href@noop {} {\bibfield  {journal} {\bibinfo  {journal}
  {Communications in Statistics-Simulation and Computation}\ }\textbf {\bibinfo
  {volume} {9}},\ \bibinfo {pages} {389} (\bibinfo {year} {1980})}\BibitemShut
  {NoStop}%
\end{thebibliography}%

\onecolumngrid
\appendix
\section{Supplementary Figures} \label{sec:supp-figures}

\setcounter{figure}{0} \renewcommand{\thefigure}{S\arabic{figure}}

\subsection{Points}

\begin{figure}[H]
\center{\includegraphics[width=17.0cm]{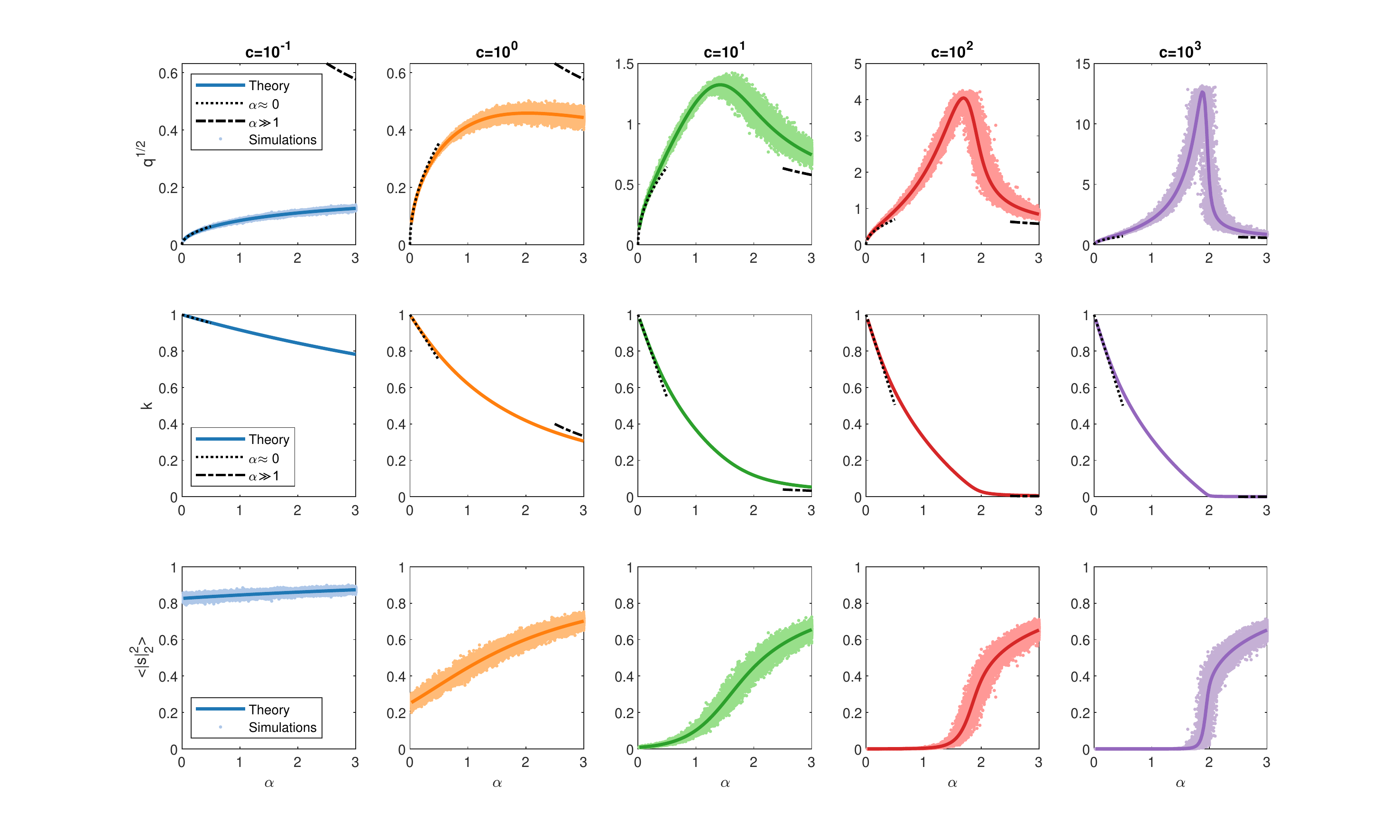}}
\caption{\textbf{Order parameters in soft classification of points.}
Top row: weights' norm $q^{1/2}$ (y-axis) at different values of $\alpha$ (x-axis) and choices of $c$ (panels). Compares theoretical predictions (solid dark lines), simulation results (light dots), and the theoretical prediction in the limits of $\alpha\to0$ and $\alpha\to\infty$ (dotted and dash-dot lines, respectively). 
Middle row: the order parameter $k$ (y-axis) at different values of $\alpha$ (x-axis) and choices of $c$ (panels). Compares theoretical predictions (solid dark lines) and the theoretical prediction in the limits of $\alpha\to0$ and $\alpha\to\infty$ (dotted and dash-dot lines, respectively). 
Bottom row: slack norm $\|\vec{s}\|^{2}$ (y-axis) at different values of $\alpha$ (x-axis) and choices of $c$ (panels). Compares theoretical predictions (solid dark lines) and simulation results (light dots).}
\label{fig:si-points-solution-props}
\end{figure}

\begin{figure}[H]
\center{\includegraphics[width=17.0cm]{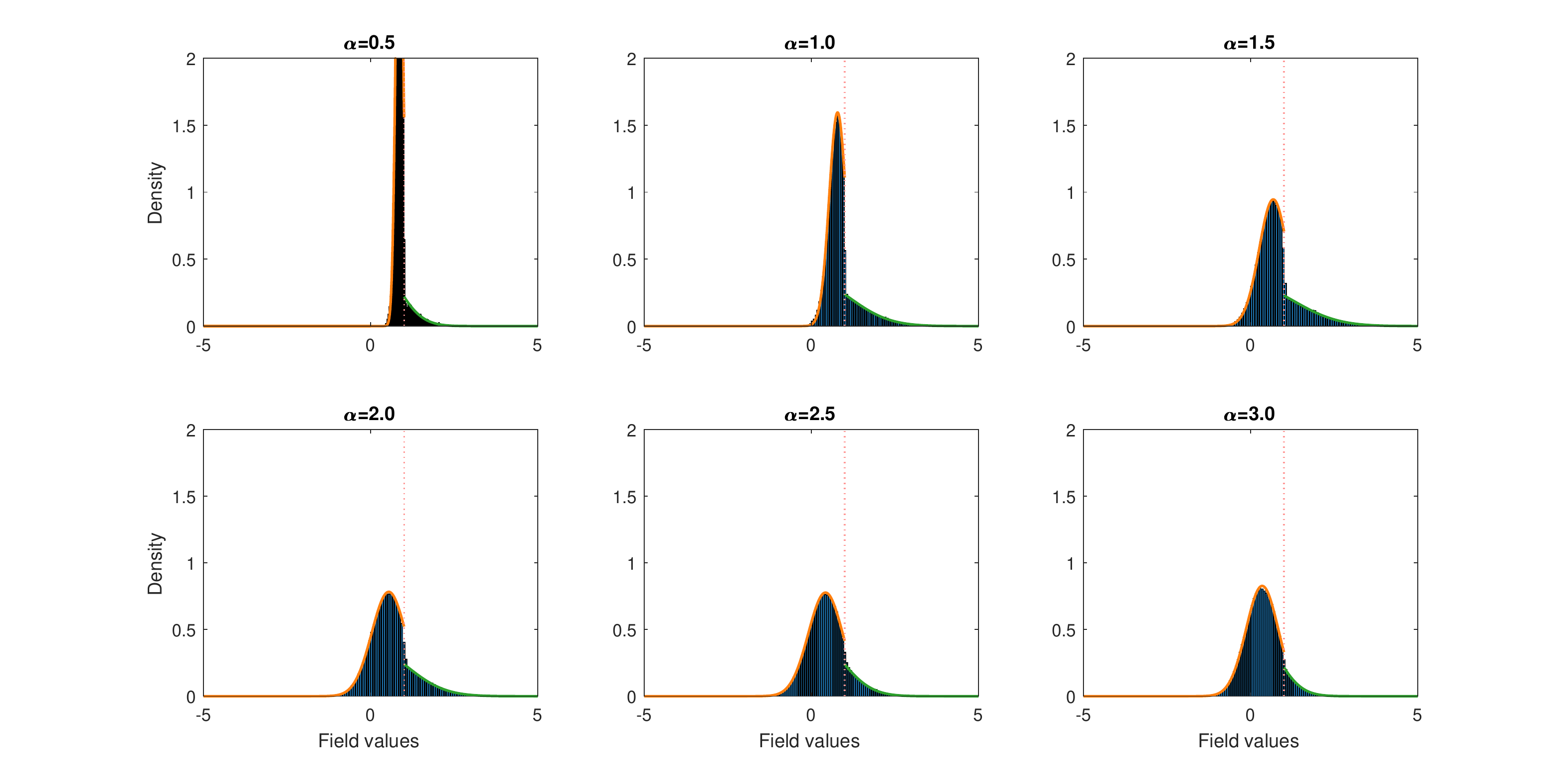}}
\caption{\textbf{Field distribution in soft classification of points.} 
The distribution predicted by theory (solid lines; orange for the ``touching'' regime, green for the ``interior'' regime) and the histogram from simulation results (black area) at different values of $\alpha$ (panels) and $c=10$.}
\label{fig:si-points-fields}
\end{figure}

\begin{figure}[H]
\center{\includegraphics[width=17.0cm]{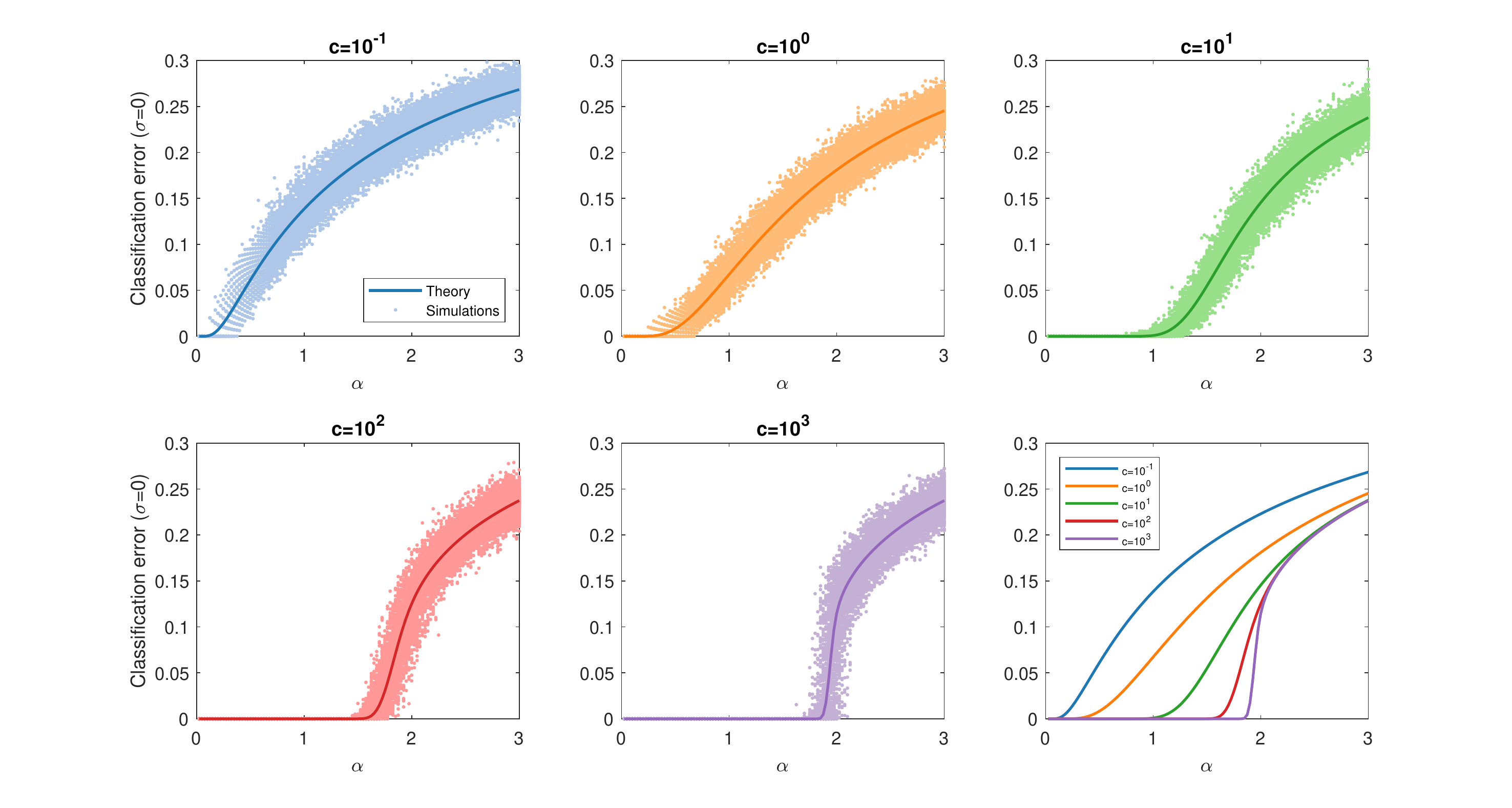}}
\caption{\textbf{Training errors in soft classification of points.} 
Classification error without noise (y-axis) at different values of $\alpha$ (x-axis) and choices of $c$ (color coded). Compares theoretical prediction (solid dark lines) and simulation results (light dots). The last panel overlays the theoretical prediction for different choices of $c$.}
\label{fig:si-points-training-error}
\end{figure}

\begin{figure}[H]
\center{\includegraphics[width=17.0cm]{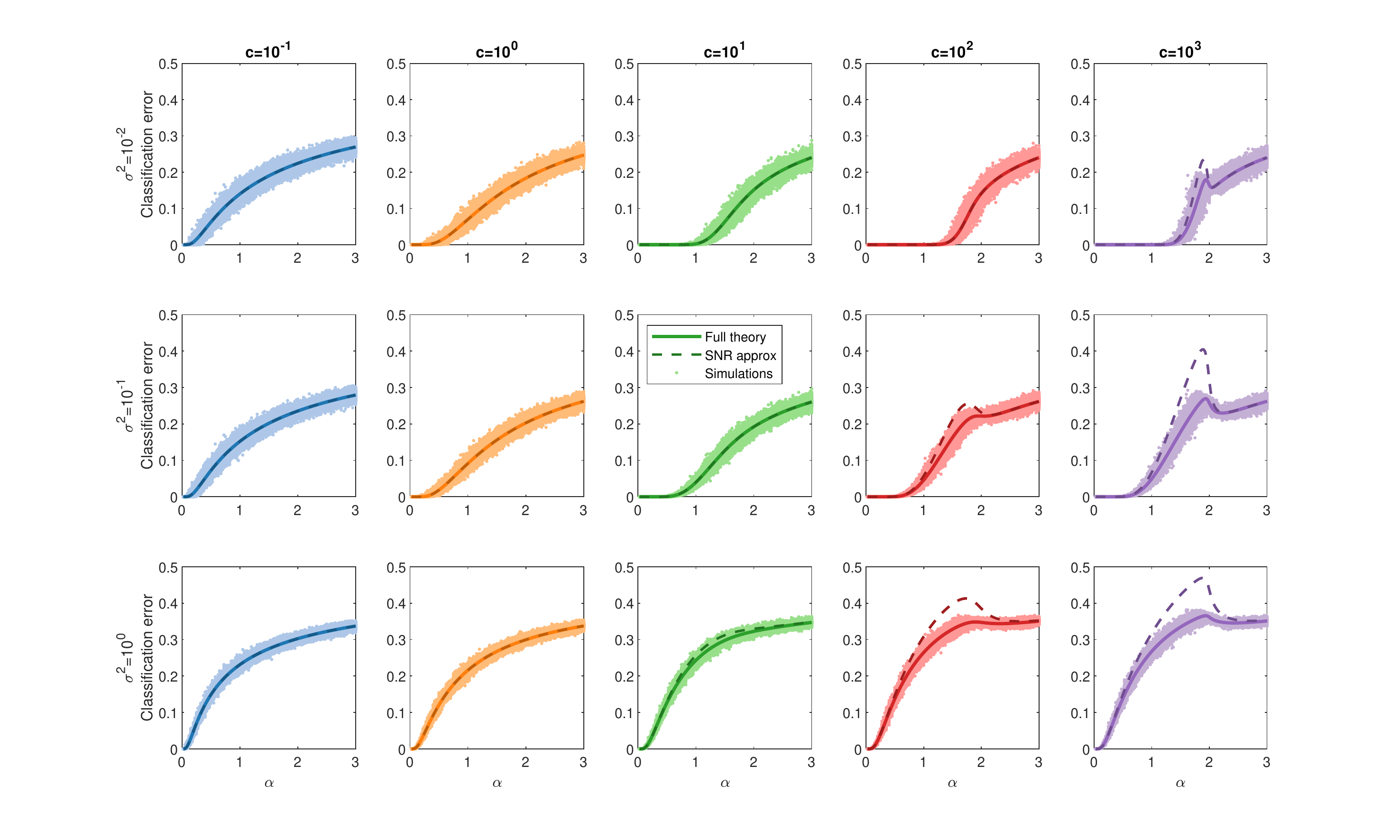}}
\caption{\textbf{Test errors in soft classification of points.}
Classification error (y-axis) at different values of $\alpha$ (x-axis) at different levels of noise $\sigma^{2}$ (rows) and choices of $c$ (columns). Compares theoretical predictions (solid dark lines), the SNR approximation (dashed lines), and simulation results (light dots).}
\label{fig:si-points-test-error}
\end{figure}

\begin{figure}[H]
\center{\includegraphics[width=17.0cm]{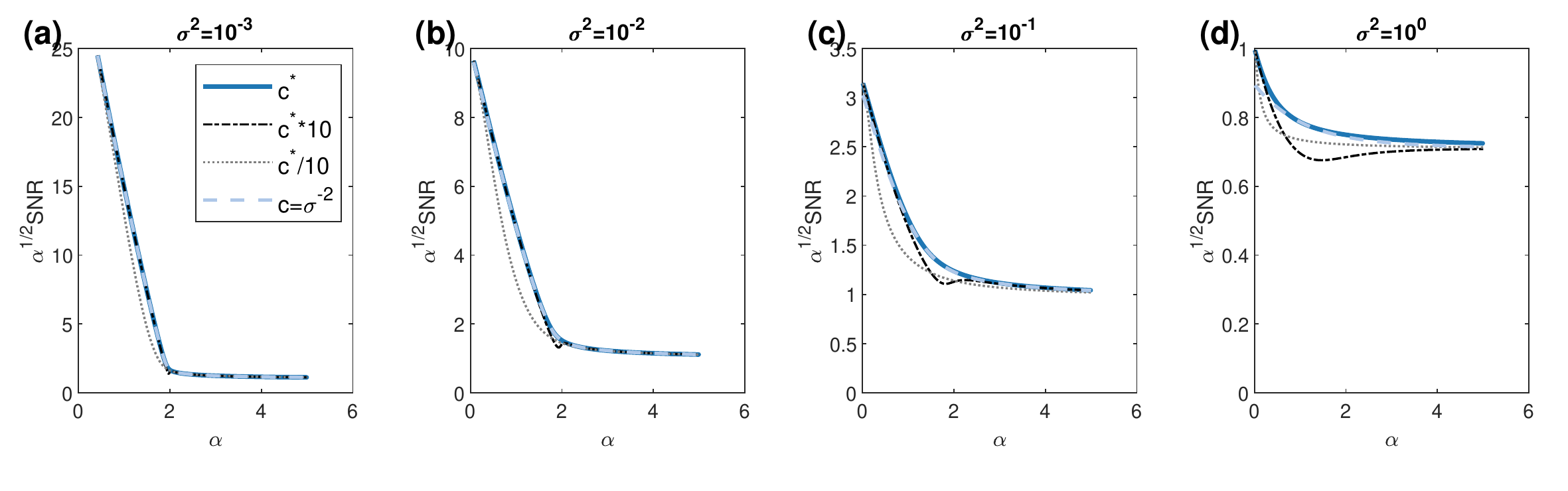}}
\caption{\textbf{Classification error using different choices of $c$ in soft classification of points.}
Classification error presented as scaled SNR $\sqrt{\alpha}{\cal S}$ (y-axis, higher values imply lower error, where ${\cal S}=H^{-1}\left(\varepsilon\right)$ using the inverse of the Gaussian tail function $H$), at different levels of noise $\sigma^{2}$ (panels), using the optimal choice $c^{*}$ (solid colored lines), the canonical choice $c=\sigma^{-2}$ (dashed colored lines), and two sub-optimal choices $10c^{*}$ and $0.1c^{*}$ (dashed and dotted black lines).}
\label{fig:si-points-optimal-c}
\end{figure}

\subsection{Spheres}

\begin{figure}[H]
\center{\includegraphics[width=17.0cm]{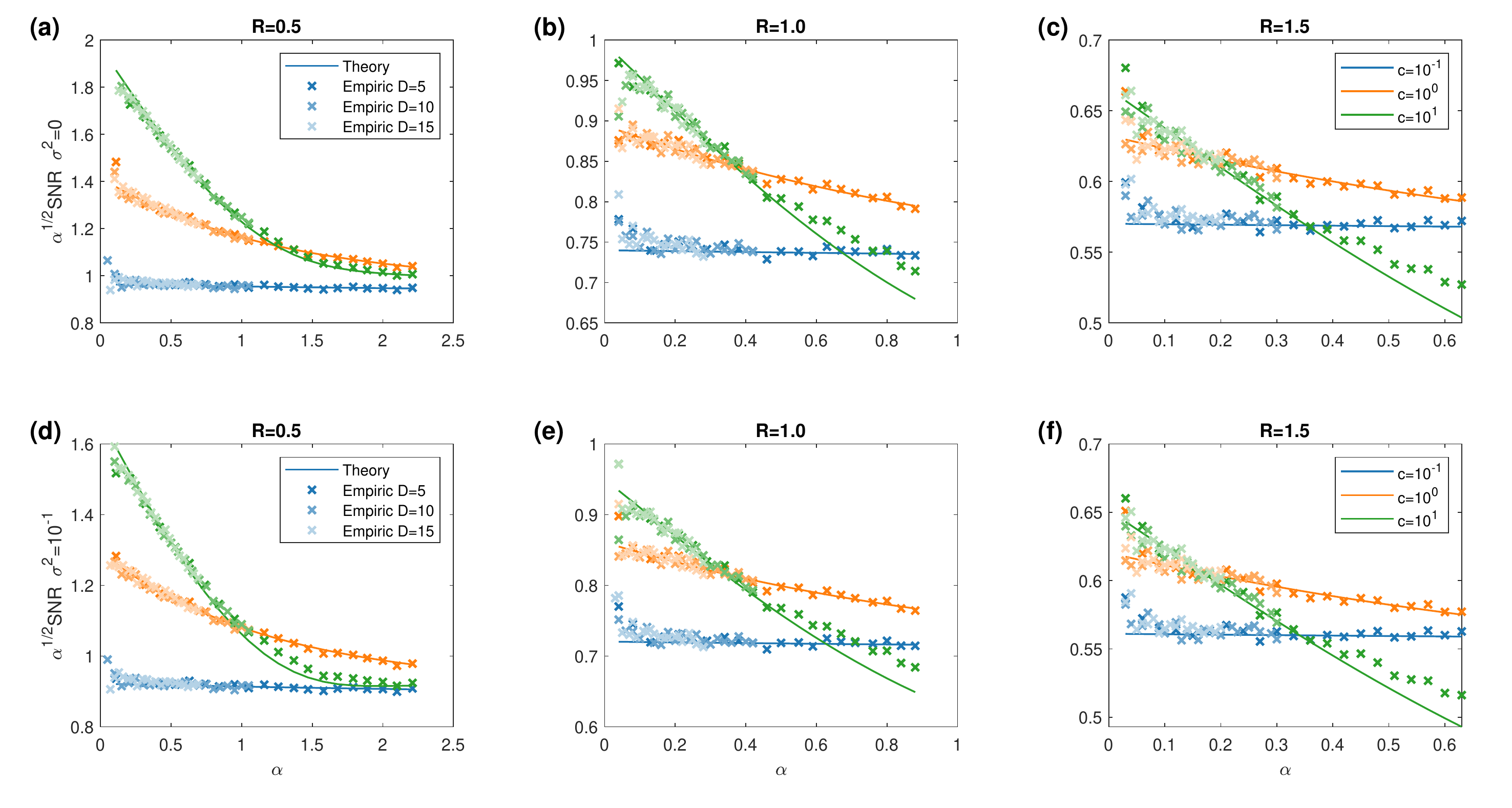}}
\caption{\textbf{Errors in soft classification of spheres using center-slacks.} 
Classification error presented as scaled SNR $\sqrt{\alpha}{\cal S}$ (y-axis, higher values imply lower error, where ${\cal S}=H^{-1}\left(\varepsilon\right)$ using the inverse of the Gaussian tail function $H$), at different values of $\alpha$ (x-axis), choices of $c$ (color coded), and values of $R$ (panels, $R$ indicated in the title). Compares theoretical predictions (solid dark lines) and simulation results (crosses) using different values of $D$ (coded in lightness). 
(a-c) Results without noise $\sigma^{2}=0$. (d-f) Results with noise $\sigma^{2}=0.1$.}
\label{fig:si-spheres-center-slack-empiric}
\end{figure}

\begin{figure}[H]
\center{\includegraphics[width=13.0cm]{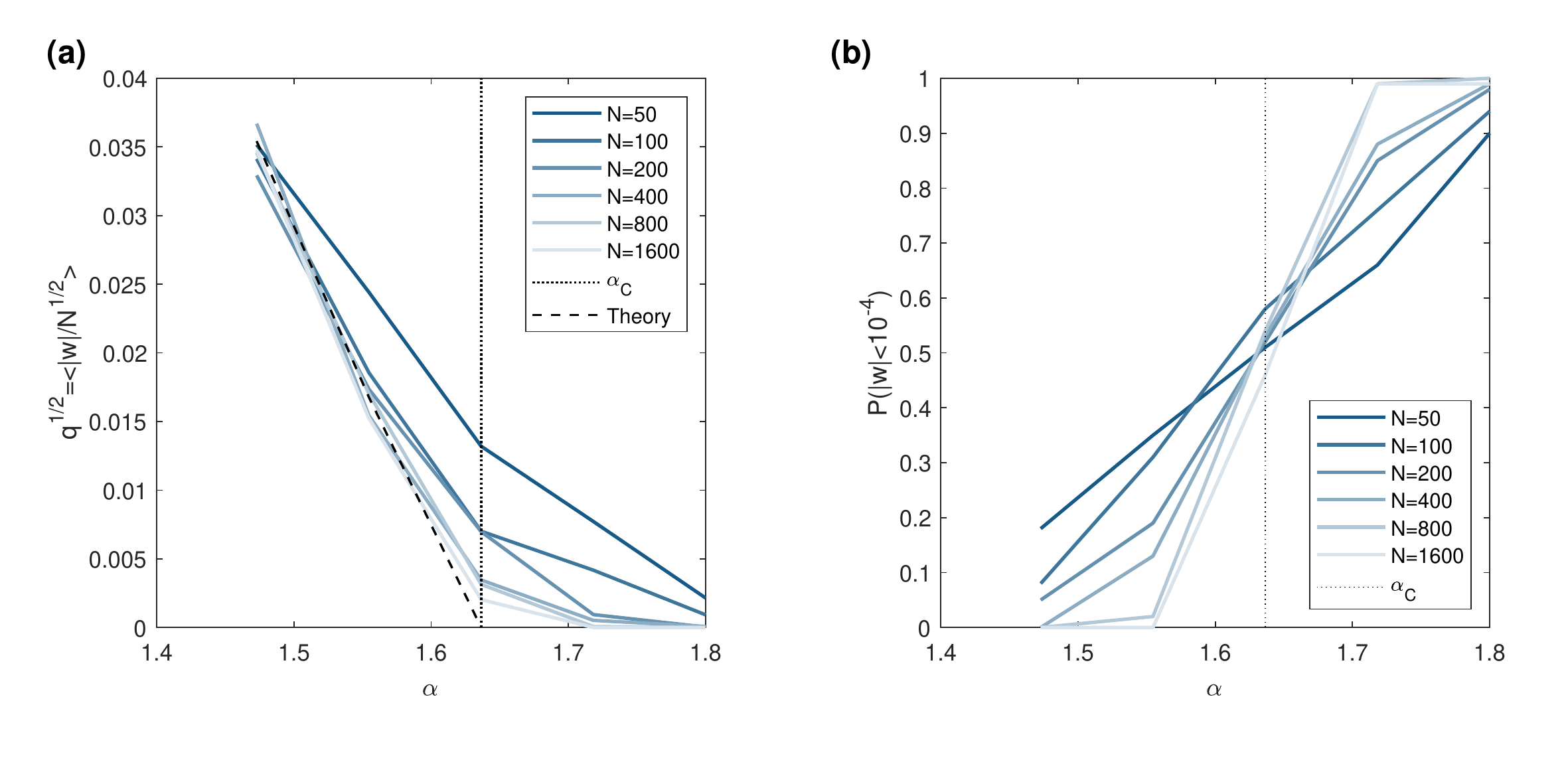}}
\caption{\textbf{Finite-size effects for the weights' norm around the phase transition using manifold-slacks.}
(a) The weights' norm $q^{1/2}$ (y-axis) at different values of $\alpha$ (y-axis) around the phase transition (dotted line). The theory (dashed line) is compared to simulation results, using different choices of $N$ (solid lines, color coded). 
(b) The fraction of simulation results below $10^{-4}$ (y-axis) at different values of $\alpha$ (y-axis) around the phase transition (dotted line), using different choices of $N$ (color coded).}
\label{fig:si-spheres-phase-transition-finite-size-effect}
\end{figure}

\begin{figure}[H]
\center{
\includegraphics[width=17.0cm]{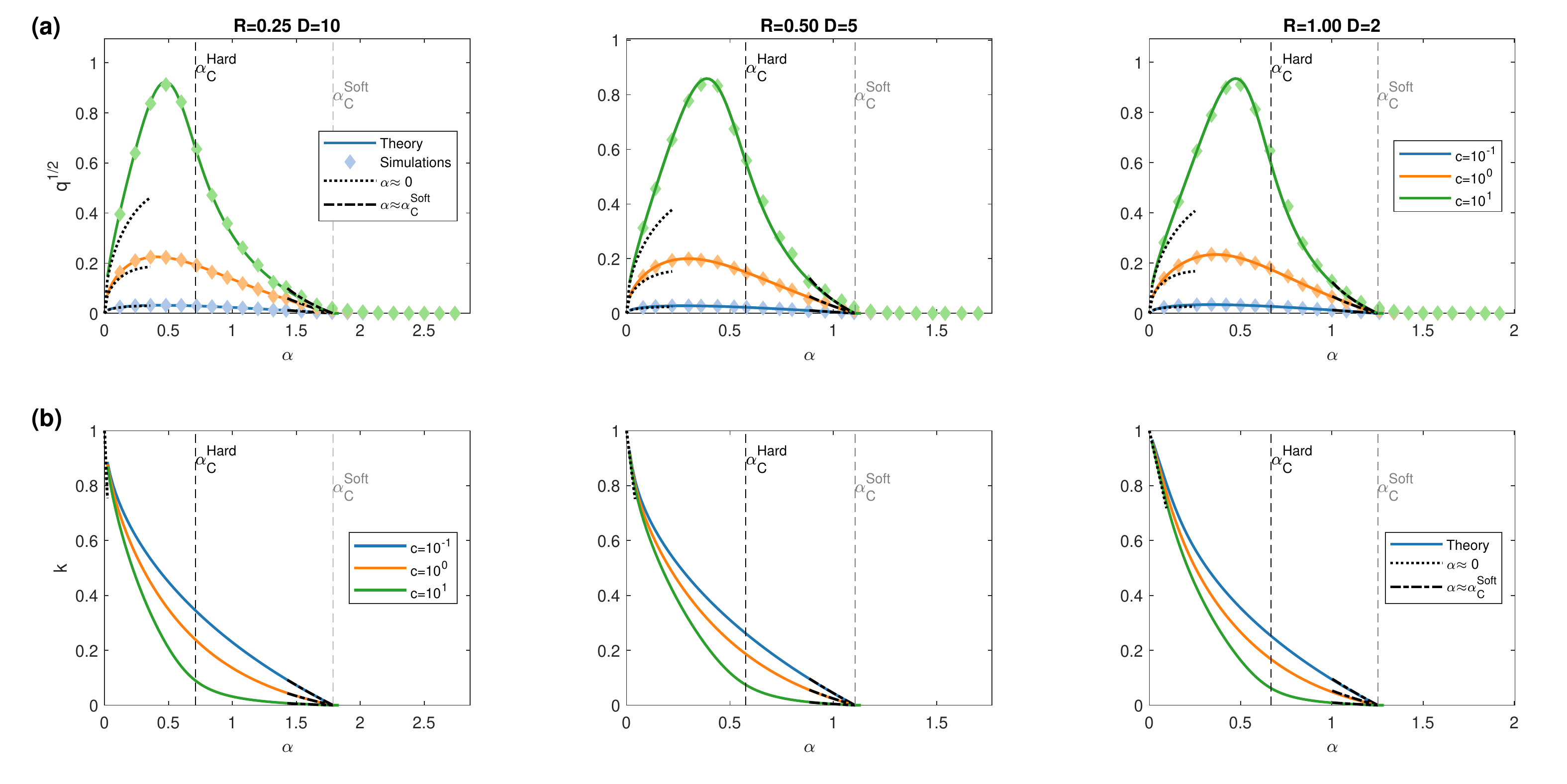}
\includegraphics[width=17.0cm]{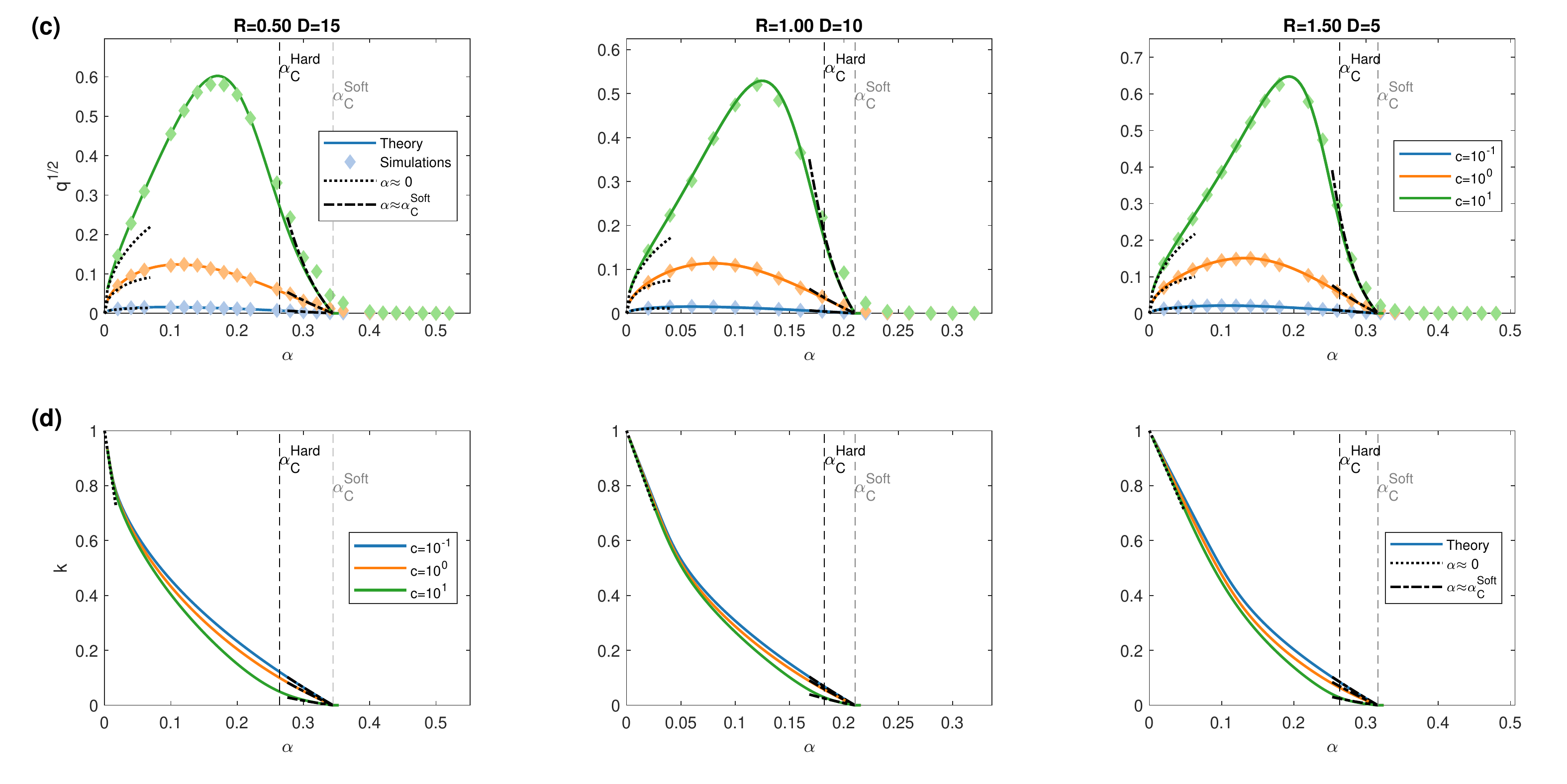}
}
\caption{\textbf{Order parameters in soft classification of spheres using manifold-slacks.}
(a,c) The weights' norm $q^{1/2}$ (y-axis) at different values of $\alpha$ (x-axis), choices of $c$ (color coded), and values of $R,D$ (indicated in title). Theory results (solid lines) are compared both to simulation results (diamonds) and the theory derived in the limits of either $\alpha\to0$ or $\alpha\to\alpha_{C}^{Soft}$ (black dotted and dash-dot lines, respectively). (b,d) The order parameter $k$ (y-axis) at different values of $\alpha$ (x-axis), choices of $c$ (color coded), and values of $R,D$ (indicated in title of the above panel). Compares full theory results (solid lines) with results derived in the limits of either $\alpha\to0$ or $\alpha\to\alpha_{C}^{Soft}$ (black dotted and dash-dot lines, respectively).}
\label{fig:si-spheres-q-k}
\end{figure}

\begin{figure}[H]
\center{\includegraphics[width=17.0cm]{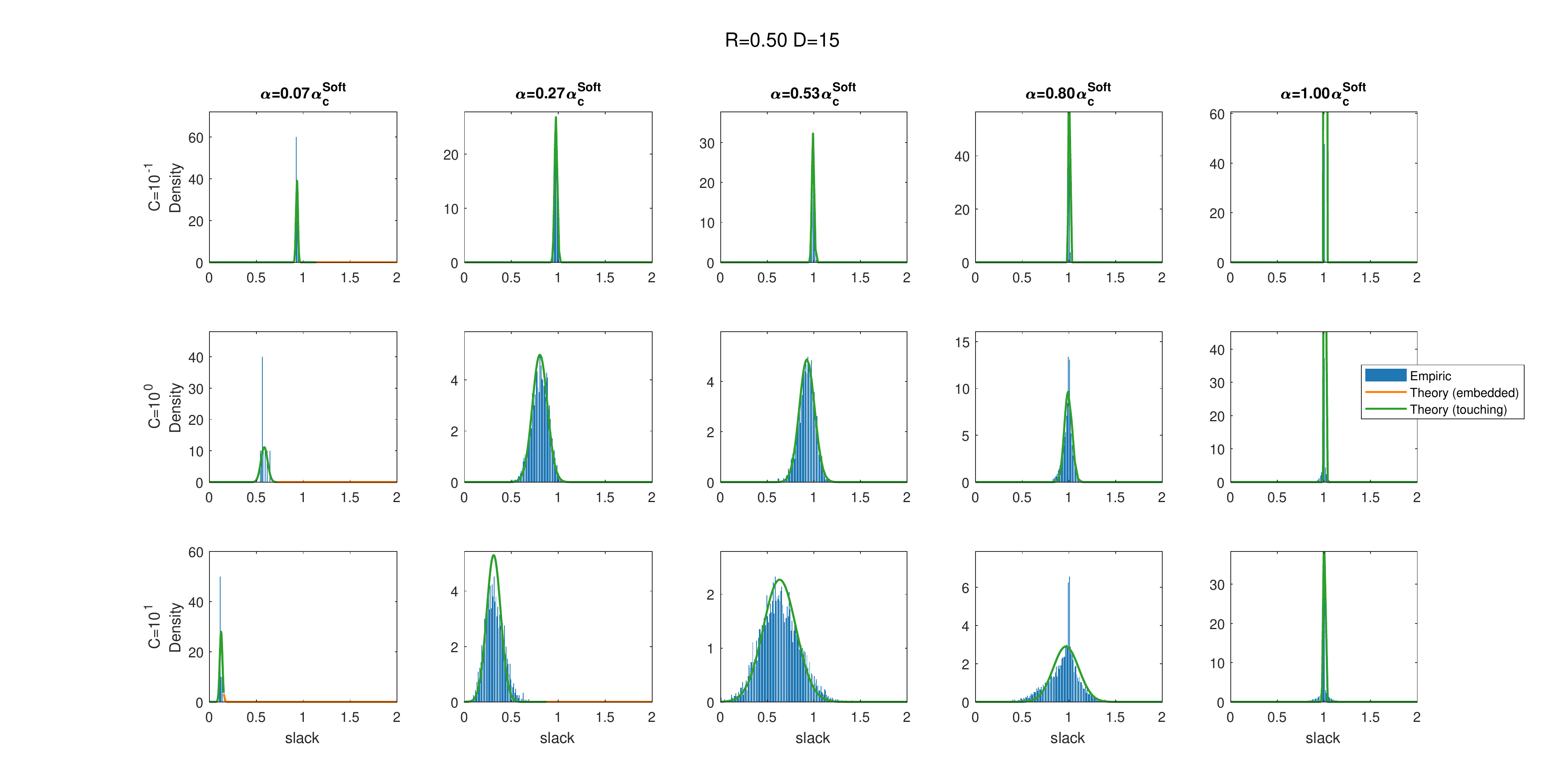}
\includegraphics[width=17.0cm]{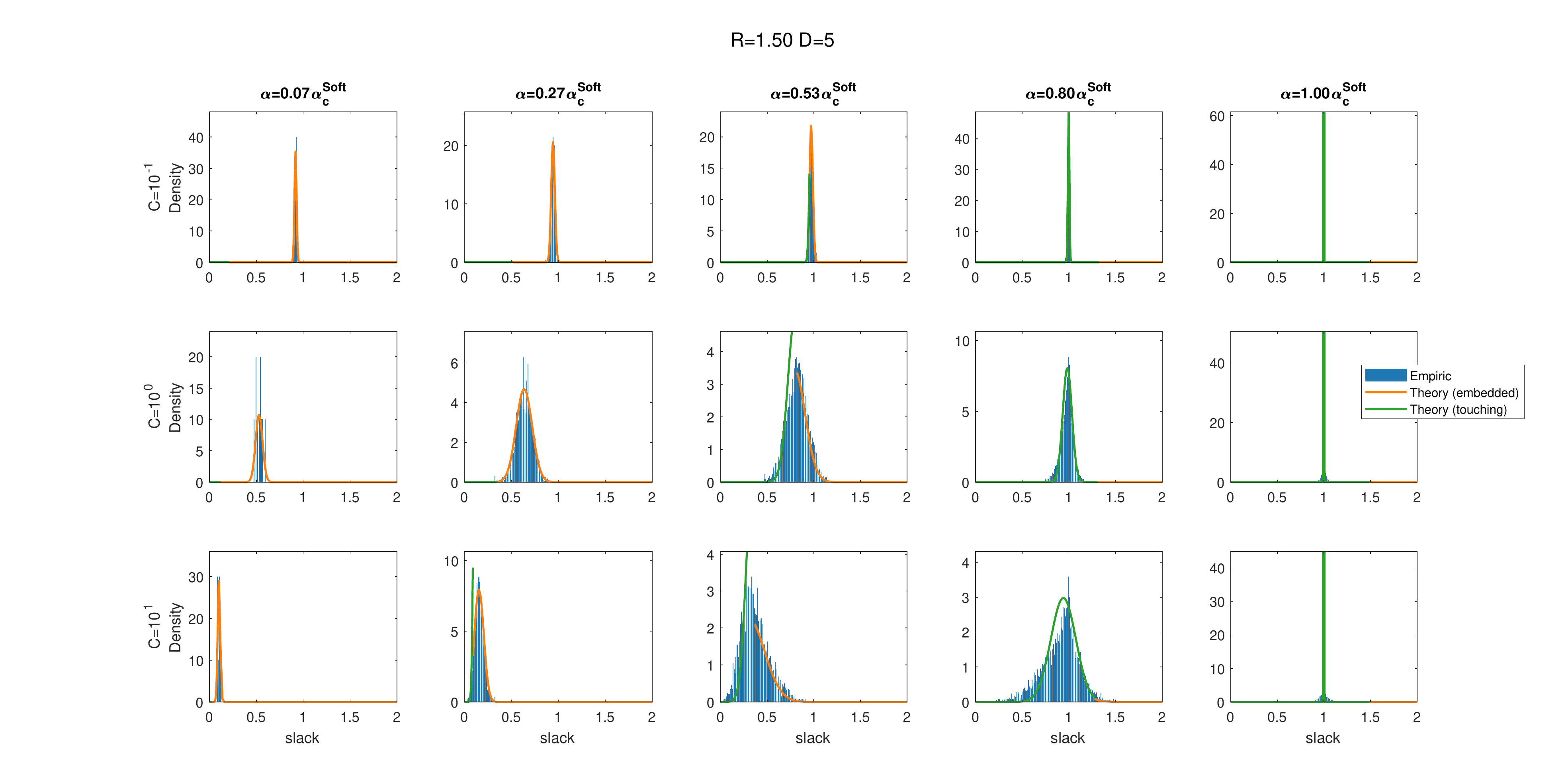}}
\caption{\textbf{Distribution of slack variables in soft classification of spheres using manifold-slacks.} 
Distribution of slack variables at different values of $\alpha$ relative to $\alpha_C^{Soft}$ (columns), choices of $c$ (rows), for $R=0.5,D=15$ (top half) and $R=1.5,D=5$ (bottom half). Compares theory (solid dark line, orange for the ``embedded'' regime, green for the ``touching'' regime), and simulation results (blue histogram).}
\label{fig:si-spheres-slack-distribution}
\end{figure}

\begin{figure}[H]
\center{\includegraphics[width=16.5cm]{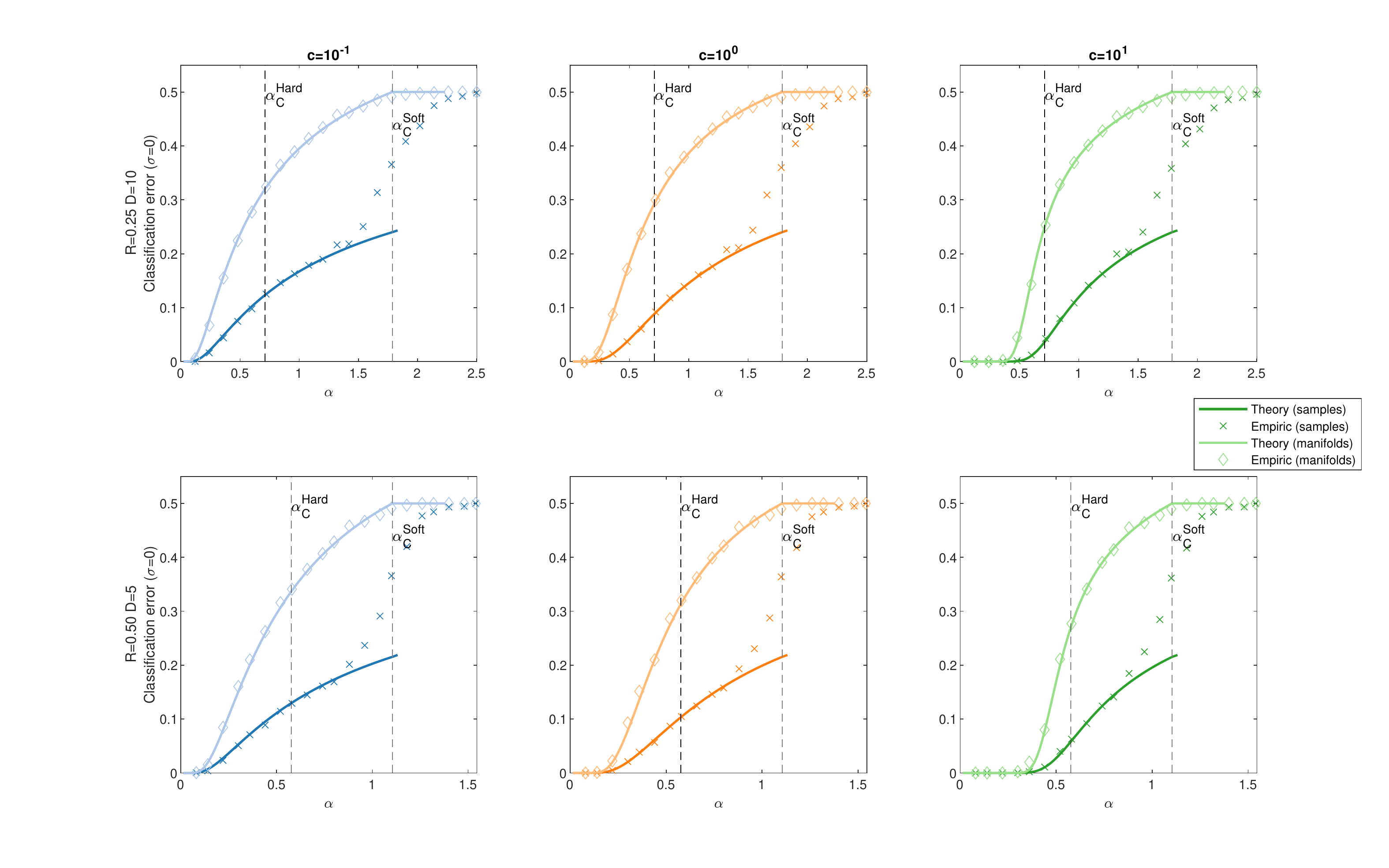}}
\caption{\textbf{Training errors in soft classification of spheres using manifold-slacks.} 
Classification error without noise (y-axis) at different values of $\alpha$ (x-axis) for several values of $R,D$ (rows) and choices of $c$ (columns). Compares samples' classification-error theory (solid dark lines) with simulation results (crosses), and entire-manifold classification-error theory (solid light lines) with simulation results (diamonds).}
\label{fig:si-spheres-training-error}
\end{figure}

\begin{figure}[H]
\center{\includegraphics[width=17.0cm]{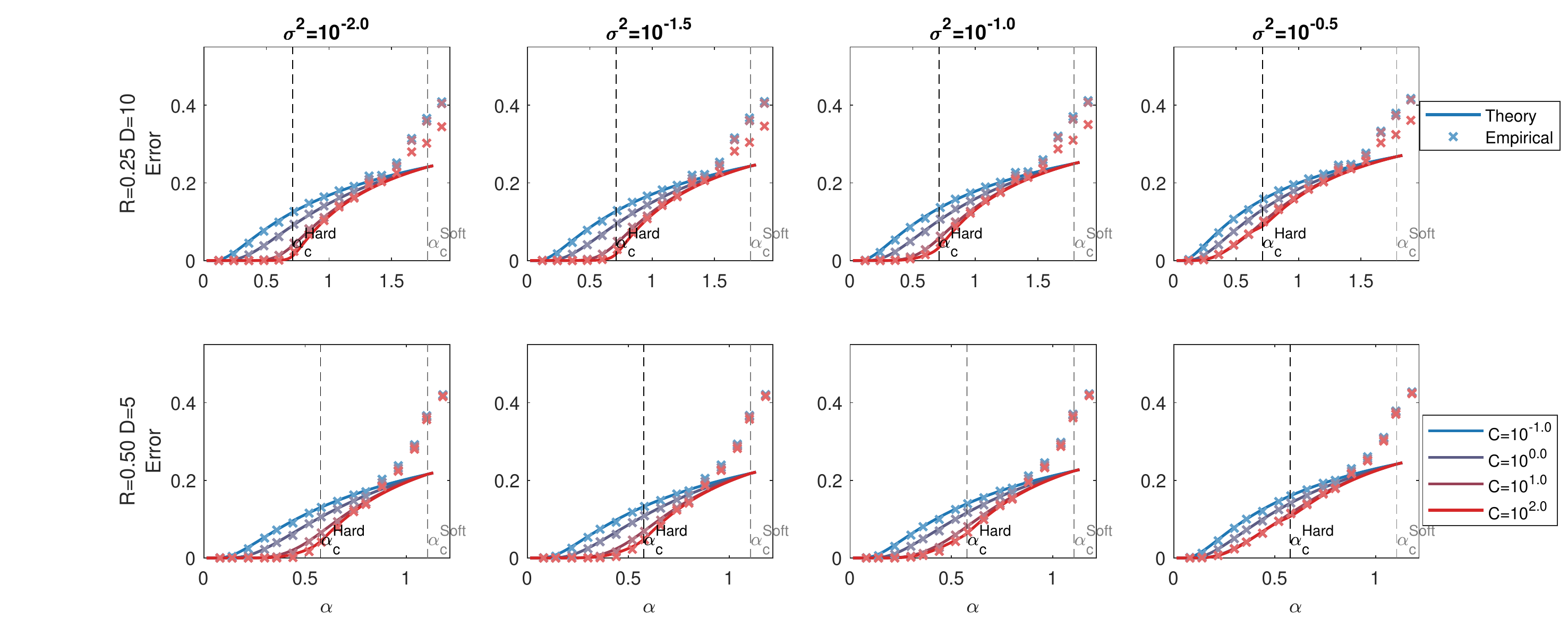}}
\caption{\textbf{Test errors in soft classification of spheres using manifold-slacks.}
Classification error (y-axis) at different values of $\alpha$ (x-axis) for several values of $R,D$ (rows) and levels of noise $\sigma^{2}$ (columns). Compares classification error theory (solid lines) with simulation results (crosses), for different choices of $c$ (color coded).}
\label{fig:si-spheres-test-error-per-sigma}
\end{figure}

\begin{figure}[H]
\center{\includegraphics[width=13.5cm]{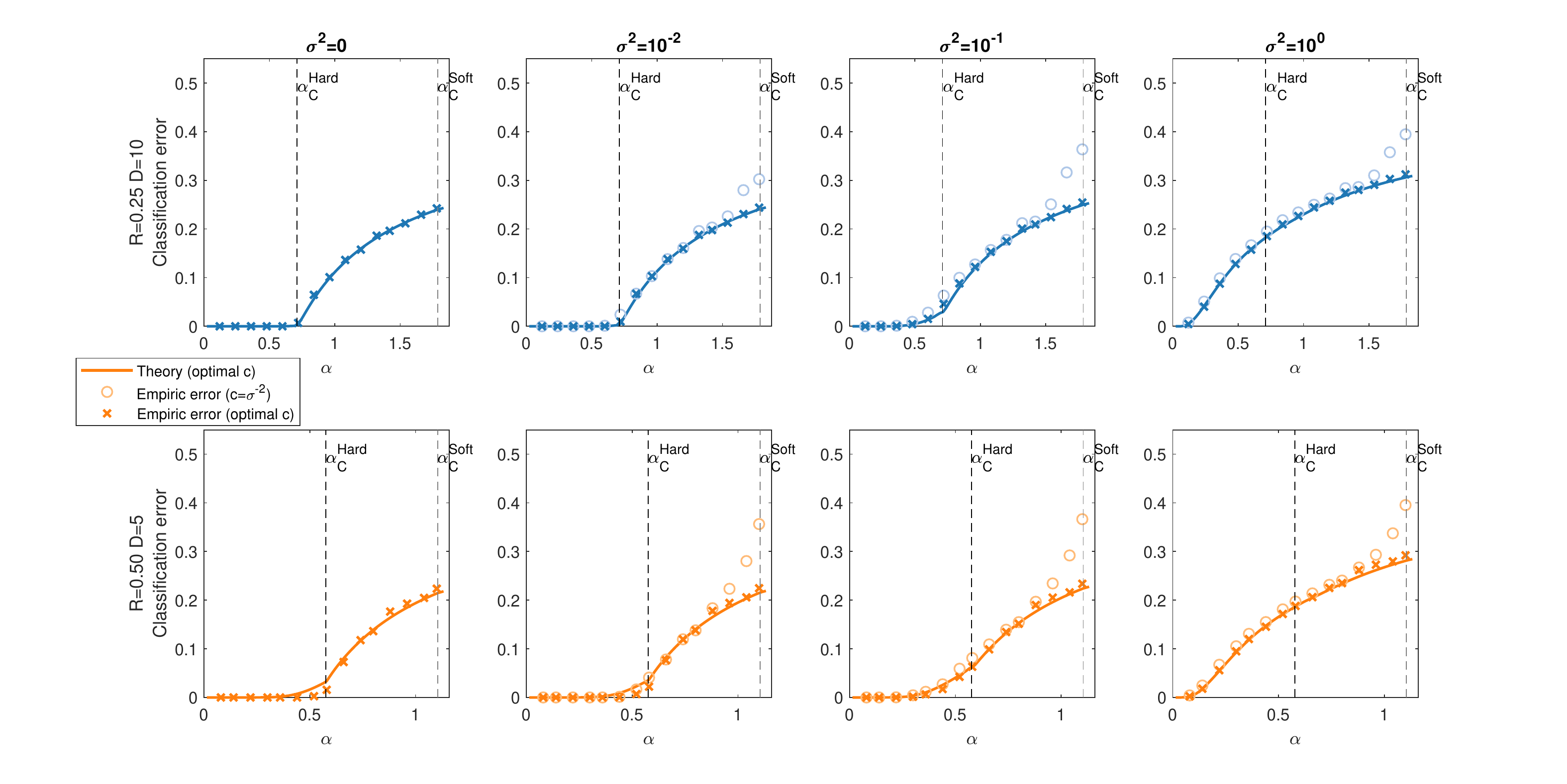}}
\caption{\textbf{Errors in soft classification of spheres using the optimal choice of $c$ using manifold-slacks.}
Classification error (y-axis) at different values of $\alpha$ (x-axis) for several values of $R,D$ (rows) and different levels of noise $\sigma^{2}$ (columns). Compares classification error theory, using the optimal $c$ (solid lines), with simulation results, using either the optimal $c$ (crosses), or the canonical choice $c=\sigma^{-2}$ (circles).}
\label{fig:si-spheres-test-error-optimal-c}
\end{figure}


\begin{figure}[H]
\center{\includegraphics[width=13.0cm]{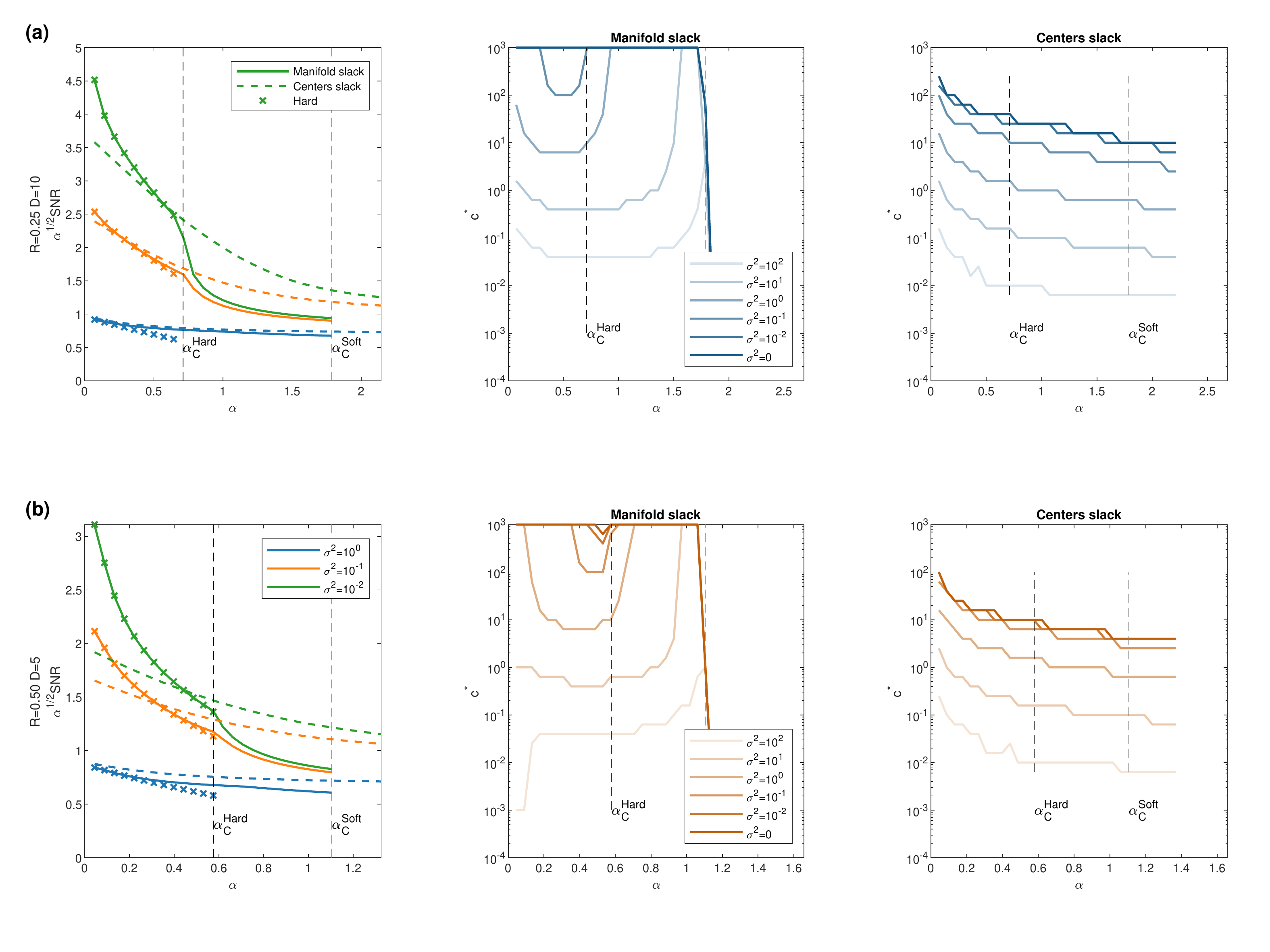}}
\caption{\textbf{Comparison of soft classification of spheres using manifold-slacks and center-slacks.}
Left column: classification error presented as scaled SNR $\sqrt{\alpha}{\cal S}$ (y-axis, higher values imply lower error, where ${\cal S}=H^{-1}\left(\varepsilon\right)$ using the inverse of the Gaussian tail function $H$) at different values of $\alpha$ (x-axis),  levels of noise $\sigma^{2}$ (color coded) and values of $R,D$ (rows). Compares simulation results of manifold-slack classifiers (solid lines), center-slack classifiers (dashed lines) and max-margin classifiers (crosses). 
Middle and right columns: the corresponding optimal choice of $c$ (y-axis, log scale) when using manifold-slacks (middle) and center-slacks (right) at different values of $\alpha$ (x-axis) and noise levels $\sigma^{2}$ (color coded).}
\label{fig:si-spheres-methods-comparison}
\end{figure}

\begin{figure}[H]
\center{\includegraphics[width=17cm]{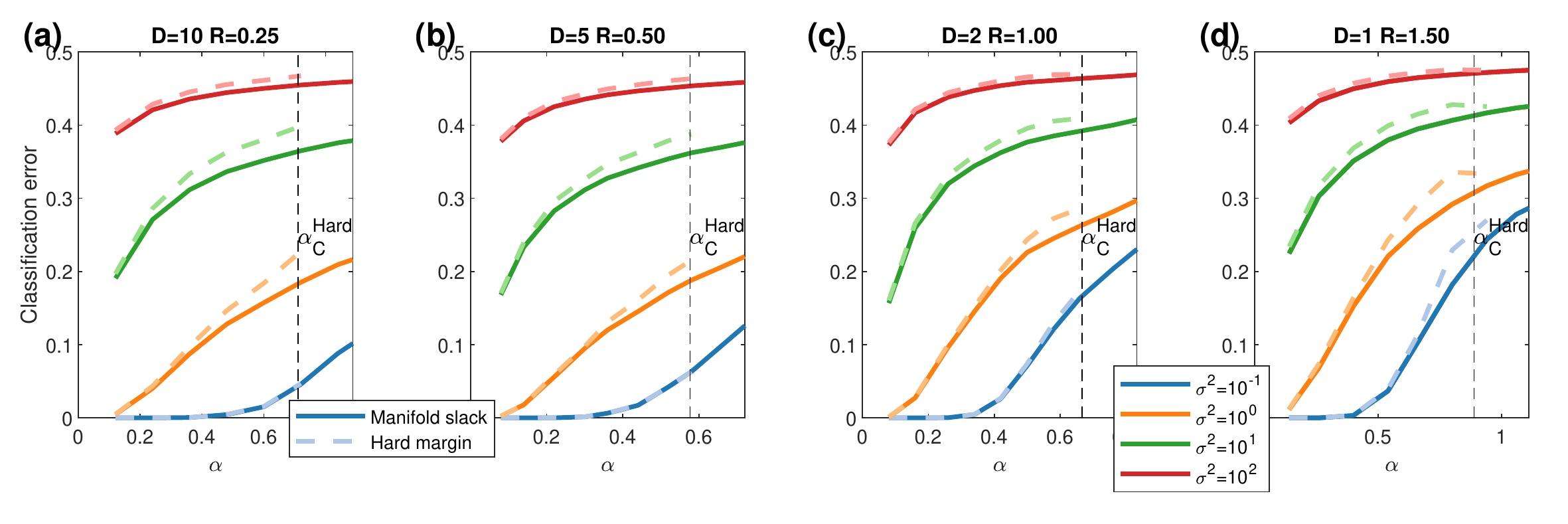}}
\caption{\textbf{Errors in classification of spheres using the manifold-slack and max-margin methods.}
Classification error (y-axis) at different values of $\alpha$ (x-axis) and different combinations of $R,D$ (panels, values indicated in title).
Compares results for manifold-slack classifiers using the optimal choice of $c$ (solid lines) and max-margin classifiers (dashed lines),
at different noise levels (color coded).}
\label{fig:si-spheres-manifold-max-comparison}
\end{figure}

\begin{figure}[H]
\center{\includegraphics[width=17cm]{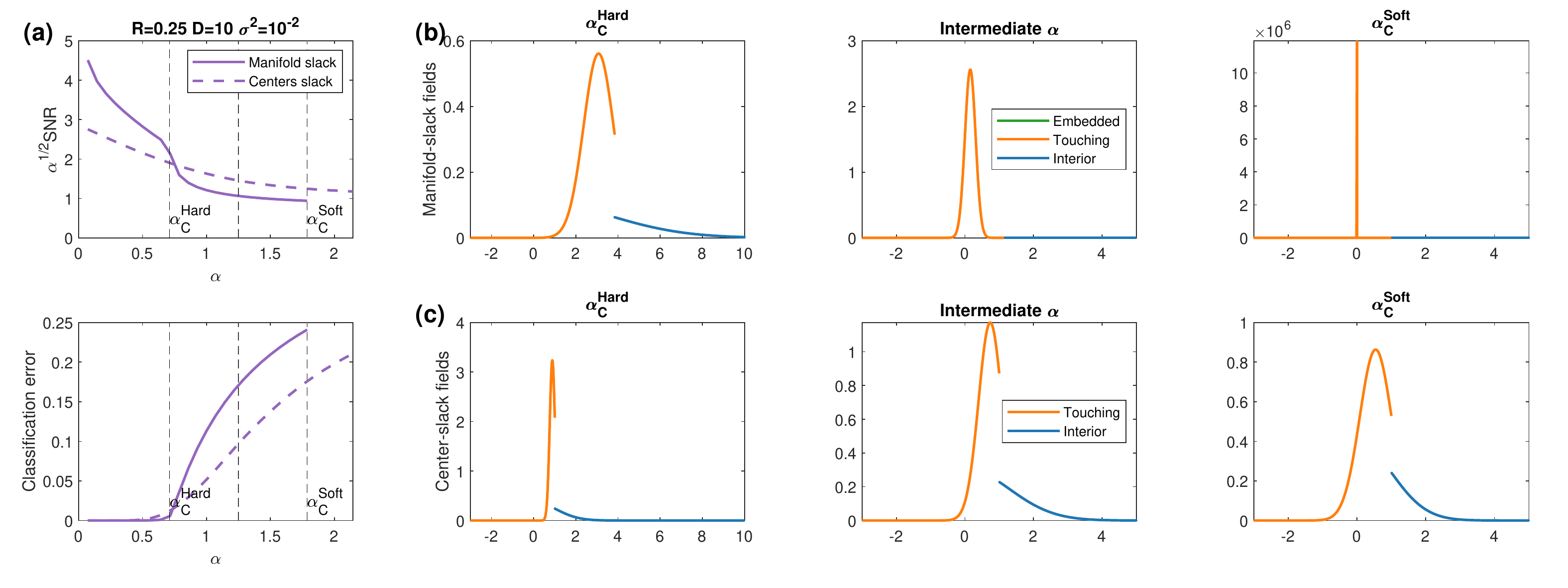}}
\caption{\textbf{Differences in errors and field distribution between manifold-slacks and center-slacks.}
(a) Classification error (y-axis) using manifold-slacks (solid lines) or center-slacks (dashed lines) at different values of $\alpha$ (x-axis) with $R=0.25,D=10$ and $\sigma^{2}=0.01$. 
Top: classification error presented as scaled SNR  $\sqrt{\alpha}{\cal S}$ (higher values imply lower error, where ${\cal S}=H^{-1}\left(\varepsilon\right)$ using the inverse of the Gaussian tail function $H$); bottom: classification error presented as $\varepsilon$. The vertical dashed lines indicate 3 values of $\alpha$: $\alpha_{C}^{Hard}$, $\alpha_{C}^{Soft}$ and their average value. (b-c) Field distribution at the manifold center using the optimal choice of $c$, at those 3 values of $\alpha$, color coded by the different regimes (see legends), using manifold-slacks (b) and center-slacks (c).}
\label{fig:si-spheres-methods-comparison-fields}
\end{figure}

\begin{figure}[H]
\center{\includegraphics[width=14.5cm]{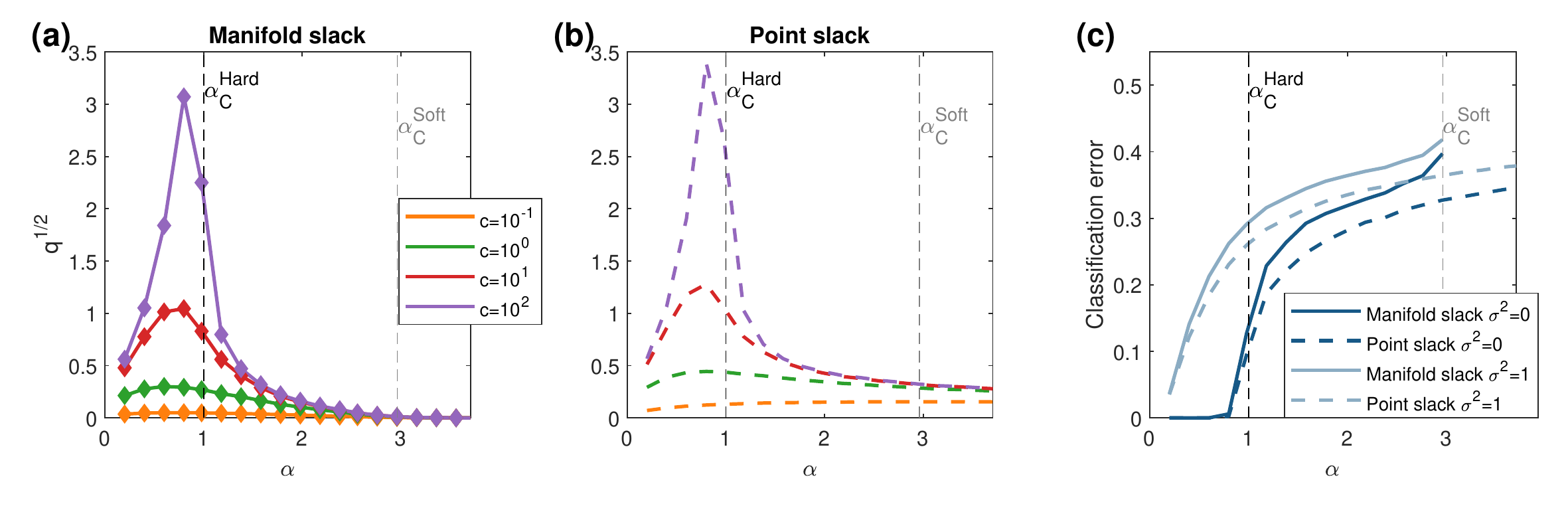}}
\caption{\textbf{Comparison of soft classification of line segments using the manifold-slack and point-slack methods.} Results for classification of line segments (spheres with dimension $D=1$ and radius $R=1$).
(a-b) The weights' norm $q^{1/2}$ (y-axis) at different values of $\alpha$ (x-axis) and choices of $c$ (color coded). (a) Compares theory results of manifold-slacks (solid lines) to simulation results (diamonds). (b) Presents simulation results using the point-slack method (dashed lines) classifying $2P$ samples of the line end-points. 
(c) Classification error (y-axis) using manifold-slack classifiers (solid lines) or point-slack classifiers (dashed lines) at different values of $\alpha$ (x-axis) and noise levels (color coded), using the optimal choice of $c$ for each method.
}
\label{fig:si-spheres-manifold-point-comparison-d1}
\end{figure}

\subsection{General manifolds}

\begin{figure}[H]
\center{\includegraphics[width=14.5cm]{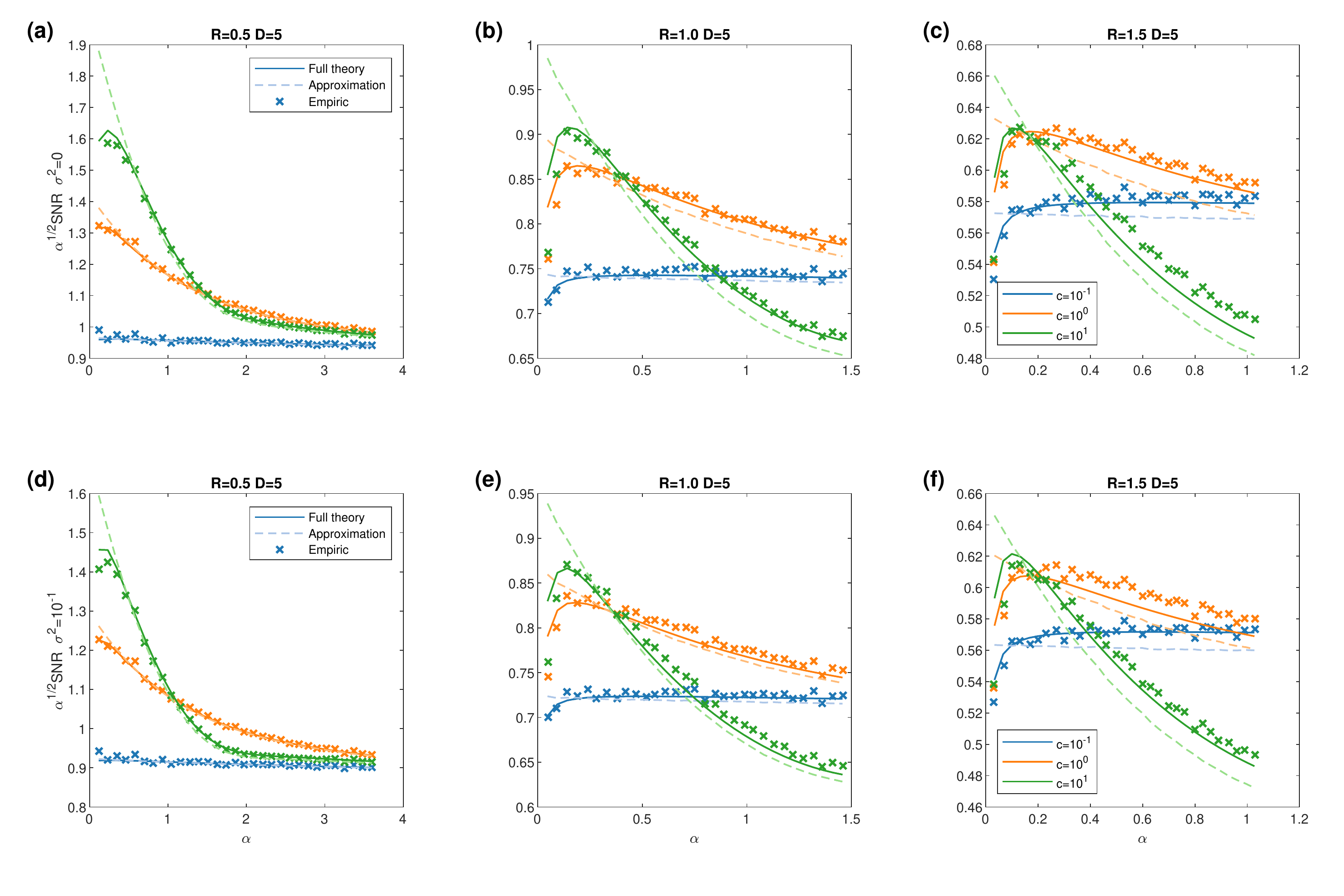}}
\caption{\textbf{Errors in soft classification of point-cloud manifolds using center-slacks.}
Classification error presented as scaled SNR $\sqrt{\alpha}{\cal S}$ (y-axis, higher values imply lower error, where ${\cal S}=H^{-1}\left(\varepsilon\right)$ using the inverse of the Gaussian tail function $H$), at different values of $\alpha$ (x-axis) and choices of $c$ (color coded). Compares the full theory (equation \ref{eq:general-manifolds-point-cloud-center-slack-error}, solid lines), an approximation (equation \ref{eq:general-manifolds-point-cloud-center-slack-error-approx}, dashed lines), and simulation results (crosses). 
Results for manifolds of $M=100$ points from an ellipsoid with $\gamma=0.5$, radius $R$ and dimension $D$ (indicated in panel title), without noise $\sigma^{2}=0$ (a-c) and with noise $\sigma^{2}=0.1$ (d-f).}
\label{fig:si-general-manifolds-center-slack-empiric}
\end{figure}

\begin{figure}[H]
\center{\includegraphics[width=17.0cm]{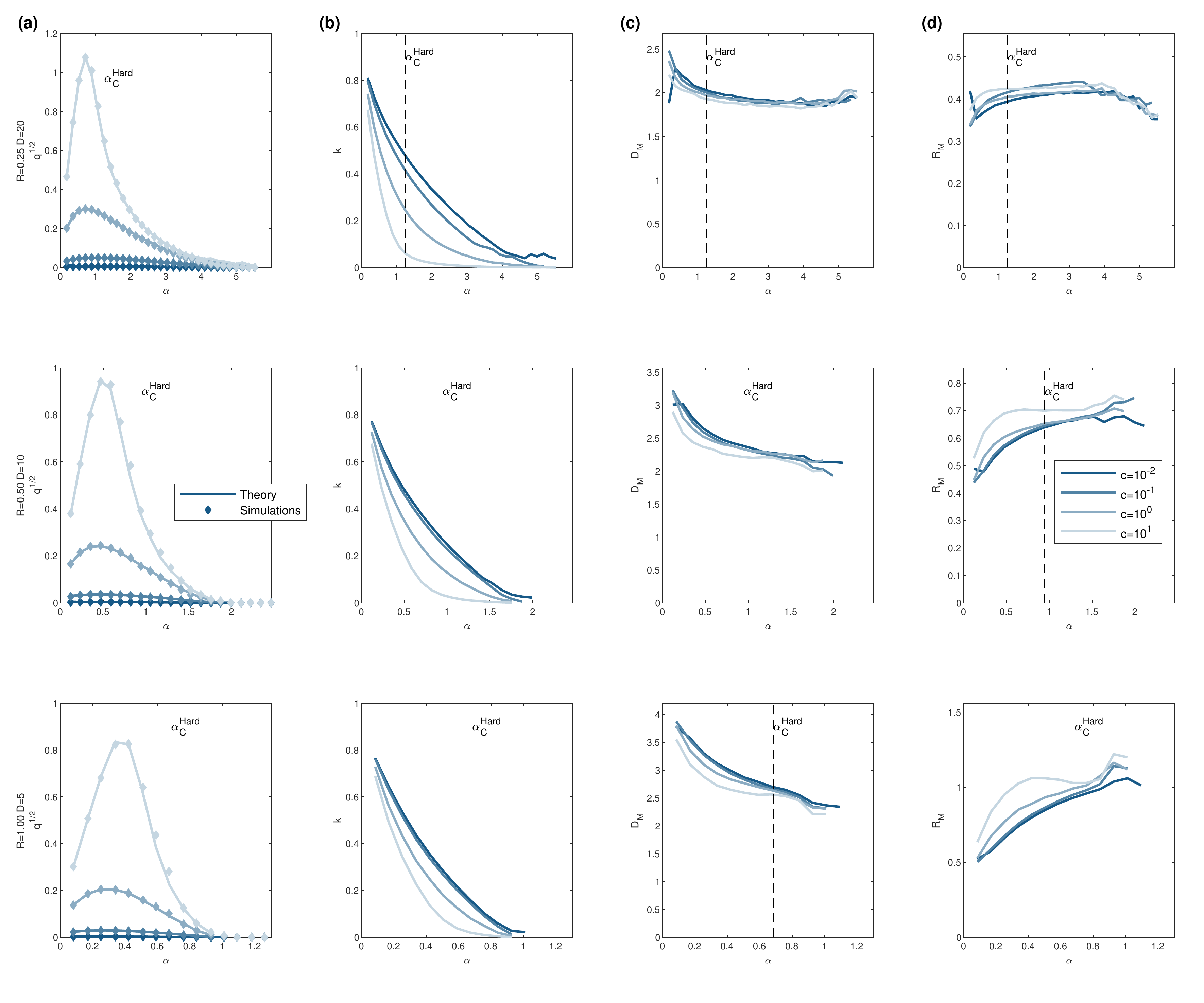}}
\caption{\textbf{Order parameters and manifold properties for point-cloud manifolds.}
Results for manifolds of $M=100$ points from an ellipsoid with $\gamma=1.5$, radius $R$, and dimension $D$ (indicated to the left of each row).
(a) The weights' norm $q^{1/2}$ (y-axis) at different values of $\alpha$ (x-axis) and choices of $c$ (color coded). Compares theory (the least-squares algorithm, solid lines) and simulation results (diamonds). 
(b-d) The corresponding values of the order parameter $k$ (b), manifold dimension $D_{M}$ (c), and manifold radius $R_{M}$ (d) at different values of $\alpha$ (x-axis) and choices of $c$ (color coded).}
\label{fig:si-general-manifolds-properties}
\end{figure}

\begin{figure}[H]
\center{\includegraphics[width=16.5cm]{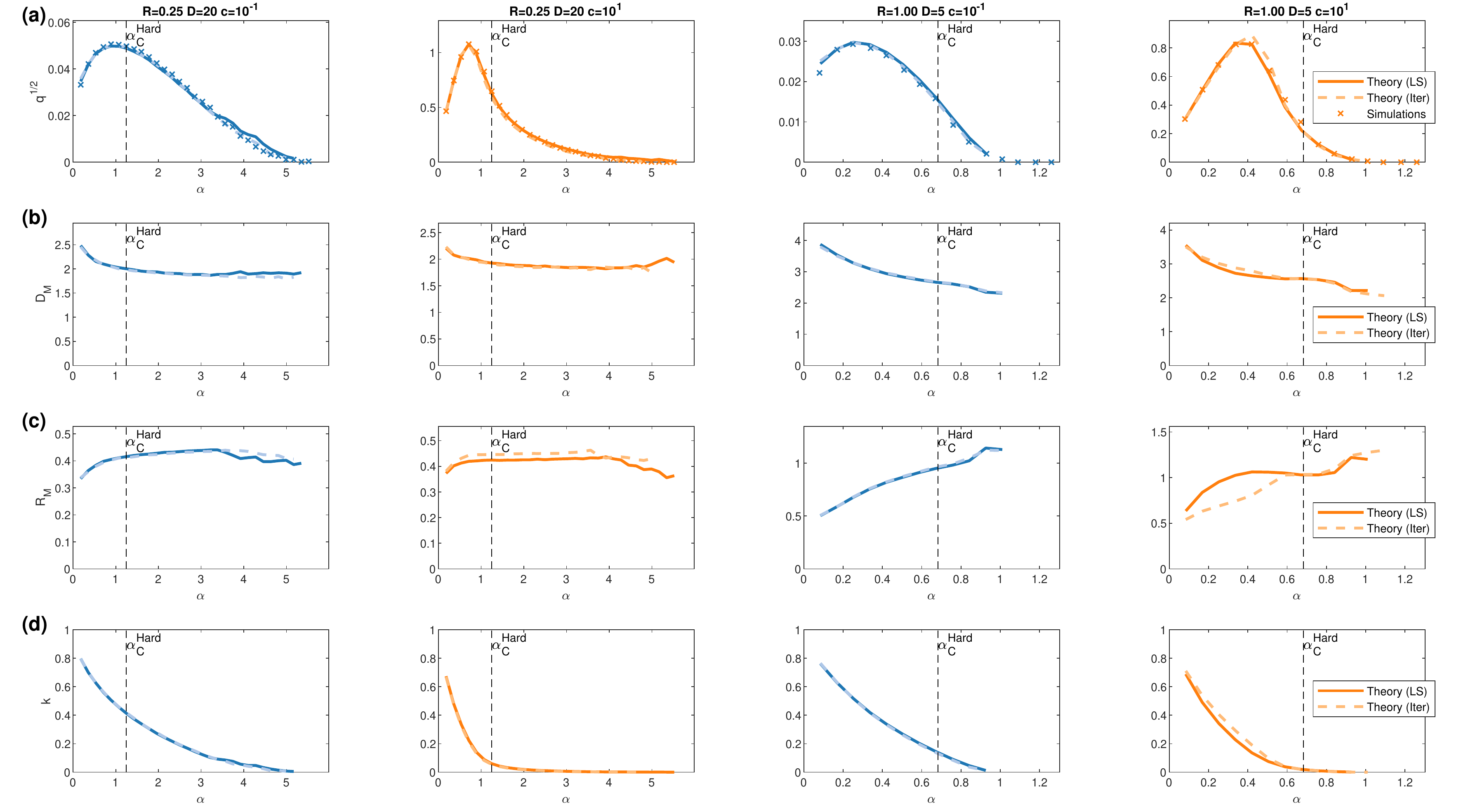}}
\caption{\textbf{Order parameters and manifold properties for point-cloud manifolds using different algorithms.}
Results for manifolds of $M=100$ points from an ellipsoid with $\gamma=1.5$, radius $R$ and dimension $D$. (a) The weights' norm $q^{1/2}$ (y-axis) at different values of $\alpha$ (x-axis) for different values of $R$, $D$ and $c$ (indicated in the top panel title). Compares simulation results (crosses) with theoretical results using the least-squares algorithm (solid lines) and the iterative algorithm (dashed lines). (b-d) The corresponding results for manifold dimension (b), manifold radius (c), and the order parameter $k$ (d) at different values of $\alpha$ (x-axis). Compares theory using the least-squares algorithm (solid lines) and the iterative algorithm (dashed lines).}
\label{fig:si-general-manifolds-properties2}
\end{figure}

\begin{figure}[H]
\center{\includegraphics[width=16.5cm]{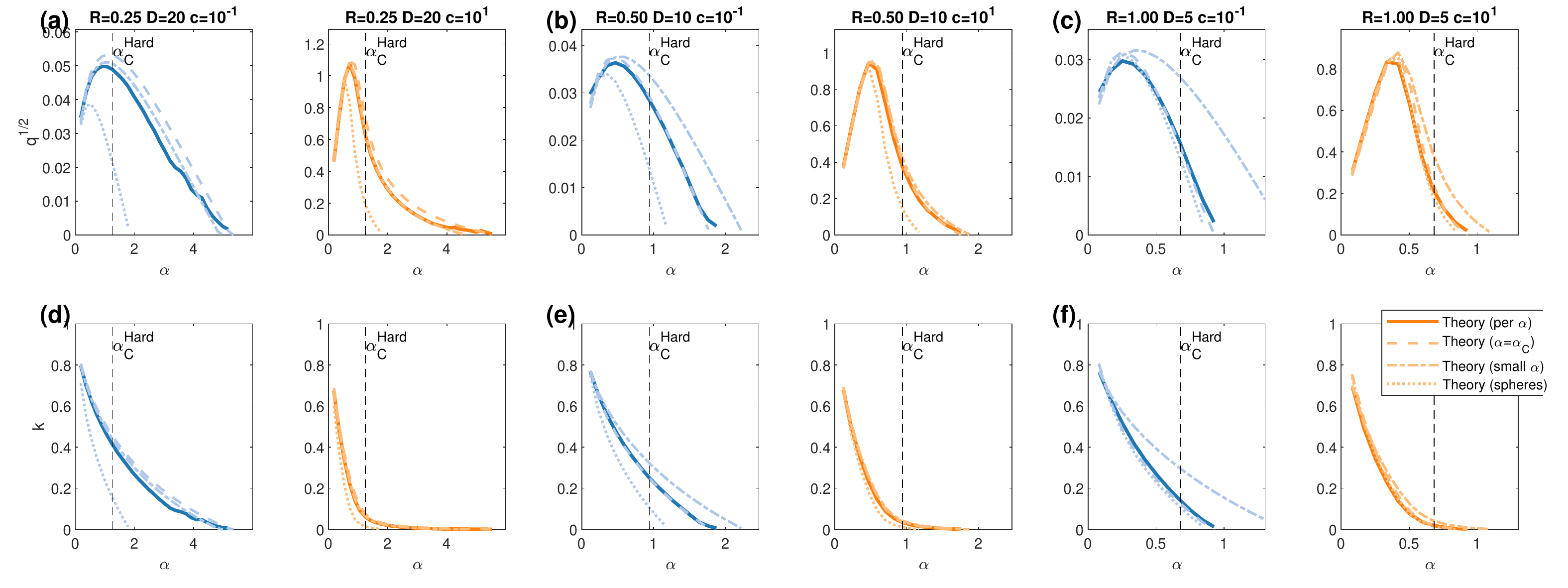}}
\caption{\textbf{Manifold properties dependence on $\alpha$ for point-cloud manifolds.}
Results for manifolds of $M=100$ points from an ellipsoid with $\gamma=1.5$, radius $R$ and dimension $D$. Compares theory calculated per $\alpha$ (the least-squares algorithm, solid lines), theory calculated at $\alpha\approx\alpha_{C}^{Soft}$ (dashed lines) or at small $\alpha$ (dash-dot lines), and naive application of the theory of spheres (dotted lines). (a-c) The weights' norm $q^{1/2}$ (y-axis) at different values of $\alpha$ (x-axis) for different values of $R$, $D$ and $c$ (indicated in the top panel title). (d-f) The corresponding order parameter $k$ (y-axis) at different values of $\alpha$ (x-axis) for different values of $R$, $D$ and $c$.}
\label{fig:si-general-manifolds-q-k-vs-alpha}
\end{figure}

\begin{figure}[H]
\center{\includegraphics[width=13.5cm]{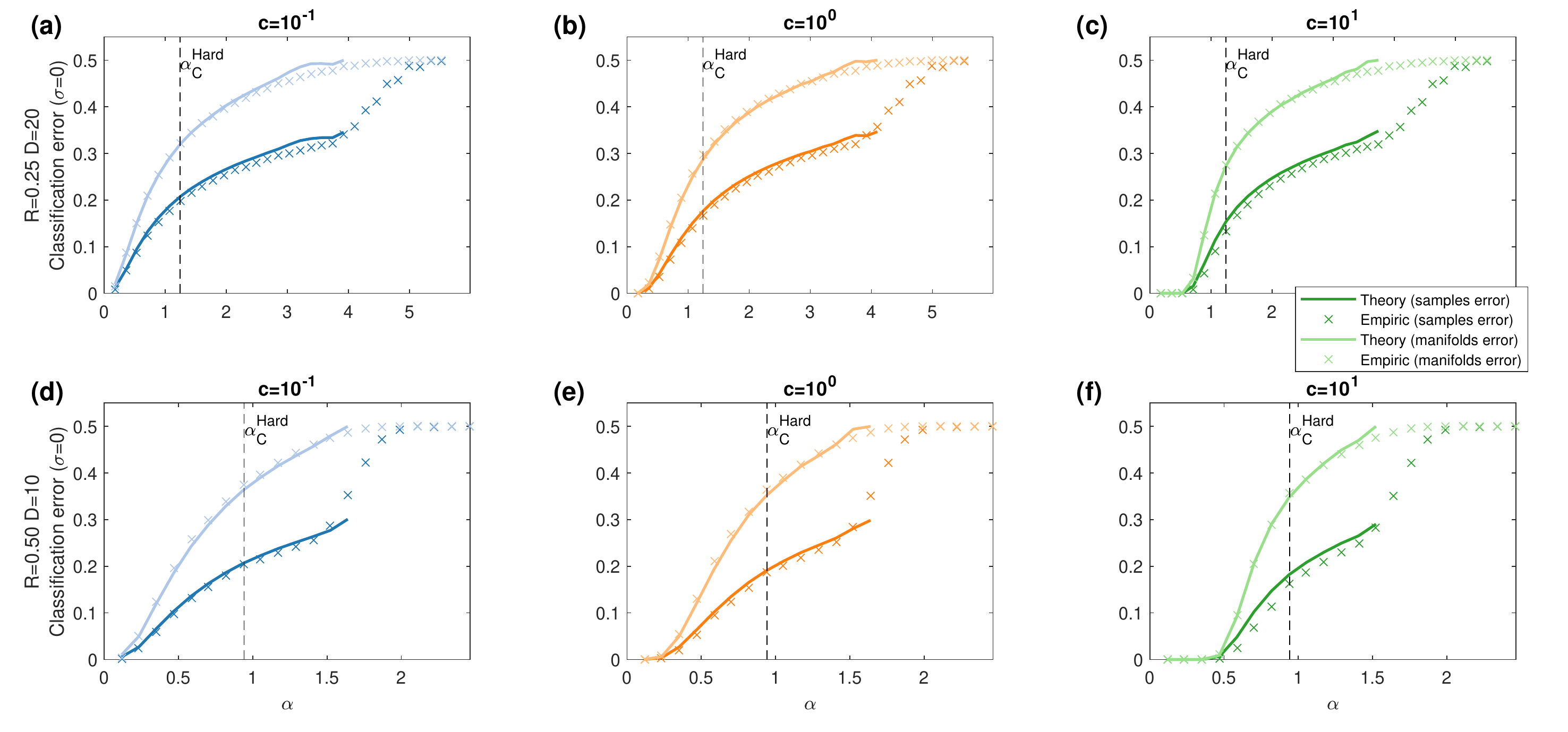}}
\caption{\textbf{Training errors for point-cloud manifolds using manifold-slacks.}
Results for manifolds of $M=100$ points from an ellipsoid with $\gamma=1.5$, radius $R$ and dimension $D$ (indicated to the left of each row). Classification error without noise $\sigma^{2}=0$ (y-axis) at different values of $\alpha$ (x-axis) and choices of $c$ (columns). Compares samples' classification-error theory (solid dark lines) with simulation results (dark crosses), and entire-manifolds classification-error theory (solid light lines) with simulation results (light crosses).}
\label{fig:si-general-manifolds-training-error}
\end{figure}

\begin{figure}[H]
\center{\includegraphics[width=14.5cm]{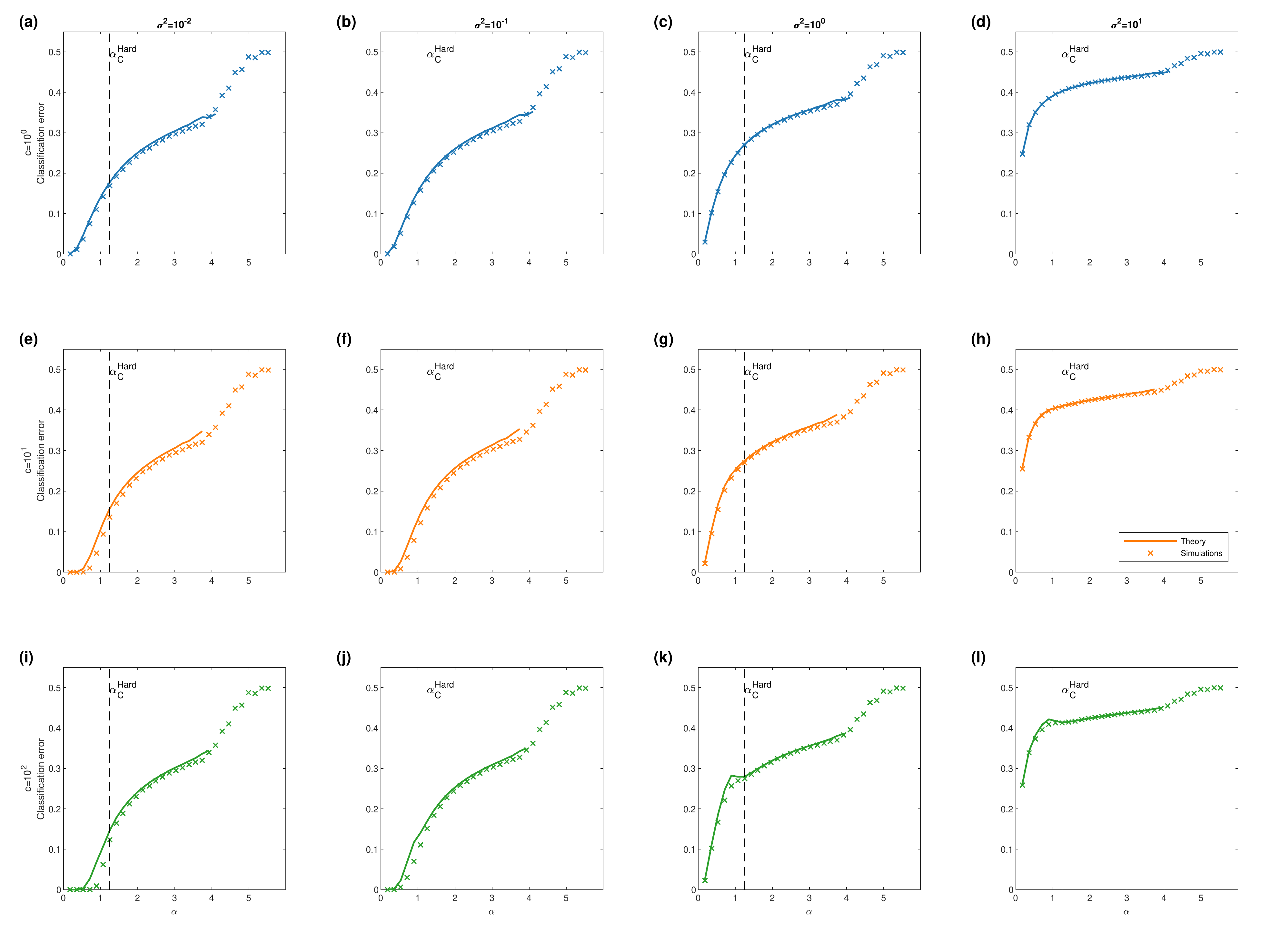}}
\caption{\textbf{Test errors for point-cloud manifolds using manifold-slacks.}
Results for manifolds of $M=100$ points from an ellipsoid with $\gamma=1.5$, radius $R=0.25$ and dimension $D=20$. Classification error (y-axis) at different values of $\alpha$ (x-axis), levels of noise $\sigma^{2}$ (columns) and choices of $c$ (rows). Compares classification error theory (solid dark lines) with simulation results (dark crosses).}
\label{fig:si-general-manifolds-generalization-error}
\end{figure}

\begin{figure}[H]
\center{\includegraphics[width=14.5cm]{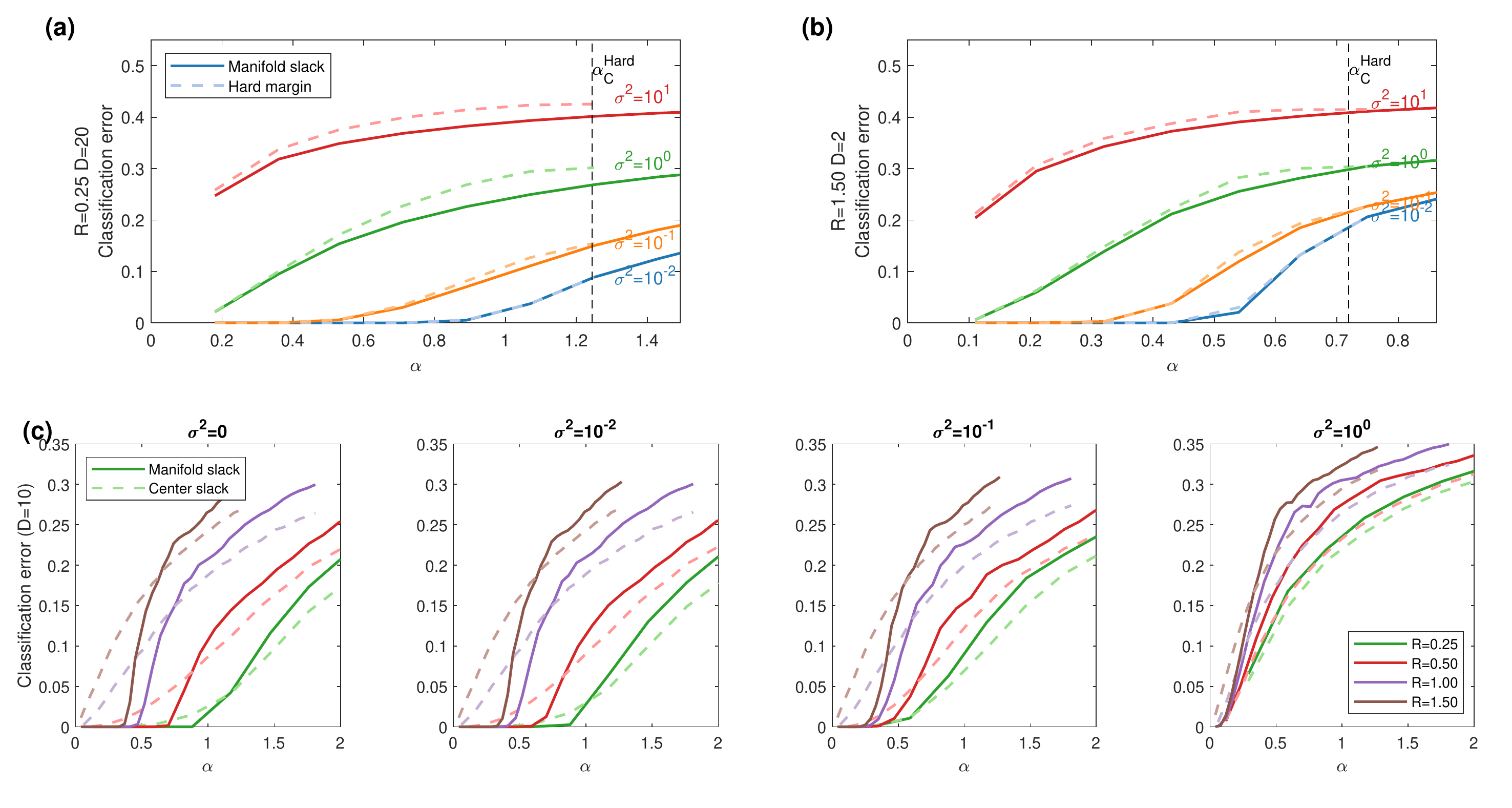}}
\caption{\textbf{Classification errors for point-cloud manifolds using different methods.}
Results for manifolds of $M=100$ points from an ellipsoid with $\gamma=1.5$, different choices for radius $R$, dimension $D$.\\
(a-b) Classification error (y-axis) at different values of $\alpha$ (x-axis) and levels of noise $\sigma^{2}$ (color coded). Compares manifold-slack classifiers using the optimal choice of $c$ (solid dark lines) and max-margin classifiers (light dashed lines) for $R=0.25$ and $D=20$ (a), $R=1.5$ and $D=2$ (b).
(c) Classification error using the optimal choice of $c$ (y-axis) at different values of $\alpha$ (x-axis), levels of noise $\sigma^{2}$ (columns) and values of $R$ (color coded, dimension is $D=10$). Compares manifold-slack classifiers (solid dark lines) and center-slack classifiers (light dashed lines).}
\label{fig:si-general-manifolds-generalization-error-optimal-c}
\end{figure}

\begin{figure}[H]
\center{\includegraphics[width=14.5cm]{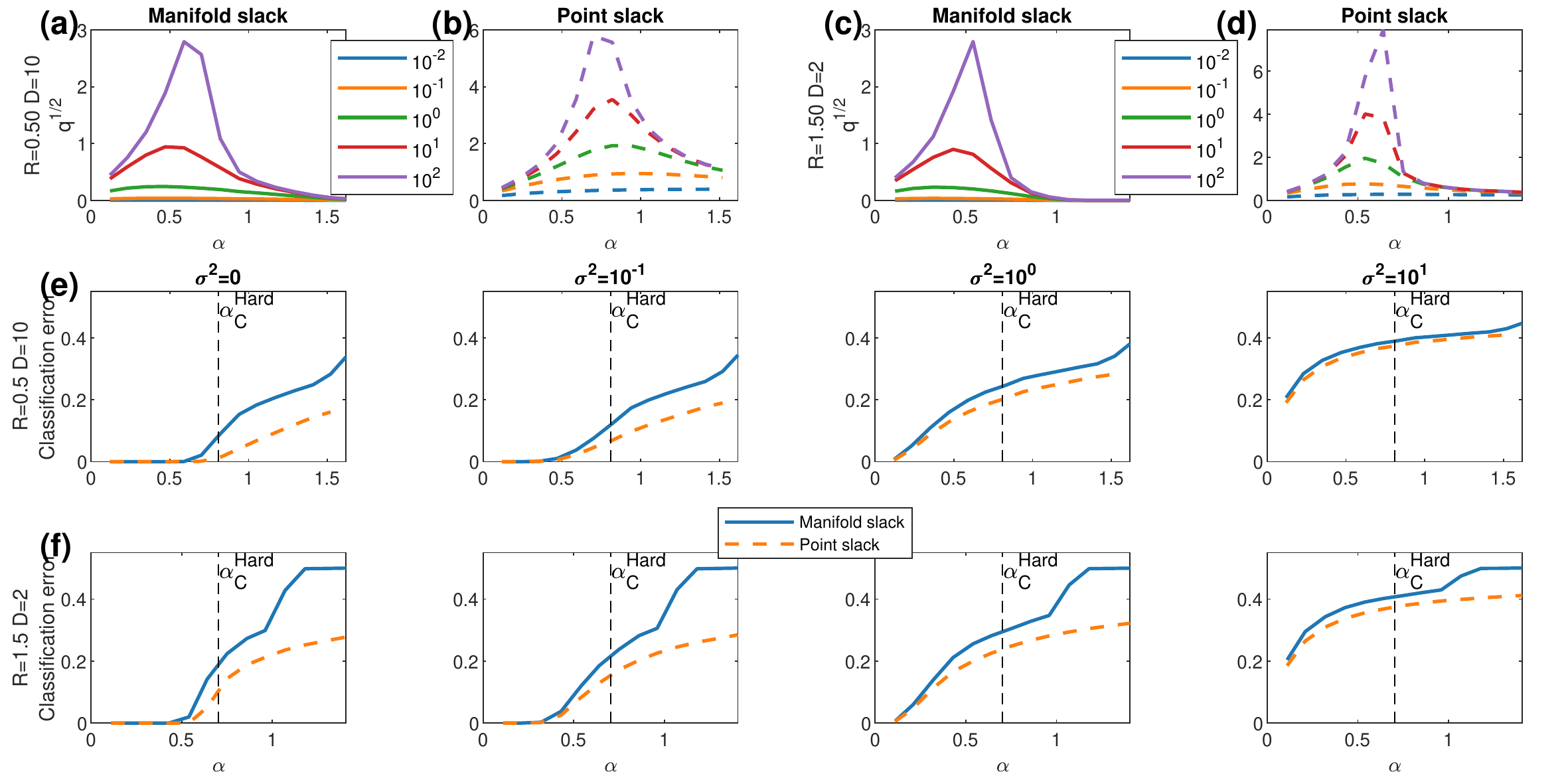}}
\caption{\textbf{Comparison of soft classification of point-cloud manifolds using the manifold-slack and point-slack methods.}
Results for manifolds of $M=100$ points from an ellipsoid with $\gamma=1.5$, different radius $R$ and dimension $D$. 
(a-d) The weights' norm $q^{1/2}$ (y-axis) using manifold-slack classifiers (solid lines, (a,c)) and point-slack classifiers (dashed lines, (b,d)) for different choices of $c$ (color coded), using $R=0.5$ and $D=10$ (a-b) or $R=1$ and $D=2$ (c-d).
(e-f) Classification error (y-axis) at different values of $\alpha$ (x-axis) using the optimal choice of $c$ at different levels of noise $\sigma^{2}$ (columns) and values of $R,D$ (indicated to the left of each row). Compares manifold-slack classifiers (solid lines) and point-slack classifiers (dashed lines). }
\label{fig:si-general-manifolds-points-slack-vs-manifold-slack-error-optimalC}
\end{figure}

\clearpage

\section{Algorithms} \label{sec:supp-algorithms}

\subsection{\label{subsec:algs-self-consistent-solution-points}Self-consistent
solution of points equations}

We rewrite the self-consistent equations for points \ref{eq:points-mf1},\ref{eq:points-mf2} as iterative update formulas. 

Given scalars $c,\alpha$:
\begin{enumerate}
\item Initialize $q_{0}=1$.
\item Update iteratively:
\begin{align}
a_{n-1} & =\alpha H\left(-1/\sqrt{q_{n-1}}\right)\\
k_{n} & =\frac{1}{2}\left(1-a_{n-1}-c^{-1}\right)+\frac{1}{2}\sqrt{\left(1+a_{n-1}+c^{-1}\right)^{2}-4a_{n-1}}\\
q_{n} & =\kappa_{0}^{-2}\left(\alpha\left(ck_{n}\right)^{2}/\left(1+ck_{n}\right)^{2}\right)
\end{align}
where $\kappa_{0}\left(\alpha\right)$ is the inverse function of
$\alpha_{0}\left(\kappa\right)$. 
\item Repeat updating until convergence,
defined as a change of less that $10^{-12}$ in $q$. 
\end{enumerate}

\subsection{\label{subsec:algs-self-consistent-solution-spheres}Self-consistent
solution of spheres equations}

The self-consistent equations for spheres \ref{eq:si-spheres-mf1},\ref{eq:si-spheres-mf2}
take the form $1=A(q,k)$ and $1=B(q,k)$ and may be solved numerically
by minimizing ${\cal L}=\frac{1}{2}\left(A\left(q,k\right)-1\right)^{2}+\frac{1}{2}\left(B\left(q,k\right)-1\right)^{2}$
with respect to $q,k$. Those non-linear equations depend on the initial conditions $q_{0},k_{0}$, which we choose as follows. 

Given a scalar $c$ and a sorted set of $\left\{\alpha_{i}\right\}$ values:
\begin{enumerate}
\item For the minimal $\alpha_{0}$, initialize $q,k$ using the expressions from the small $\alpha$ regime in the theory for points: $k_{0}=1-\alpha c/\left(1+c\right)$ and $q_{0}=\alpha c^{2}/\left(1+c\right)^{2}$.
\item Minimize ${\cal L}\left(q,k\right)$ from initial conditions $q_0,k_0$.
\item If this optimization finds a solution with ${\cal L}>10^{-4}$ interpret this as a failure and exit.
\item Use the previous solution
as initial conditions for the next $\alpha_{i}$: $q_{0}\left(\alpha_{i}\right)=q\left(\alpha_{i-1}\right)$,
$k_{0}\left(\alpha_{i}\right)=k\left(\alpha_{i-1}\right)$
\item Go back to (2).
\end{enumerate}

\subsection{\label{subsec:algs-numeric-classification-spheres}Classification of synthetic sphere manifolds}

Simulation results for spheres use a cutting-plane approach to find the classifier which optimally separates the entire spheres, i.e., corresponds to the case of an infinite number of samples, and is brought here for completeness.

Given $P$ spheres defined by their centers $\left\{ \boldsymbol{x}_{0}^{\mu}\in\mathbb{R}^{N}\right\}$ and axes $\left\{ U^{\mu}\in\mathbb{R}^{N\times D}\right\}$, with the radius absorbed into each axis, classification of the entire spheres with respect to target labels $\left\{y^{\mu}\right\}$ is done such that at stage $K$, exactly $K$ points from each sphere $X^{\mu}\in\mathbb{R}^{N\times K}$ are used: 
\begin{enumerate}
\item At $K=1$ we initialize $X^{\mu}=\left\{ \boldsymbol{x}_{0}^{\mu}\right\} $.
\item Get an optimal classifier:
\begin{equation}
\boldsymbol{w}^{*}=\arg\min_{\forall\boldsymbol{x}^{\mu}\in X^{\mu}\ y^{\mu}\boldsymbol{w}\cdot\boldsymbol{x}^{\mu}\ge1-s^{\mu}}\|\boldsymbol{w}\|^{2}+c\|\vec{s}\|^{2}
\end{equation}
and denote the achieved margin $\kappa_{0}=\min_{\mu}\min_{\boldsymbol{x}^{\mu}\in X^{\mu}}y^{\mu}\boldsymbol{w}\cdot\boldsymbol{x}^{\mu}+s^{\mu}$ (which is usually 1).
\item Find the worst point with respect to the current $\boldsymbol{w}^{*}$, $\boldsymbol{x}_{worst}^{\mu}=\boldsymbol{x}_{0}^{\mu}-U^{\mu}\hat{S}^{\mu}$ where $\vec{S}^{\mu}=y^{\mu}\boldsymbol{w}^{*T}U^{\mu}$ and $\hat{S}^{\mu}=\vec{S}^{\mu}/\|\vec{S}^{\mu}\|$ and add it to each set $X^{\mu}=X^{\mu}\cup\left\{ \boldsymbol{x}_{worst}^{\mu}\right\}$. Denote the margin on the worst points $\kappa_{1}=\min_{\mu}y^{\mu}\boldsymbol{w}^{*}\cdot\boldsymbol{x}_{worst}^{\mu}+s^{\mu}.$
\item Go back to $\left(2\right)$ as long as $\kappa_{0}\ge0$ and $\kappa_{1}\le\left(1-\epsilon\right)\kappa_{0}$ (where $\epsilon=10^{-4}$ was used).
\end{enumerate}

\subsection{\label{subsec:algs-least-square-method-v-s-general-manifolds}Least-squares calculation of anchor points for point-cloud manifolds}

Given $\vec{t}\in\mathbb{R}^{N},t_{0}\in\mathbb{R}$ and scalars $c,k,q$:
\begin{enumerate}
\item Use least-squares optimization with linear constraints: 
\begin{align}
\vec{v}^{*},v_{0}^{*} & =\arg\min\left\{ \|\vec{v}-\vec{t}\|^{2}+\frac{ck}{1+ck}\left(v_{0}-t_{0}\right)^{2}\right\} \\
 & \cdots s.t.\ \forall\vec{S}\in M\ v_{0}+\vec{v}\cdot\vec{S}\ge1/\sqrt{q}
\end{align}
\item Calculate $\tilde{S}$:
\begin{align}
\tilde{S} & =\frac{\vec{v}^{*}-\vec{t}}{\frac{ck}{1+ck}\left(v_{0}^{*}-t_{0}\right)}
\end{align}
\end{enumerate}

\subsection{\label{subsec:algs-iterative-method-v-s-general-manifolds}Iterative
calculation of anchor points for point-cloud manifolds}

Given $\vec{t}\in\mathbb{R}^{N},t_{0}\in\mathbb{R}$ and scalars $c,k,q$:
\begin{enumerate}
\item Initialize: $\vec{v}=\vec{t}$
\item Update iteratively:
\begin{align}
\tilde{S} & =\arg\min_{\vec{S}\in M}\vec{S}\cdot\vec{v}\\
\vec{v} & =\vec{t}+\frac{ck}{1+ck}\left(1/\sqrt{q}-\vec{v}\cdot\tilde{S}-t_{0}\right)\tilde{S}\\
v_{0} & =t_{0}+\frac{ck}{1+ck}\left(1/\sqrt{q}-\vec{v}\cdot\tilde{S}-t_{0}\right)
\end{align}
\item Repeat until convergence, defined by a change of less than $\max{\Delta\vec{v}}<10^{-6}$ or reaching $100$ iterations.
\end{enumerate}

\subsection{\label{subsec:algs-iterative-general-manifolds}Iterative algorithm for the properties of point-cloud manifolds}

An iterative algorithm for the calculation of $R_{M},D_{M},k,q$ at
specific values of $\alpha$ and $c$.
\begin{enumerate}
\item Initialize $k^{(0)}=1$ and $q^{(0)}=\alpha$.
\item For each iteration $n$ use $k^{(n)},q^{(n)}$. Sample $T$ values of standard Gaussian variables $\vec{t}\in\mathbb{R}^{N},t_{0}\in\mathbb{R}$:
\begin{enumerate}
\item Use one of the above methods (from section \ref{subsec:algs-least-square-method-v-s-general-manifolds}
or section \ref{subsec:algs-iterative-method-v-s-general-manifolds})
to sample anchor points $\tilde{S}\left(\vec{t},t_{0}\right)$.
\item Collect the resulting statistics $\left(\vec{t}\cdot\tilde{S}\right)^{2}/\tilde{S}^{2},\tilde{S}^{2}$. 
\end{enumerate}
\item Calculate current estimation of manifold geometry:
\begin{align}
R_{M}^{2} & =\left\langle \|\delta\tilde{S}\|^{2}\right\rangle _{\vec{t},t_{0}}\\
D_{M} & =\left\langle \left(\vec{t}\cdot\delta\tilde{S}\right)^{2}/\|\delta\tilde{S}\|^{2}\right\rangle _{\vec{t},t_{0}}
\end{align}
\item Find the order parameters $k^{(n+1)},q^{(n+1)}$ which solve the self-consistent
equations using the current estimation of $R_{M},D_{M}$, as described
in section \ref{subsec:algs-self-consistent-solution-spheres}.
\item Repeat until convergence, defined by a change of less than $10^{-3}$ in both $k$ and $q$ or reaching $50$ iterations.
\end{enumerate}

\clearpage
\section{Detailed derivations} \label{sec:supp-derivations}

\subsection{\label{subsec:appendix-points-replica-theory}Replica theory for points}

Consider $P$ points $\boldsymbol{x}^{\mu}\in\mathbb{R}^{N}$ and labels $y^{\mu}\in\{\pm1\}$; soft-margin classification
is defined as solving:

\begin{align}
\boldsymbol{w}^{*},\vec{s}^{*} & =\arg\min_{\boldsymbol{w}}\|\boldsymbol{w}\|^{2}+c\|\vec{s}\|^{2}\ s.t.\ h^{\mu}\ge1-s^{\mu}
\end{align}
where $\boldsymbol{w}^{*}\in\mathbb{R}^{N}$ and $\vec{s}^{*}\in\mathbb{R}_{+}^{P}$ and the fields:
\begin{equation}
h^{\mu}=y^{\mu}\left(\boldsymbol{w}\cdot\boldsymbol{x}^{\mu}+b\right)
\end{equation}
where we will assume $b=0$ for brevity. Denote the optimal loss:

\begin{align}
L^{*} & =\min_{\boldsymbol{w}}\|\boldsymbol{w}\|^{2}/N+c\|\vec{s}\|^{2}/N\ s.t.\ h^{\mu}\ge1-s^{\mu}
\end{align}

We write an expression for the volume $V\left(L,c\right)$ for $L=\|\boldsymbol{w}\|^{2}/N+c\|\vec{s}\|^{2}/N$
which would vanish for $L<L^{*}$:
\begin{align}
V\left(L,c\right) & =\int d^{N}\boldsymbol{w}\int d^{P}\vec{s}\prod_{\mu}^{P}\Theta\left(h^{\mu}-1+s^{\mu}\right)\delta\left(\|\boldsymbol{w}\|^{2}+c\|\vec{s}\|^{2}-NL\right)\\
 & =\int d^{N}\boldsymbol{w}\int d^{P}\vec{s}\int_{1-s^{\mu}}^{\infty}d^{P}h^{\mu}\int\frac{d^{P}\hat{h}^{\mu}}{2\pi}e^{i\left(y^{\mu}\boldsymbol{w}\cdot\boldsymbol{x}^{\mu}-h^{\mu}\right)\hat{h}^{\mu}}\int\frac{d\hat{l}}{2\pi}e^{i\left(\|\boldsymbol{w}\|^{2}+c\|\vec{s}\|^{2}-NL\right)\hat{l}}
\end{align}

We wish to calculate the values for which the volume vanishes assuming random (Gaussian) points $\boldsymbol{x}^\mu$ and random (binary) labels $y^\mu$. Using the replica identity:
\begin{equation}
\left[\log V\right]_{x,y}=\lim_{n\to0}\left[\frac{1}{n}\left(V^{n}-1\right)\right]_{x,y}
\end{equation}
it is enough to find $G$ which satisfies $\left[V^{n}\right]=e^{nG}$, to have that $\left[\log V\right]\approx G$.
Thus we consider $V^{n}$:
\begin{align}
V^{n} & =\int d^{n\times N}\boldsymbol{w}^{\alpha}\int d^{n\times P}\vec{s}^{\alpha}\int_{1-s^{\alpha,\mu}}^{\infty}d^{n\times P}h^{\alpha,\mu}\int\frac{d^{n\times P}\hat{h}^{\alpha,\mu}}{2\pi}\int\frac{d^{n}\hat{l}^{\alpha}}{2\pi}\\
 & \cdots e^{\sum_{\alpha}^{n}\sum_{\mu}^{P}i\left(y^{\mu}\sum_{i}^{N}w_{i}^{\alpha}x_{i}^{\mu}-h^{\alpha,\mu}\right)\hat{h}^{\alpha,\mu}+\sum_{\alpha}^{n}i\left(\|\boldsymbol{w}^{\alpha}\|^{2}+c\|\vec{s}^{\alpha}\|^{2}-NL\right)\hat{l}^{\alpha}}
\end{align}
Gaussian integral on $x_{i}^{\mu}\sim{\cal N}\left(0,1/N\right)$:
\begin{align}
\left[e^{\sum_{\alpha}^{n}\sum_{\mu}^{P}iy^{\mu}\sum_{i}^{N}w_{i}^{\alpha}x_{i}^{\mu}\hat{h}^{\alpha,\mu}}\right]_{x} & =e^{-\frac{1}{2}\sum_{\mu}^{P}\left(\sum_{\alpha,\beta}^{n}\left(\frac{1}{N}\sum_{i}^{N}w_{i}^{\alpha}w_{i}^{\beta}\right)\hat{h}^{\alpha,\mu}\hat{h}^{\beta,\mu}\right)}
\end{align}
so denote:
\begin{align}
q_{\alpha\beta} & =\frac{1}{N}\sum_{i}^{N}w_{i}^{\alpha}w_{i}^{\beta}
\end{align}
we have:
\begin{align}
\left[V^{n}\right]_{x} & =\int d^{n\times n}q_{\alpha\beta}\int\frac{d^{n\times n}\hat{q}_{\alpha\beta}}{2\pi}\int d^{n\times N}\boldsymbol{w}^{\alpha}\int d^{n\times P}\vec{s}^{\alpha}\int_{1-s^{\alpha,\mu}}^{\infty}d^{n\times P}h^{\alpha,\mu}\int\frac{d^{n\times P}\hat{h}^{\alpha,\mu}}{2\pi}\int\frac{d^{n}\hat{l}^{\alpha}}{2\pi}\\
 & \cdots e^{-i\sum_{\alpha,\beta}^{n}Nq_{\alpha\beta}\hat{q}_{\alpha\beta}-i\sum_{\alpha}^{n}\sum_{\mu}^{P}h^{\alpha,\mu}\hat{h}^{\alpha,\mu}-iN\sum_{\alpha}^{n}L\hat{l}^{\alpha}}\\
 & \cdots e^{-\frac{1}{2}\sum_{\mu}^{P}\sum_{\alpha,\beta}^{n}q_{\alpha\beta}\hat{h}^{\alpha,\mu}\hat{h}^{\beta,\mu}+i\sum_{\alpha}^{n}c\sum_{\mu}^{P}\left(s^{\alpha,\mu}\right)^{2}\hat{l}^{\alpha}+i\sum_{\alpha}^{n}\sum_{i}^{N}w_{i}^{\alpha}w_{i}^{\alpha}\hat{l}^{\alpha}+i\sum_{\alpha,\beta}^{n}\sum_{i}^{N}w_{i}^{\alpha}w_{i}^{\beta}\hat{q}_{\alpha\beta}}
\end{align}
Integration over $\hat{h}^{\alpha,\mu}$, $w_{i}^{\alpha}$ using
$\int\frac{d^{n}x}{\sqrt{2\pi}}e^{-x^{T}Ax/2+b^{T}x}=e^{-\frac{1}{2}\log\det A+b^{T}A^{-1}b/2}$:
\begin{align}
I_{1} & \doteq\int\frac{d^{n\times P}\hat{h}^{\alpha,\mu}}{\sqrt{2\pi}}e^{-\frac{1}{2}\sum_{\mu}^{P}\sum_{\alpha,\beta}^{n}q_{\alpha\beta}\hat{h}^{\alpha,\mu}\hat{h}^{\beta,\mu}-i\sum_{\alpha}^{n}\sum_{\mu}^{P}h^{\alpha,\mu}\hat{h}^{\alpha,\mu}}\\
 & =e^{-\frac{1}{2}\sum_{\mu}^{P}\sum_{\alpha,\beta}^{n}q_{\alpha\beta}^{-1}h^{\alpha,\mu}h^{\beta,\mu}-\frac{P}{2}\log\det q}\\
I_{2} & \doteq\int d^{n\times N}\boldsymbol{w}^{\alpha}e^{i\sum_{\alpha,\beta}^{n}\sum_{i}^{N}w_{i}^{\alpha}w_{i}^{\beta}\hat{q}_{\alpha\beta}+i\sum_{\alpha}^{n}\sum_{i}^{N}w_{i}^{\alpha}w_{i}^{\alpha}\hat{l}^{\alpha}}\\
 & =\int d^{n\times N}\boldsymbol{w}^{\alpha}e^{-\frac{1}{2}\sum_{\alpha,\beta}^{n}\sum_{i}^{N}w_{i}^{\alpha}w_{i}^{\beta}\left(-2i\hat{q}_{\alpha\beta}-\delta_{\alpha\beta}2i\hat{l}^{\alpha}\right)}\\
 & =e^{-\frac{N}{2}\log\det\left(-2i\hat{q}_{\alpha\beta}-\delta_{\alpha\beta}2i\hat{l}^{\alpha}\right)}
\end{align}
so that:

\begin{align}
\left[V^{n}\right]_{x} & =\int d^{n\times n}q_{\alpha\beta}\int\frac{d^{n\times n}\hat{q}_{\alpha\beta}}{2\pi}\int d^{n\times P}\vec{s}^{\alpha}\int_{1-s^{\alpha,\mu}}^{\infty}d^{n\times P}h^{\alpha,\mu}\int\frac{d^{n}\hat{l}^{\alpha}}{2\pi}\\
 & \cdots e^{-\frac{N}{2}\log\det\left(-2i\hat{q}_{\alpha\beta}-\delta_{\alpha\beta}2i\hat{l}^{\alpha}\right)-iN\sum_{\alpha,\beta}^{n}q_{\alpha\beta}\hat{q}_{\alpha\beta}-iN\sum_{\alpha}^{n}L\hat{l}^{\alpha}}\\
 & \cdots e^{i\sum_{\alpha}^{n}c\sum_{\mu}^{P}\left(s^{\alpha,\mu}\right)^{2}\hat{l}^{\alpha}-\frac{1}{2}\sum_{\mu}^{P}\sum_{\alpha,\beta}^{n}q_{\alpha\beta}^{-1}h^{\alpha,\mu}h^{\beta,\mu}-\frac{P}{2}\log\det q}
\end{align}

Rewrite it such that the all the $\mu$-s are decoupled:

\begin{align}
I & \doteq e^{-\frac{P}{2}\log\det q}\int\frac{d^{n\times P}\vec{s}^{\alpha}}{\sqrt{2\pi}}\int_{1-s^{\alpha,\mu}}^{\infty}e^{-\frac{1}{2}\sum_{\mu}^{P}\sum_{\alpha,\beta}^{n}q_{\alpha\beta}^{-1}h^{\alpha,\mu}h^{\beta,\mu}+i\sum_{\alpha}^{n}c\sum_{\mu}^{P}\left(s^{\alpha,\mu}\right)^{2}\hat{l}^{\alpha}}\\
 & =\left(e^{-\frac{1}{2}\log\det q}\int\frac{d^{n}s^{\alpha}}{\sqrt{2\pi}}\int_{1-s^{\alpha}}^{\infty}e^{-\frac{1}{2}\sum_{\alpha,\beta}^{n}q_{\alpha\beta}^{-1}h^{\alpha}h^{\beta}+i\sum_{\alpha}^{n}c\left(s^{\alpha}\right)^{2}\hat{l}^{\alpha}}\right)^{P}
\end{align}
we have:

\begin{align}
\left[V^{n}\right]_{x} & =\int d^{n\times n}q_{\alpha\beta}\int\frac{d^{n\times n}\hat{q}_{\alpha\beta}}{2\pi}\int\frac{d^{n}\hat{l}^{\alpha}}{\sqrt{2\pi}}e^{-nNG_{0}-nNG_{1}}\\
G_{0} & =\frac{i}{n}\sum_{\alpha,\beta}^{n}q_{\alpha\beta}\hat{q}_{\alpha\beta}+\frac{1}{2n}\log\det\left(-2i\hat{q}_{\alpha\beta}-\delta_{\alpha\beta}2i\hat{l}^{\alpha}\right)+\frac{i}{n}\sum_{\alpha}^{n}L\hat{l}^{\alpha}\\
G_{1} & =\frac{\alpha}{2n}\log\det q-\frac{\alpha}{n}\log\int\frac{d^{n}s^{\alpha}}{\sqrt{2\pi}}\int_{1-s^{\alpha}}^{\infty}d^{n}h^{\alpha}e^{iC\sum_{\alpha}^{n}\left(s^{\alpha}\right)^{2}\hat{l}^{\alpha}-\frac{1}{2}\sum_{\alpha,\beta}^{n}q_{\alpha\beta}^{-1}h^{\alpha}h^{\beta}}
\end{align}

Assuming replica symmetry:
\begin{align}
q_{\alpha\beta} & =q+(q_{0}-q)\delta_{\alpha\beta}\\
\hat{q}_{\alpha\beta} & =\hat{q}+(\hat{q}_{0}-\hat{q})\delta_{\alpha\beta}\\
\hat{l}^{\alpha} & =\hat{l}
\end{align}
we have:

\begin{align}
\frac{1}{n}\sum_{\alpha,\beta}^{n}q_{\alpha\beta}\hat{q}_{\alpha\beta} & \approx\hat{q}_{0}q_{0}-\hat{q}q\\
\log\det\left[q\right] & \approx n\log\left(q_{0}-q\right)+n\frac{q}{q_{0}-q}\\
\log\det|-2i\hat{q}-2i\hat{l}| & \approx n\log\left(-2i\hat{l}-2i\hat{q}_{0}+2i\hat{q}\right)+n\frac{-2i\hat{q}}{-2i\hat{l}-2i\hat{q}_{0}+2i\hat{q}}\\
q_{\alpha\beta}^{-1} & \approx\frac{1}{\left(q_{0}-q\right)}\delta_{\alpha\beta}-\frac{q}{\left(q_{0}-q\right)^{2}}
\end{align}

so that we get, changing all $-i\hat{l}\to\hat{l}$,$-i\hat{q}\to\hat{q}$,$-i\hat{q}_{0}\to\hat{q}_{0}$: 

\begin{align}
\left[V^{n}\right]_{x} & =\int dq\int dq_{0}\int d\hat{q}\int d\hat{q}_{0}\int\frac{d\hat{l}}{\sqrt{2\pi}}e^{-nNG_{0}-nNG_{1}}\\
G_{0} & =-\left[\hat{q}_{0}q_{0}-\hat{q}q\right]+\frac{1}{2}\left[\log\left(2\hat{l}+2\hat{q}_{0}-2\hat{q}\right)+\frac{2\hat{q}}{2\hat{l}+2\hat{q}_{0}-2\hat{q}}\right]-L\hat{l}\\
G_{1} & =\frac{\alpha}{2}\left[\log\left(q_{0}-q\right)+\frac{q}{q_{0}-q}\right]-\frac{\alpha}{n}\log\int\frac{d^{n}s^{\alpha}}{\sqrt{2\pi}}\int_{1-s^{\alpha}}^{\infty}\frac{d^{n}h^{\alpha}}{\sqrt{2\pi}}e^{-\frac{1}{2}\frac{1}{q_{0}-q}\sum_{\alpha}^{n}\left(h^{\alpha}\right)^{2}+\frac{1}{2}\frac{q}{\left(q_{0}-q\right)^{2}}\left(\sum_{\alpha}^{n}h^{\alpha}\right)^{2}-c\sum_{\alpha}^{n}\left(s^{\alpha}\right)^{2}\hat{l}}
\end{align}

Assuming the behavior in the thermodynamic limit $N\to\infty$ is dominated by the maximum of the integral, we calculate the derivatives of $G_0$:
\begin{align}
0 & =\frac{\partial G_{0}}{\partial\hat{q}}=q+\frac{1}{2}\left[-\frac{1}{\hat{l}+\hat{q}_{0}-\hat{q}}+\frac{\left(\hat{l}+\hat{q}_{0}-\hat{q}\right)+\hat{q}}{\left(\hat{l}+\hat{q}_{0}-\hat{q}\right)^{2}}\right]\\
0 & =\frac{\partial G_{0}}{\partial\hat{q}_{0}}=-q_{0}+\frac{1}{2}\left[\frac{1}{\hat{l}+\hat{q}_{0}-\hat{q}}-\frac{\hat{q}}{\left(\hat{l}+\hat{q}_{0}-\hat{q}\right)^{2}}\right]
\end{align}
so that:
\begin{align}
q & =-\frac{1}{2}\frac{\hat{q}}{\left(\hat{l}+\hat{q}_{0}-\hat{q}\right)^{2}}\\
q_{0} & =\frac{1}{2}\frac{1}{\hat{l}+\hat{q}_{0}-\hat{q}}+q\\
q_{0}-q & =\frac{1}{2}\frac{1}{\hat{l}+\hat{q}_{0}-\hat{q}}\\
\frac{q}{\left(q_{0}-q\right)^{2}} & =-\frac{1}{2}\frac{\hat{q}}{\left(\hat{l}+\hat{q}_{0}-\hat{q}\right)^{2}}/\frac{1}{4}\frac{1}{\left(\hat{l}+\hat{q}_{0}-\hat{q}\right)^{2}}=-2\hat{q}\\
q_{0}\hat{q}_{0}-q\hat{q} & =\frac{1}{2}\frac{\hat{q}_{0}}{\hat{l}+\hat{q}_{0}-\hat{q}}-\frac{1}{2}\frac{\hat{q}\hat{q}_{0}}{\left(\hat{l}+\hat{q}_{0}-\hat{q}\right)^{2}}+\frac{1}{2}\frac{\hat{q}\hat{q}}{\left(\hat{l}+\hat{q}_{0}-\hat{q}\right)^{2}}=\frac{1}{2}-\hat{l}q_{0}
\end{align}
and $G_{0}$ becomes:

\begin{align}
G_{0} & =-\frac{1}{2}+\left(q_{0}-L\right)\hat{l}-\frac{1}{2}\log\left(q_{0}-q\right)-\frac{1}{2}\frac{q}{q_{0}-q}
\end{align}

For $G_{1}$ we have:

\begin{align}
G_{1} & =\frac{\alpha}{2}\log\left(q_{0}-q\right)+\frac{\alpha}{2}\frac{q}{q_{0}-q}-\frac{\alpha}{n}\log\int\frac{d^{n}s^{\alpha}}{\sqrt{2\pi}}\int_{1-s^{\alpha}}^{\infty}\frac{d^{n}h^{\alpha}}{\sqrt{2\pi}}e^{-\frac{1}{2}\frac{1}{q_{0}-q}\sum_{\alpha}^{n}\left(h^{\alpha}\right)^{2}+\frac{1}{2}\frac{q}{\left(q_{0}-q\right)^{2}}\left(\sum_{\alpha}^{n}h^{\alpha}\right)^{2}-c\sum_{\alpha}^{n}\left(s^{\alpha}\right)^{2}\hat{l}}
\end{align}
and using the Hubbard-Stratonovich transform $e^{ay^{2}/2}=\int\frac{dt}{\sqrt{2\pi}}e^{-t^{2}/2+\sqrt{a}yt}$:
\begin{align}
I & =\int\frac{d^{n}s^{\alpha}}{\sqrt{2\pi}}\int_{1-s^{\alpha}}^{\infty}\frac{d^{n}h^{\alpha}}{\sqrt{2\pi}}e^{-\frac{1}{2}\frac{1}{q_{0}-q}\sum_{\alpha}^{n}\left(h^{\alpha}\right)^{2}+\frac{1}{2}\frac{q}{\left(q_{0}-q\right)^{2}}\left(\sum_{\alpha}^{n}h^{\alpha}\right)^{2}-c\sum_{\alpha}^{n}\left(s^{\alpha}\right)^{2}\hat{l}}\\
 & =\int Dt\int\frac{d^{n}s^{\alpha}}{\sqrt{2\pi}}\int_{1-s^{\alpha}}^{\infty}\frac{d^{n}h^{\alpha}}{\sqrt{2\pi}}e^{-\frac{1}{2}\frac{1}{q_{0}-q}\sum_{\alpha}^{n}\left(h^{\alpha}\right)^{2}+t\frac{\sqrt{q}}{q_{0}-q}\sum_{\alpha}^{n}h_{\mu}^{\alpha}-c\sum_{\alpha}^{n}\left(s^{\alpha}\right)^{2}\hat{l}}
\end{align}
so that $G_{1}$ decouples into $n$ terms:

\begin{align}
G_{1} & =\frac{\alpha}{2}\log\left(q_{0}-q\right)+\frac{\alpha}{2}\frac{q}{q_{0}-q}-\frac{\alpha}{n}\log\int\frac{d^{n}s^{\alpha}}{\sqrt{2\pi}}\int_{1-s^{\alpha}}^{\infty}\frac{d^{n}h^{\alpha}}{\sqrt{2\pi}}e^{-\frac{1}{2}\frac{1}{q_{0}-q}\sum_{\alpha}^{n}\left(h^{\alpha}\right)^{2}+t\frac{\sqrt{q}}{q_{0}-q}\sum_{\alpha}^{n}h_{\mu}^{\alpha}-c\sum_{\alpha}^{n}\left(s^{\alpha}\right)^{2}\hat{l}}\\
 & =\frac{\alpha}{2}\log\left(q_{0}-q\right)+\frac{\alpha}{2}\frac{q}{q_{0}-q}-\frac{\alpha}{n}\log\int Dt\left(\int\frac{ds}{\sqrt{2\pi}}\int_{1-s}^{\infty}\frac{dh}{\sqrt{2\pi}}e^{-\frac{1}{2}\frac{1}{q_{0}-q}h^{2}+t\frac{\sqrt{q}}{q_{0}-q}h-Cs^{2}\hat{l}}\right)^{n}
\end{align}
and using the replica identity $\log\int Dt\,z(t)^{n}\approx\log\left[1+n\int Dt\log z(t)\right]\approx n\int Dt\log z(t)$
for $n\to0$: 

\begin{align}
G_{1} & =\frac{\alpha}{2}\log\left(q_{0}-q\right)+\frac{\alpha}{2}\frac{q}{q_{0}-q}-\alpha\int Dt\log\int\frac{ds}{\sqrt{2\pi}}\int_{1-s}^{\infty}\frac{dh}{\sqrt{2\pi}}e^{-\frac{1}{2}\frac{1}{q_{0}-q}h^{2}+t\frac{\sqrt{q}}{q_{0}-q}h-Cs^{2}\hat{l}}
\end{align}
by changing $h\to h-s$ and integrating over $s$ using $\int\frac{dx}{\sqrt{2\pi}}e^{-x^{2}/2a+bx}=e^{\frac{1}{2}\log a+\frac{1}{2}b^{2}a}$:

\begin{align}
I & \doteq\int\frac{ds}{\sqrt{2\pi}}\int_{1-s}^{\infty}\frac{dh}{\sqrt{2\pi}}e^{-\frac{1}{2}\frac{1}{q_{0}-q}h^{2}+t\frac{\sqrt{q}}{q_{0}-q}h-Cs^{2}\hat{l}}\\
 & =\int\frac{ds}{\sqrt{2\pi}}\int_{1}^{\infty}\frac{dh}{\sqrt{2\pi}}e^{-\frac{1}{2}\frac{1}{q_{0}-q}h^{2}+\frac{1}{q_{0}-q}hs-\frac{1}{2}\frac{1}{q_{0}-q}s^{2}+t\frac{\sqrt{q}}{q_{0}-q}h-t\frac{\sqrt{q}}{q_{0}-q}s-c\hat{l}s^{2}}\\
 & =\int_{1}^{\infty}\frac{dh}{\sqrt{2\pi}}e^{-\frac{1}{2}\frac{1}{q_{0}-q}h^{2}+t\frac{\sqrt{q}}{q_{0}-q}h+\frac{1}{2}\log\frac{q_{0}-q}{1+2C\hat{l}\left(q_{0}-q\right)}+\frac{1}{2}\frac{q_{0}-q}{1+2C\hat{l}\left(q_{0}-q\right)}\left(\frac{h-t\sqrt{q}}{q_{0}-q}\right)^{2}}
\end{align}
and by completion to square:
\begin{align}
G_{1} & =\frac{\alpha}{2}\log\left(q_{0}-q\right)+\frac{\alpha}{2}\frac{q}{q_{0}-q}-\alpha\int Dt\log\int_{1}^{\infty}\frac{dh}{\sqrt{2\pi}}e^{-\frac{1}{2}\frac{1}{q_{0}-q}h^{2}+t\frac{\sqrt{q}}{q_{0}-q}h+\frac{1}{2}\log\frac{q_{0}-q}{1+2C\hat{l}\left(q_{0}-q\right)}+\frac{1}{2}\frac{q_{0}-q}{1+2C\hat{l}\left(q_{0}-q\right)}\left(\frac{h-t\sqrt{q}}{q_{0}-q}\right)^{2}}\\
 & =\frac{\alpha}{2}\frac{q}{q_{0}-q}-\alpha\int Dt\log e^{\frac{1}{2}\frac{t^{2}q}{q_{0}-q}}\int_{1}^{\infty}\frac{dh}{\sqrt{2\pi}}e^{-\frac{1}{2}\frac{\left(h-t\sqrt{q}\right)^{2}}{q_{0}-q}\left(1-\frac{1}{1+2C\hat{l}\left(q_{0}-q\right)}\right)-\frac{1}{2}\log\left(q_{0}-q\right)-\frac{1}{2}\log\frac{1+2C\hat{l}\left(q_{0}-q\right)}{q_{0}-q}}\\
 & =\frac{\alpha}{2}\frac{q}{q_{0}-q}-\alpha\int Dt\left[\frac{1}{2}\frac{t^{2}q}{q_{0}-q}+\log\int_{1}^{\infty}\frac{dh}{\sqrt{2\pi}}e^{-\frac{1}{2}\frac{\left(h-t\sqrt{q}\right)^{2}2C\hat{l}}{1+2C\hat{l}\left(q_{0}-q\right)}-\frac{1}{2}\log\left(1+2C\hat{l}\left(q_{0}-q\right)\right)}\right]\\
 & =-\alpha\int Dt\log\int_{1}^{\infty}\frac{dh}{\sqrt{2\pi}}e^{-\frac{1}{2}\frac{\left(h-t\sqrt{q}\right)^{2}2C\hat{l}}{1+2C\hat{l}\left(q_{0}-q\right)}-\frac{1}{2}\log\left(1+2C\hat{l}\left(q_{0}-q\right)\right)}
\end{align}
and by a change of variable $h=h_{0}\sqrt{q}$:
\begin{align}
G_{1} & =-\alpha\int Dt\log\int_{1/\sqrt{q}}^{\infty}\frac{dh_{0}\sqrt{q}}{\sqrt{2\pi}}e^{-\frac{1}{2}\frac{\left(h_{0}\sqrt{q}-t\sqrt{q}\right)^{2}2C\hat{l}}{1+2C\hat{l}\left(q_{0}-q\right)}-\frac{1}{2}\log\left(1+2C\hat{l}\left(q_{0}-q\right)\right)}\\
 & =-\alpha\int Dt\log\int_{1/\sqrt{q}}^{\infty}\frac{dh_{0}}{\sqrt{2\pi}}e^{-\frac{1}{2}\frac{\left(h_{0}-t\right)^{2}2C\hat{l}q}{1+2C\hat{l}\left(q_{0}-q\right)}-\frac{1}{2}\log\left(1+2C\hat{l}\left(q_{0}-q\right)\right)/q}
\end{align}

Thus we conclude:
\begin{align}
\left[V^{n}\right]_{x} & =\int dq\int dq_{0}\int d\hat{q}\int d\hat{q}_{0}\int\frac{d\hat{l}}{\sqrt{2\pi}}e^{-nNG_{0}-nNG_{1}}\\
G_{0} & =-\frac{1}{2}+\left(q_{0}-L\right)\hat{l}-\frac{1}{2}\log\left(q_{0}-q\right)-\frac{1}{2}\frac{q}{q_{0}-q}\\
G_{1} & =-\alpha\int Dt\log\int_{1/\sqrt{q}}^{\infty}\frac{dh_{0}}{\sqrt{2\pi}}e^{-\frac{1}{2}\frac{\left(h_{0}-t\right)^{2}2C\hat{l}q}{1+2C\hat{l}\left(q_{0}-q\right)}-\frac{1}{2}\log\left(1+2C\hat{l}\left(q_{0}-q\right)\right)/q}
\end{align}

Now $G=G_{0}+G_{1}$ is thus, renaming $\hat{l}=\hat{l}_{0}/\left(q_{0}-q\right)$ and $h_0$ to $h$:

\begin{align}
G & =-\frac{1}{2}+\left(q_{0}-L\right)\frac{\hat{l}_{0}}{q_{0}-q}-\frac{1}{2}\log\left(q_{0}-q\right)-\frac{1}{2}\frac{q}{q_{0}-q}\\
 & \cdots-\alpha\int Dt\log\int_{1/\sqrt{q}}^{\infty}\frac{dh}{\sqrt{2\pi}}e^{-\frac{1}{2}\frac{\left(h-t\right)^{2}}{q_{0}-q}\frac{2C\hat{l}_{0}q}{1+2C\hat{l}_{0}}-\frac{1}{2}\log\left(1+2C\hat{l}_{0}\right)/q}
\end{align}
so the limit of $q\to q_{0}$: 

\begin{align}
\lim_{q\to q_{0}}\left(q_{0}-q\right)G_{0} & =\left(q_{0}-L\right)\hat{l}_{0}-\frac{1}{2}q_{0}\\
\lim_{q\to q_{0}}\left(q_{0}-q\right)G_{1} & =-\alpha\lim_{q\to q_{0}}\left(q_{0}-q\right)\int Dt\log\int_{1/\sqrt{q}}^{\infty}\frac{dh}{\sqrt{2\pi}}e^{-\frac{1}{2}\frac{\left(h-t\right)^{2}}{q_{0}-q}\frac{2C\hat{l}_{0}q}{1+2C\hat{l}_{0}}-\frac{1}{2}\log\left(1+2C\hat{l}_{0}\right)/q}\\
 & =-\alpha\int Dt\lim_{q\to q_{0}}\left(q_{0}-q\right)\log\max_{h>1/\sqrt{q}}e^{-\frac{1}{2}\frac{\left(h-t\right)^{2}}{q_{0}-q}\frac{2C\hat{l}_{0}q}{1+2C\hat{l}_{0}}-\frac{1}{2}\log\left(1+2C\hat{l}_{0}\right)/q}\\
 & =-\alpha\int Dt\lim_{q\to q_{0}}\left(q_{0}-q\right)\max_{h>1/\sqrt{q}}-\frac{1}{2}\frac{\left(h-t\right)^{2}}{q_{0}-q}\frac{2C\hat{l}_{0}q}{1+2C\hat{l}_{0}}-\frac{1}{2}\log\left(1+2C\hat{l}_{0}\right)/q\\
 & =\frac{\alpha}{2}\frac{2C\hat{l}_{0}q_{0}}{1+2C\hat{l}_{0}}\int Dt\min_{h>1/\sqrt{q}}\left(h-t\right)^{2}
\end{align}
So to sum up:
\begin{equation}
\lim_{q\to q_{0}}\left(q_{0}-q\right)G=\left(q_{0}-L\right)\hat{l}_{0}-\frac{1}{2}q_{0}+\frac{\alpha}{2}\frac{2C\hat{l}_{0}q_{0}}{1+2C\hat{l}_{0}}\alpha_{0}^{-1}\left(1/\sqrt{q}\right)
\end{equation}
for Gardner's $\alpha_{0}$:
\begin{equation}
\alpha_{0}^{-1}\left(\kappa\right)=\int_{-\infty}^{\kappa}Dt\left(\kappa-t\right)^{2}
\end{equation}
Thus we have:

\begin{equation}
1=\left(1-L/q\right)2\hat{l}_{0}+\alpha\frac{2C\hat{l}_{0}}{1+2C\hat{l}_{0}}\alpha_{0}^{-1}\left(1/\sqrt{q}\right)
\end{equation}
and denoting $k=2\hat{l}_{0}$ we have the mean-field equation:

\begin{align}
L/q & =1-\frac{1}{k}+\alpha\frac{c}{1+ck}\alpha_{0}^{-1}\left(1/\sqrt{q}\right)
\end{align}

\bigskip

\subsection{\label{subsec:appendix-points-self-consistent}Self-consistent equations for points}

The self-consistent equations for points, equations \ref{eq:points-mf1}-\ref{eq:points-mf2}, are derived directly from the mean-field equation \ref{eq:points-L-mf} by assuming that for the optimal loss we expect saddle-point conditions on $L\left(q,k\right)$, namely that $0=\frac{\partial L}{\partial q}=\frac{\partial L}{\partial k}$:
\begin{align}
0=\frac{\partial}{\partial k}L	& =-q\alpha\frac{1}{\left(c+k\right)^{2}}\alpha_{0}^{-1}\left(1/\sqrt{q}\right)+q/k^{2}\\
1	& =\frac{\left(ck\right)^{2}}{\left(1+ck\right)^{2}}\alpha\alpha_{0}^{-1}\left(1/\sqrt{q}\right)\\
0=\frac{\partial}{\partial q}L	& =1-1/k+\frac{c\alpha}{1+ck}\frac{\partial}{\partial q}\left[q\alpha_{0}^{-1}\left(1/\sqrt{q}\right)\right]\\
	& =\frac{k-1}{k}+\frac{c\alpha}{1+ck}\int_{-\infty}^{1/\sqrt{q}}Dt\left(1/\sqrt{q}-t\right)t\\
	& =\frac{k-1}{k}+\frac{c\alpha}{1+ck}H\left(-1/\sqrt{q}\right)\\
1-k	& =\frac{ck}{1+ck}\alpha H\left(-1/\sqrt{q}\right)
\end{align}

Those self-consistent equations can be evaluated for the limits $\alpha\to0$ and $\alpha\to\infty$. 

When $\alpha\to0$ we have $k\to1$ and $q\to0$ so that $\alpha_{0}^{-1}\left(1/\sqrt{q}\right)\approx1/q$ and thus:
\begin{align}
k	&=1-\frac{ck}{1+ck}\alpha\approx =1-\frac{c}{1+c}\alpha\\
\sqrt{q}	&=\frac{ck}{1+ck}\sqrt{\alpha}
	\approx\frac{c}{1+c}\sqrt{\alpha}
\end{align}

When $\alpha\to\infty$ we have $k\to0$ and $q\to0$ so that scaling $k=k_{0}/\alpha$ we have:
\begin{align}
1	&=\frac{ck_{0}}{\left(1-k_{0}/\alpha\right)\left(1+ck_{0}/\alpha\right)}
	\approx ck_{0}\\
k	&=1/c\alpha\\
q	&=\frac{\left(ck\right)^{2}}{\left(1+ck\right)^{2}}\alpha
	=\frac{\left(1/\alpha\right)^{2}}{\left(1+1/\alpha\right)^{2}}\alpha
	=\frac{\alpha}{\left(\alpha+1\right)^{2}}
	\approx1/\alpha
\end{align}

So the limits:
\begin{align}
k	=\begin{cases}
1-\frac{c}{1+c}\alpha & \alpha\to0\\
1/c\alpha & \alpha\to\infty
\end{cases}\\
q	=\begin{cases}
\frac{c^2}{\left(1+c\right)^{2}}\alpha & \alpha\to0\\
1/\alpha & \alpha\to\infty
\end{cases}
\end{align}

The limit $c\to\infty$ exhibit different behavior for $\alpha<2$ and $\alpha>2$. For $\alpha<2$ the solution satisfies 
\begin{equation}
1	=\alpha\alpha_{0}^{-1}\left(1/\sqrt{q}\right)
\end{equation}
which is the max-margin solution. On the other hand, for $\alpha\ge2$ and $c\to\infty$ we have that $\lim_{c\to\infty}k=0$ with finite $q,K=\lim_{c\to\infty}ck$, which obey the self-consistent equations:
\begin{align}
1	& =\frac{K^{2}}{\left(1+K\right)^2}\alpha\alpha_{0}^{-1}\left(1/\sqrt{q}\right)\\
1	& =\frac{K}{1+K}\alpha H\left(-1/\sqrt{q}\right)
\end{align}
The relation between $\alpha$ and $q$ thus becomes:
\begin{equation}
\alpha =\alpha_{0}^{-1}\left(1/\sqrt{q}\right)/ H^2\left(-1/\sqrt{q}\right)
\end{equation}

\bigskip

\subsection{\label{subsec:appendix-points-fields-slack-distribution}Field and slack distribution for points}

Note we have in the theory the inner integral given as, using our
notation $k=2\hat{l}_{0}=2\hat{l}\left(q_{0}-q\right)$:

\begin{align}
I & \doteq\int\frac{ds}{\sqrt{2\pi}}\int_{1-s}^{\infty}\frac{dh}{\sqrt{2\pi}}e^{-\frac{1}{2}\frac{1}{q_{0}-q}h^{2}+\frac{\sqrt{qt}}{q_{0}-q}h-Cs^{2}\hat{l}}\\
 & =\int\frac{ds}{\sqrt{2\pi}}\int_{1-s}^{\infty}\frac{dh}{\sqrt{2\pi}}e^{-\frac{1}{2}\frac{1}{q_{0}-q}\left(h-\sqrt{qt}\right)^{2}+\frac{1}{2}\frac{qt^{2}}{q_{0}-q}-\frac{1}{2}\frac{1}{q_{0}-q}cks^{2}}
\end{align}
So that:

\begin{align}
G_{1} & =\frac{\alpha}{2}\log\left(q_{0}-q\right)+\frac{\alpha}{2}\frac{q}{q_{0}-q}-\alpha\int Dt\log I\\
 & =\frac{\alpha}{2}\frac{q}{q_{0}-q}-\alpha\int Dt\log e^{\frac{1}{2}\frac{qt^{2}}{q_{0}-q}}\int\frac{ds}{\sqrt{2\pi}}\int_{1-s}^{\infty}\frac{dh}{\sqrt{2\pi}}e^{-\frac{1}{2}\log\left(q_{0}-q\right)-\frac{1}{2}\frac{1}{q_{0}-q}\left(h-\sqrt{qt}\right)^{2}-\frac{1}{2}\frac{1}{q_{0}-q}cks^{2}}\\
 & =-\alpha\int Dt\log\int\frac{ds}{\sqrt{2\pi}}\int_{1-s}^{\infty}\frac{dh}{\sqrt{2\pi}}e^{-\frac{1}{2}\log\left(q_{0}-q\right)-\frac{1}{2}\frac{1}{q_{0}-q}\left(h-\sqrt{qt}\right)^{2}-\frac{1}{2}\frac{1}{q_{0}-q}cks^{2}}
\end{align}
so the limit:
\begin{align}
\lim_{q_{0}\to q}\left(q_{0}-q\right)G_{1} & =-\alpha\int Dt\lim_{q_{0}\to q}\left(q_{0}-q\right)\log\int\frac{ds}{\sqrt{2\pi}}\int_{1-s}^{\infty}\frac{dh}{\sqrt{2\pi}}e^{-\frac{1}{2}\log\left(q_{0}-q\right)-\frac{1}{2}\frac{1}{q_{0}-q}\left(h-\sqrt{qt}\right)^{2}-\frac{1}{2}\frac{1}{q_{0}-q}cks^{2}}\\
 & =-\alpha\int Dt\lim_{q_{0}\to q}\left(q_{0}-q\right)\max_{s>1-h}\left\{ -\frac{1}{2}\log\left(q_{0}-q\right)-\frac{1}{2}\frac{1}{q_{0}-q}\left(h-\sqrt{qt}\right)^{2}-\frac{1}{2}\frac{1}{q_{0}-q}cks^{2}\right\} \\
 & =\alpha\int Dt\min_{s>1-h}\left\{ \frac{1}{2}\left(h-\sqrt{qt}\right)^{2}+\frac{1}{2}cks^{2}\right\} 
\end{align}
with Lagrangian:
\begin{align}
h^{*},s^{*} & =\arg\min_{s>1-h}\frac{1}{2}\left(h-t\sqrt{q}\right)^{2}+\frac{1}{2}cks^{2}\\
{\cal L} & =\frac{1}{2}\left(h-t\sqrt{q}\right)^{2}+\frac{1}{2}cks^{2}+\lambda\left(1-h-s\right)
\end{align}
and derivatives:

\begin{align}
\frac{\partial{\cal L}}{\partial h} & =\left(h-t\sqrt{q}\right)-\lambda=0\\
\frac{\partial{\cal L}}{\partial s} & =cks-\lambda=0
\end{align}

so from KKT conditions:

\begin{align}
0 & =\lambda\left(1-h-s\right)\\
\lambda & =h-t\sqrt{q}\\
\lambda & =cks
\end{align}
so either $\lambda=0$ and then:

\begin{align}
s & =0\\
1\le h & =t\sqrt{q}
\end{align}
or $\lambda\ge0$ and then:

\begin{align}
s & \ge0\\
1 \ge h & =\frac{ck}{1+ck}+\frac{\sqrt{q}}{1+ck}t
\end{align}
so that we can write:

\begin{equation}
h=\begin{cases}
\frac{ck}{1+ck}+\frac{\sqrt{q}}{1+ck}t_{0} & -\infty\le t_{0}\le1/\sqrt{q}\\
\sqrt{q}t_{0} & 1/\sqrt{q}\le t_{0}
\end{cases}
\end{equation}
which can be written equivalently as equation \ref{eq:point-fields}.
The slack variables satisfy $s=\max\left\{ 1-h,0\right\} $ which yields 

\begin{equation}
s=\begin{cases}
\frac{1}{1+ck}-\frac{\sqrt{q}}{1+ck}t_{0} & -\infty\le t_{0}\le1/\sqrt{q}\\
0 & 1/\sqrt{q}\le t_{0}
\end{cases}
\end{equation}

Interestingly, the slack distribution allows to derive the self-consistent
equations \ref{eq:points-mf1}-\ref{eq:points-mf2} without saddle-point
assumption (i.e., without taking derivatives of $L$). From the definition
of $L$ we have that:

\begin{align}
L & =\|\boldsymbol{w}\|^{2}/N+c\|\vec{s}\|^{2}/N=q+\alpha c\left\langle s^{2}\right\rangle \label{eq:si-points-L-def}\\
\left\langle s^{2}\right\rangle  & =\frac{q}{\left(1+ck\right)^{2}}\int_{-\infty}^{1/\sqrt{q}}Dt_{0}\left(1/\sqrt{q}-t_{0}\right)^{2}
\end{align}

and the following equation is true for the optimal loss $L^{*}$ (see details in section \ref{subsec:appendix-optimal-loss}):

\begin{align}
L & =\frac{c}{N}\sum_{\mu}^{P}s_{\mu}=\alpha c\left\langle s\right\rangle \label{eq:si-points-L-opt}\\
\left\langle s\right\rangle  & =\frac{\sqrt{q}}{1+ck}\int_{-\infty}^{1/\sqrt{q}}Dt_{0}\left(1/\sqrt{q}-t_{0}\right)
\end{align}
Combining equations \ref{eq:si-points-L-def},\ref{eq:points-L-mf} and
\ref{eq:si-points-L-opt} yields equations \ref{eq:points-mf1}-\ref{eq:points-mf2}.

Furthermore, from the expression for $\left<s^2\right>$ and the self-consistent equation $\frac{1}{\left(ck\right)^{2}}=\frac{1}{\left(1+ck\right)^2}\alpha\alpha_{0}^{-1}\left(1/\sqrt{q}\right)$ we have:
\begin{align}
\frac{q}{c^{2}k^{2}} & =\alpha\left\langle s^{2}\right\rangle 
\end{align}

\bigskip

\subsection{\label{subsec:appendix-optimal-loss}Optimal loss at classification of points and spheres}

We show the optimal loss satisfies $L^{*}=\alpha c\left\langle s\right\rangle $
for both points (equation \ref{eq:si-points-L-opt}) and spheres (equation
\ref{eq:si-spheres-L-opt}).

We write a Lagrangian for the problem, assuming no bias for brevity,
i.e. $v_{0}^{\mu}=y^{\mu}\boldsymbol{w}\cdot\boldsymbol{u}_{0}^{\mu}$.
For spheres we have the constraint on the minimal field $h^{\mu}_{min}= v_{0}^{\mu}-R\|v^{\mu}\|\ge1-s^\mu$ so that both cases are captured by the same Lagrangian (with $R=0$
for points):

\begin{align}
L^{*} & =\min_{\boldsymbol{w}}\|\boldsymbol{w}\|^{2}/N+c\|\vec{s}\|^{2}/N\ s.t.\ v_{0}^{\mu}-Rv^{\mu}\ge1-s^{\mu}\ \forall\mu\\
{\cal L} & =\frac{1}{2}\|\boldsymbol{w}\|^{2}/N+\frac{1}{2}c\|\vec{s}\|^{2}/N+\sum_{\mu}^{P}\lambda_{\mu}\left(1-s^{\mu}-v_{0}^{\mu}+R\|v^{\mu}\|\right)
\end{align}

KKT conditions yield 3 equations:

\begin{align}
0=\frac{\partial{\cal L}}{\partial w_{i}} & =w_{i}+\sum_{\mu}\lambda_{\mu}\left(-y^{\mu}u_{0i}^{\mu}+y^{\mu}R\sum_{l}^{D}\frac{v_{l}^{\mu}u_{li}^{\mu}}{\|v^{\mu}\|}\right)\\
0=\frac{\partial{\cal L}}{\partial s^{\mu}}&=2cs^{\mu}/N+2\lambda_{\mu}\\
\forall\mu\ 0 & =\lambda_{\mu}\left(1-s^{\mu}-v_{0}^{\mu}+R\|v^{\mu}\|\right)
\end{align}
Denoting $h_{\mu}=v_{0}^{\mu}-R\|v^{\mu}\|$ those can be written
as:

\begin{align}
0 & =\vec{\lambda}\left(1-\vec{s}-\vec{h}\right)\\
\vec{\lambda} & =\vec{s}c/N\\
\boldsymbol{w} & =\sum_{\mu}\lambda_{\mu}y^{\mu}\left(\boldsymbol{u}_{0}^{\mu}-R\sum_{l}^{D}\frac{v_{l}^{\mu}}{\|v^{\mu}\|}\boldsymbol{u}_{l}^{\mu}\right)\\
\boldsymbol{w}^{T}\boldsymbol{w} & =\sum_{\mu}\lambda_{\mu}\left(v_{0}^{\mu}-R\|v^{\mu}\|\right)=\vec{\lambda}^{T}\vec{h}
\end{align}
so that we have that the optimal solution $L^{*}$ satisfies:

\begin{align}
L^{*} & =\boldsymbol{w}^{T}\boldsymbol{w}+c\|\vec{s}\|^{2}/N+2\sum_{\mu}^{P}\lambda_{\mu}\left(1-s^{\mu}-h_{\mu}\right)\\
 & =\vec{\lambda}^{T}\left(1-\vec{s}-\vec{h}\right)+\sum_{\mu}^{P}\lambda_{\mu}=\frac{c}{N}\sum_{\mu}^{P}s_{\mu}
\end{align}
Yielding the result:
\begin{equation}
L^{*}=c\alpha\left\langle s\right\rangle 
\end{equation}

\bigskip

\subsection{\label{subsec:appendix-points-classfication-error}Classification error for points}

We seek to derive the classification error for training $\varepsilon_{tr}$ and testing $\varepsilon_{g}\left(\sigma\right)$ with respect to Gaussian noise ${\cal N}\left(0,\sigma^{2}/N\right)$ applied to the training set. 

The training error has contribution only from the ``touching'' regime of the field distribution (equation \ref{eq:point-fields}):
\begin{align}
\varepsilon_{tr} & =\int_{-\infty}^{0}dhe^{-\left(h-\mu_{1}\right)^{2}/2\sigma_{1}^{2}}/\sqrt{2\pi\sigma_{1}^{2}}=H\left(\mu_{1}/\sigma_{1}\right)
\end{align}
with $\mu_1=\frac{ck}{1+ck}$ and $\sigma_1^2=\frac{q}{\left(1+ck\right)^{2}}$, so that:

\begin{align}
\varepsilon_{tr} & =H\left(ck/\sqrt{q}\right)
\end{align}

When i.i.d Gaussian noise ${\cal N}\left(0,\sigma^{2}/N\right)$ is applied to each input component, as the weights are independent of this noise, the fields are affected
by i.i.d Gaussian noise ${\cal N}\left(0,\sigma^{2}q\right)$: 
\begin{align}
h & =h^{\mu}+\sigma\sqrt{q}\eta^{\mu}
\end{align}
when $\eta^\mu$ are standard Gaussian variables. The noisy field distribution can be written explicitly, denoted
$P_{g}\left(h\right)$:

\begin{align}
P_{g}\left(h\right) & =e^{-\left(h-\mu_{1}\right)^{2}/2\left(\sigma_{1}^{2}+\sigma^{2}q\right)}/\sqrt{2\pi\left(\sigma_{1}^{2}+\sigma^{2}q\right)}H\left(\frac{\left(h-1\right)\left(\sigma^{2}q+\sigma_{1}^{2}\right)-\sigma^{2}q\left(h-\mu_{1}\right)}{\sigma\sqrt{q}\sigma_{1}\sqrt{\sigma^{2}q+\sigma_{1}^{2}}}\right)\\
 & +e^{-\left(h-\mu_{2}\right)^{2}/2\left(\sigma_{2}^{2}+\sigma^{2}q\right)}/\sqrt{2\pi\left(\sigma_{2}^{2}+\sigma^{2}q\right)}H\left(-\frac{\left(h-1\right)\left(\sigma^{2}q+\sigma_{2}^{2}\right)-\sigma^{2}q\left(h-\mu_{2}\right)}{\sigma\sqrt{q}\sigma_{2}\sqrt{\sigma^{2}q+\sigma_{2}^{2}}}\right)
\end{align}
so the error is given by:
\begin{align}
\varepsilon_{g} & =\int_{-\infty}^{0}dhP_{g}\left(h\right)
\end{align}
which is useful for numerical evaluation. 
For the analytic derivation below it is more useful to consider a different formalism, using the field distribution (equation \ref{eq:point-fields}):

\begin{align}
\varepsilon_{g} & =\int_{-\infty}^{\infty}dhP_{tr}\left(h\right)H\left(h/\sigma\sqrt{q}\right)\\
 & =\int_{-\infty}^{1}dhe^{-\left(h-ck/\left(1+ck\right)\right)^{2}/2q/\left(1+ck\right)^{2}}/\sqrt{2\pi q/\left(1+ck\right)^{2}}\int_{h/\sigma\sqrt{q}}^{\infty}dxe^{-x^{2}/2}/\sqrt{2\pi}\\
 & +\int_{1}^{\infty}dhe^{-h^{2}/2q}/\sqrt{2\pi q}\int_{h/\sigma\sqrt{q}}^{\infty}dxe^{-x^{2}/2}/\sqrt{2\pi}
\end{align}
so that replacing $g=h\left(1+ck\right)/\sqrt{q}-ck/\sqrt{q}$ and
$g=h/\sqrt{q}$:

\begin{align}
\varepsilon_{g} & =\int_{-\infty}^{1/\sqrt{q}}Dg_{1}H\left(\frac{g_{1}\sqrt{q}+ck}{\sigma\sqrt{q}\left(1+ck\right)}\right)+\int_{1/\sqrt{q}}^{\infty}Dg_{2}H\left(g_{2}/\sigma\right)
\end{align}
so using identity 10,010.4 from \cite{owen1980table}:
\begin{align}
\int_{h}^{k}DxH\left(a+bx\right) & =\int_{a/\sqrt{b^{2}+1}}^{\infty}DxH\left(h\sqrt{b^{2}+1}+bx\right)-\int_{a/\sqrt{b^{2}+1}}^{\infty}DxH\left(k\sqrt{b^{2}+1}+bx\right)
\end{align}
we have an exact expression:
\begin{align}
\varepsilon_{g} & =H\left(1/\sqrt{q/\left(ck\right)^{2}+q\sigma^{2}\left(1+ck\right)^{2}/\left(ck\right)^{2}}\right)+\int_{0}^{\infty}DxH\left(\frac{\sqrt{1+\sigma^{2}}}{\sqrt{q}\sigma}+\frac{x}{\sigma}\right)\\
 & -\int_{1/\sqrt{q/\left(ck\right)^{2}+q\sigma^{2}\left(1+ck\right)^{2}/\left(ck\right)^{2}}}^{\infty}DxH\left(\frac{\sqrt{1+\sigma^{2}\left(1+ck\right)^{2}}}{\sqrt{q}\sigma\left(1+ck\right)}+\frac{x}{\sigma\left(1+ck\right)}\right)
\end{align}
where in the second and third integrals we have only positive terms,
with $x\ge0$ and $\sqrt{1+\sigma^{2}},\sqrt{1+\sigma^{2}\left(1+ck\right)^{2}}\ge1$.
Thus if $\sqrt{q}\sigma\ll1$ we expect a good approximation:

\begin{align}
\varepsilon_{g} & \approx H\left({\cal S}\right)\\
{\cal S} & =ck/\sqrt{q\left(1+\left(1+ck\right)^{2}\sigma^{2}\right)}
\end{align}

\bigskip

\subsection{\label{subsec:appendix-points-optimal-c}Optimal choice of c for points}

The optimal SNR ${\cal S}$ should be optimized with respect to $c$:
\begin{align}
c^{*} & =\arg\min_{c}{\cal S}^{-2}
\end{align}
so that taking its derivative should satisfy $0=\frac{\partial{\cal S}^{-2}}{\partial c}$:

\begin{align}
\frac{\partial q}{\partial c}\frac{\left(\sigma^{2}\left(1+ck\right)^{2}+1\right)}{\left(ck\right)^{2}} & =2q\left(k+c\frac{\partial k}{\partial c}\right)\frac{\left(\sigma^{2}\left(1+ck\right)^{2}+1\right)-\left(\sigma^{2}\left(1+ck\right)\right)\left(ck\right)}{\left(ck\right)^{3}}\\
\frac{\partial q}{\partial c} & =2q\left(k+c\frac{\partial k}{\partial c}\right)\frac{\left(\sigma^{2}\left(1+ck\right)+1\right)}{\left(ck\right)\left(\sigma^{2}\left(1+ck\right)^{2}+1\right)}
\end{align}

Starting from the self-consistent equations \ref{eq:points-mf1}, \ref{eq:points-mf2}
we rewrite them as:
\begin{align}
q\left(1+ck\right)^{2} & =\left(ck\right)^{2}\alpha q\alpha_{0}^{-1}\left(1/\sqrt{q}\right)\\
\left(1-k\right)\left(1+ck\right) & =\left(ck\right)\alpha H\left(-1/\sqrt{q}\right)
\end{align}
where we have multiplied the first equation by $q$, so that we may
use the identity $\frac{\partial}{\partial c}q\alpha_{0}^{-1}\left(1/\sqrt{q}\right)=H\left(-1/\sqrt{q}\right)\frac{\partial q}{\partial c}$
when taking its derivative:

\begin{align}
\frac{\partial q}{\partial c}\left(1+ck\right)^{2}+q2\left(1+ck\right)\left(k+c\frac{\partial k}{\partial c}\right) & =2\left(ck\right)\left(k+c\frac{\partial k}{\partial c}\right)\alpha q\alpha_{0}^{-1}\left(1/\sqrt{q}\right)+\left(ck\right)^{2}\alpha H\left(-1/\sqrt{q}\right)\frac{\partial q}{\partial c}
\end{align}
in order to get:

\begin{align}
\frac{\partial q}{\partial c} & =2q\left(k+c\frac{\partial k}{\partial c}\right)\frac{\left(ck\right)\alpha\alpha_{0}^{-1}\left(1/\sqrt{q}\right)-\left(1+ck\right)}{\left(1+ck\right)^{2}-\left(ck\right)^{2}\alpha H\left(-1/\sqrt{q}\right)}
\end{align}
which yields for the optimal SNR an expression without $\frac{\partial q}{\partial c}$
or $\frac{\partial k}{\partial c}$:

\begin{align}
\frac{\left(\sigma^{2}\left(1+ck\right)+1\right)}{\left(ck\right)\left(\sigma^{2}\left(1+ck\right)^{2}+1\right)} & =\frac{\left(ck\right)\alpha\alpha_{0}^{-1}\left(1/\sqrt{q}\right)-\left(1+ck\right)}{\left(1+ck\right)^{2}-\left(ck\right)^{2}\alpha H\left(-1/\sqrt{q}\right)}
\end{align}

Using the self-consistent equations again to substitute $\alpha\alpha_{0}^{-1}\left(1/\sqrt{q}\right)$
and $\alpha H\left(-1/\sqrt{q}\right)$ we get:

\begin{align}
\frac{\left(\sigma^{2}\left(1+ck\right)+1\right)}{\left(ck\right)\left(\sigma^{2}\left(1+ck\right)^{2}+1\right)} & =\frac{\frac{\left(1+ck\right)^{2}}{\left(ck\right)}-\left(1+ck\right)}{\left(1+ck\right)^{2}-\left(ck\right)\left(1-k\right)\left(1+ck\right)}
\end{align}
so that for $\sigma=0$ we have no solution with both $c>0$ and $q>0$
(thus proving the $\varepsilon_{tr}$ is monotonic in $c$ for any
$\alpha$) while for $\sigma>0$ we have:

\begin{align}
\sigma^{-2}ck^{2} & =\left(1+ck\right)^{2}-\left(1+ck\right)\left[1+ck^{2}\right]
\end{align}
and the optimal $c$ should satisfy:

\begin{equation}
c=\frac{\sigma^{-2}}{1-k}-\frac{1}{k}
\end{equation}
which needs to be solved self-consistently as $k$ depends on $c$.

\bigskip

\subsection{\label{subsec:appendix-spheres-replica-theory}Replica theory for spheres}

We write an expression for the volume $V\left(L,c\right)$ for $L=\|\boldsymbol{w}\|^{2}/N+c\|\vec{s}\|^{2}/N$
which would vanish for $L<L^{*}$:
\begin{align}
V\left(L,c\right) & =\int d^{N}\boldsymbol{w}\int d^{P}\vec{s}\int d^{P\times D}\vec{v}\prod_{\mu}^{P}\prod_{l}^{D}\delta\left(y^{\mu}\boldsymbol{w}\cdot\boldsymbol{u}_{l}^{\mu}-v_{l}^{\mu}\right)\times\\
 & \cdots\prod_{\mu}^{P}\Theta\left(v_{0}^{\mu}-R\|\vec{v}^{\mu}\|\ge1-s^{\mu}\right)\delta\left(\|\boldsymbol{w}\|^{2}+c\|\vec{s}\|^{2}-NL\right)\\
 & =\int d^{N}\boldsymbol{w}\int d^{P}\vec{s}\int d^{P\times D}\vec{v}_{l}^{\mu}\int\frac{d^{P\times D}\hat{v}_{l}^{\mu}}{2\pi}e^{i\sum_{\mu}^{P}\sum_{l}^{D}\left(y^{\mu}\boldsymbol{w}\cdot\boldsymbol{u}_{l}^{\mu}-v_{l}^{\mu}\right)\hat{v}_{l}^{\mu}}\times\\
 & \cdots\int_{1-s^{\mu}}^{\infty}d^{P}h^{\mu}\int\frac{d^{P}\hat{h}^{\mu}}{2\pi}e^{i\left(v_{0}^{\mu}-R\|\vec{v}^{\mu}\|-h^{\mu}\right)\hat{h}^{\mu}}\int\frac{d\hat{l}}{2\pi}e^{i\left(\|\boldsymbol{w}\|^{2}+c\|\vec{s}\|^{2}-NL\right)\hat{l}}
\end{align}

We wish to calculate the values for which the volume vanishes assuming random (Gaussian) axes $\boldsymbol{u}_l^\mu$ and random (binary) labels $y^\mu$. 
Using the replica identity:
\begin{equation}
\left[\log V\right]_{x,y}=\lim_{n\to0}\left[\frac{1}{n}\left(V^{n}-1\right)\right]_{x,y}
\end{equation}
it is enough to find $G$ which satisfies $\left[V^{n}\right]=e^{nG}$, to have that $\left[\log V\right]\approx G$.
Thus we consider $V^{n}$:
\begin{align}
V^{n} & =\int d^{n\times N}\boldsymbol{w}^{\alpha}\int d^{n\times P}\vec{s}^{\alpha}\int d^{n\times P\times D}\vec{v}_{l}^{\alpha,\mu}\int\frac{d^{n\times P\times D}\hat{v}_{l}^{\alpha,\mu}}{2\pi}\int_{1-s^{\alpha,\mu}}^{\infty}d^{n\times P}h^{\alpha,\mu}\int\frac{d^{n\times P}\hat{h}^{\alpha,\mu}}{2\pi}\int\frac{d^{n}\hat{l}^{\alpha}}{2\pi}\\
 & \cdots e^{i\sum_{\alpha}^{n}\sum_{\mu}^{P}\sum_{l}^{D}\left(y^{\mu}\boldsymbol{w}^{\alpha}\cdot\boldsymbol{u}_{l}^{\mu}-v_{l}^{\alpha,\mu}\right)\hat{v}_{l}^{\alpha,\mu}+\sum_{\alpha}^{n}\sum_{\mu}^{P}i\left(v_{0}^{\alpha,\mu}-R\|\vec{v}^{\alpha,\mu}\|-h^{\alpha,\mu}\right)\hat{h}^{\alpha,\mu}+\sum_{\alpha}^{n}i\left(\|\boldsymbol{w}^{\alpha}\|^{2}+c\|\vec{s}^{\alpha}\|^{2}-NL\right)\hat{l}^{\alpha}}
\end{align}
Using Gaussian integral on the exes $u_{li}^{\mu}\sim{\cal N}\left(0,1/N\right)$
for $l=0..D$ and $\mu=1..P$:

\begin{align}
\left[e^{\sum_{\alpha}^{n}\sum_{\mu}^{P}iy^{\mu}\sum_{i}^{N}w_{i}^{\alpha}u_{li}^{\mu}\hat{v}_{l}^{\alpha,\mu}}\right]_{u} & =e^{-\frac{1}{2}\sum_{\mu}^{P}\left(\sum_{\alpha,\beta}^{n}\left(\frac{1}{N}\sum_{i}^{N}w_{i}^{\alpha}w_{i}^{\beta}\right)\hat{v}_{l}^{\alpha,\mu}\hat{v}_{l}^{\beta,\mu}\right)}
\end{align}
so denoting as usual:
\begin{align}
q_{\alpha\beta} & =\frac{1}{N}\sum_{i}^{N}w_{i}^{\alpha}w_{i}^{\beta}
\end{align}
we have:
\begin{align}
\left[V^{n}\right]_{x} & =\int d^{n\times n}q_{\alpha\beta}\int\frac{d^{n\times n}\hat{q}_{\alpha\beta}}{2\pi}\int d^{n\times N}\boldsymbol{w}^{\alpha}\int d^{n\times P}\vec{s}^{\alpha}\int d^{n\times P\times D}\vec{v}_{l}^{\alpha,\mu}\times\\
 & \cdots\int\frac{d^{n\times P\times D}\hat{v}_{l}^{\alpha,\mu}}{2\pi}\int_{1-s^{\alpha,\mu}}^{\infty}d^{n\times P}h^{\alpha,\mu}\int\frac{d^{n\times P}\hat{h}^{\alpha,\mu}}{2\pi}\int\frac{d^{n}\hat{l}^{\alpha}}{2\pi}\times\\
 & \cdots e^{-i\sum_{\alpha}^{n}\sum_{\mu}^{P}\sum_{l}^{D}v_{l}^{\alpha,\mu}\hat{v}_{l}^{\alpha,\mu}-i\sum_{\alpha,\beta}^{n}Nq_{\alpha\beta}\hat{q}_{\alpha\beta}-i\sum_{\alpha}^{n}\sum_{\mu}^{P}h^{\alpha,\mu}\hat{h}^{\alpha,\mu}-iN\sum_{\alpha}^{n}L\hat{l}^{\alpha}}\\
 & \cdots e^{-\frac{1}{2}\sum_{\mu}^{P}\sum_{\alpha,\beta}^{n}q_{\alpha\beta}\hat{v}_{0}^{\alpha,\mu}\hat{v}_{0}^{\beta,\mu}-\frac{1}{2}\sum_{\mu}^{P}\sum_{l}^{D}\sum_{\alpha,\beta}^{n}q_{\alpha\beta}\hat{v}_{l}^{\alpha,\mu}\hat{v}_{l}^{\beta,\mu}+i\sum_{\alpha}^{n}\sum_{\mu}^{P}\left(v_{0}^{\alpha,\mu}-R\|\vec{v}^{\alpha,\mu}\|\right)\hat{h}^{\alpha,\mu}}\\
 & \cdots e^{+i\sum_{\alpha}^{n}c\sum_{\mu}^{P}\left(s^{\alpha,\mu}\right)^{2}\hat{l}^{\alpha}+i\sum_{\alpha}^{n}\sum_{i}^{N}w_{i}^{\alpha}w_{i}^{\alpha}\hat{l}^{\alpha}+i\sum_{\alpha,\beta}^{n}\sum_{i}^{N}w_{i}^{\alpha}w_{i}^{\beta}\hat{q}_{\alpha\beta}}
\end{align}
Integration over $\hat{v}_{l}^{\alpha,\mu}$, $w_{i}^{\alpha}$ using
$\int\frac{d^{n}x}{\sqrt{2\pi}}e^{-x^{T}Ax/2+b^{T}x}=e^{-\frac{1}{2}\log\det A+b^{T}A^{-1}b/2}$:
\begin{align}
\int\frac{d^{n\times P}\hat{h}^{\alpha,\mu}}{\sqrt{2\pi}}e^{-\frac{1}{2}\sum_{\mu}^{P}\sum_{\alpha,\beta}^{n}q_{\alpha\beta}\hat{v}_{0}^{\alpha,\mu}\hat{v}_{0}^{\beta,\mu}-i\sum_{\alpha}^{n}\sum_{\mu}^{P}v_{0}^{\alpha,\mu}\hat{v}_{0}^{\alpha,\mu}} & =e^{-\frac{1}{2}\sum_{\mu}^{P}\sum_{\alpha,\beta}^{n}q_{\alpha\beta}^{-1}v_{0}^{\alpha,\mu}v_{0}^{\beta,\mu}-\frac{P}{2}\log\det q}\\
\int\frac{d^{n\times P}\hat{h}^{\alpha,\mu}}{\sqrt{2\pi}}e^{-\frac{1}{2}\sum_{\mu}^{P}\sum_{l}^{D}\sum_{\alpha,\beta}^{n}q_{\alpha\beta}\hat{v}_{l}^{\alpha,\mu}\hat{v}_{l}^{\beta,\mu}-i\sum_{\alpha}^{n}\sum_{\mu}^{P}\sum_{l}^{D}v_{l}^{\alpha,\mu}\hat{v}_{l}^{\alpha,\mu}} & =e^{-\frac{1}{2}\sum_{\mu}^{P}\sum_{l}^{D}\sum_{\alpha,\beta}^{n}q_{\alpha\beta}^{-1}v_{l}^{\alpha,\mu}v_{l}^{\beta,\mu}-\frac{DP}{2}\log\det q}\\
\int d^{n\times N}\boldsymbol{w}^{\alpha}e^{i\sum_{\alpha,\beta}^{n}\sum_{i}^{N}w_{i}^{\alpha}w_{i}^{\beta}\hat{q}_{\alpha\beta}+i\sum_{\alpha}^{n}\sum_{i}^{N}w_{i}^{\alpha}w_{i}^{\alpha}\hat{l}^{\alpha}} & =e^{-\frac{N}{2}\log\det\left(-2i\hat{q}_{\alpha\beta}-\delta_{\alpha\beta}2i\hat{l}^{\alpha}\right)}
\end{align}
so that:
\begin{align}
\left[V^{n}\right]_{x} & =\int d^{n\times n}q_{\alpha\beta}\int\frac{d^{n\times n}\hat{q}_{\alpha\beta}}{2\pi}\int d^{n\times P}\vec{s}^{\alpha}\int d^{n\times P\times D}\vec{v}_{l}^{\alpha,\mu}\int_{1-s^{\alpha,\mu}}^{\infty}d^{n\times P}h^{\alpha,\mu}\int\frac{d^{n\times P}\hat{h}^{\alpha,\mu}}{2\pi}\int\frac{d^{n}\hat{l}^{\alpha}}{2\pi}\\
 & \cdots e^{-i\sum_{\alpha,\beta}^{n}Nq_{\alpha\beta}\hat{q}_{\alpha\beta}-i\sum_{\alpha}^{n}\sum_{\mu}^{P}h^{\alpha,\mu}\hat{h}^{\alpha,\mu}-iN\sum_{\alpha}^{n}L\hat{l}^{\alpha}-\frac{1}{2}\sum_{\mu}^{P}\sum_{\alpha,\beta}^{n}q_{\alpha\beta}^{-1}v_{0}^{\alpha,\mu}v_{0}^{\beta,\mu}-\frac{1}{2}\sum_{\mu}^{P}\sum_{l}^{D}\sum_{\alpha,\beta}^{n}q_{\alpha\beta}^{-1}v_{l}^{\alpha,\mu}v_{l}^{\beta,\mu}}\\
 & \cdots e^{+i\sum_{\alpha}^{n}\sum_{\mu}^{P}\left(v_{0}^{\alpha,\mu}-R\|\vec{v}^{\alpha,\mu}\|\right)\hat{h}^{\alpha,\mu}+i\sum_{\alpha}^{n}c\sum_{\mu}^{P}\left(s^{\alpha,\mu}\right)^{2}\hat{l}^{\alpha}-\frac{N}{2}\log\det\left(-2i\hat{q}_{\alpha\beta}-\delta_{\alpha\beta}2i\hat{l}^{\alpha}\right)-\frac{\left(D+1\right)P}{2}\log\det q}
\end{align}

This can be rewritten such that all the $\mu$-s are decoupled:

\begin{align}
I & \doteq\int d^{n\times P}\vec{s}^{\alpha}\int d^{n\times P\times D}\vec{v}_{l}^{\alpha,\mu}\int_{1-s^{\alpha,\mu}}^{\infty}d^{n\times P}h^{\alpha,\mu}\int\frac{d^{n\times P}\hat{h}^{\alpha,\mu}}{2\pi}\\
 & \cdots e^{-i\sum_{\alpha}^{n}\sum_{\mu}^{P}\sum_{l}^{D}v_{l}^{\alpha,\mu}\hat{v}_{l}^{\alpha,\mu}-i\sum_{\alpha,\beta}^{n}Nq_{\alpha\beta}\hat{q}_{\alpha\beta}-i\sum_{\alpha}^{n}\sum_{\mu}^{P}h^{\alpha,\mu}\hat{h}^{\alpha,\mu}-\frac{1}{2}\sum_{\mu}^{P}\sum_{\alpha,\beta}^{n}q_{\alpha\beta}^{-1}v_{0}^{\alpha,\mu}v_{0}^{\beta,\mu}-\frac{1}{2}\sum_{\mu}^{P}\sum_{l}^{D}\sum_{\alpha,\beta}^{n}q_{\alpha\beta}^{-1}v_{l}^{\alpha,\mu}v_{l}^{\beta,\mu}}\\
 & \cdots e^{+i\sum_{\alpha}^{n}\sum_{\mu}^{P}\left(v_{0}^{\alpha,\mu}-R\|\vec{v}^{\alpha,\mu}\|\right)\hat{h}^{\alpha,\mu}+i\sum_{\alpha}^{n}c\sum_{\mu}^{P}\left(s^{\alpha,\mu}\right)^{2}\hat{l}^{\alpha}-\frac{\left(D+1\right)P}{2}\log\det q}\\
 & =\left(\int d^{n}\vec{s}^{\alpha}\int d^{n\times D}\vec{v}_{l}^{\alpha}\int_{1-s^{\alpha}}^{\infty}d^{n}h^{\alpha}\int\frac{d^{n}\hat{h}^{\alpha}}{2\pi}e^{F}\right)^{P}\\
F & \doteq-i\sum_{\alpha}^{n}h^{\alpha}\hat{h}^{\alpha}-\frac{1}{2}\sum_{l=0}^{D}\sum_{\alpha,\beta}^{n}q_{\alpha\beta}^{-1}v_{l}^{\alpha}v_{l}^{\beta}+i\sum_{\alpha}^{n}\left(v_{0}^{\alpha}-R\|\vec{v}^{\alpha}\|\right)\hat{h}^{\alpha}+i\sum_{\alpha}^{n}c\left(s^{\alpha}\right)^{2}\hat{l}^{\alpha}-\frac{D+1}{2}\log\det q
\end{align}
note that replacing $v_{0}^{\alpha}=h^{\alpha,\mu}+R\|\vec{v}^{\alpha}\|$
we can write:
\begin{align}
I & \doteq\int d^{n}\vec{v}_{0}^{\alpha}\int_{1-s^{\alpha}}^{\infty}d^{n}h^{\alpha}\int\frac{d^{n}\hat{h}^{\alpha}}{2\pi}e^{-i\sum_{\alpha}^{n}h^{\alpha}\hat{h}^{\alpha}+i\sum_{\alpha}^{n}\left(v_{0}^{\alpha}-R\|\vec{v}^{\alpha}\|\right)\hat{h}^{\alpha}-\frac{1}{2}\sum_{\alpha,\beta}^{n}q_{\alpha\beta}^{-1}v_{0}^{\alpha}v_{0}^{\beta}}\\
 & =\int_{1-s^{\alpha}+R\|\vec{v}^{\alpha}\|}^{\infty}d^{n}v_{0}^{\alpha}e^{-\frac{1}{2}\sum_{\alpha,\beta}^{n}q_{\alpha\beta}^{-1}v_{0}^{\alpha}v_{0}^{\beta}}
\end{align}
so that we have:

\begin{align}
\left[V^{n}\right]_{x} & =\int d^{n\times n}q_{\alpha\beta}\int\frac{d^{n\times n}\hat{q}_{\alpha\beta}}{2\pi}\int\frac{d^{n}\hat{l}^{\alpha}}{\sqrt{2\pi}}e^{-nNG_{0}-nNG_{1}}\\
G_{0} & =\frac{i}{n}\sum_{\alpha,\beta}^{n}q_{\alpha\beta}\hat{q}_{\alpha\beta}+\frac{1}{2n}\log\det\left(-2i\hat{q}_{\alpha\beta}-\delta_{\alpha\beta}2i\hat{l}^{\alpha}\right)+\frac{i}{n}\sum_{\alpha}^{n}L\hat{l}^{\alpha}\\
G_{1} & =\frac{\alpha}{2n}\left(D+1\right)\log\det q-\frac{\alpha}{n}\log\int\frac{d^{n}\vec{s}^{\alpha}}{\sqrt{2\pi}}\int d^{n\times D}\vec{v}_{l}^{\alpha}\int_{1-s^{\alpha}+R\|\vec{v}^{\alpha}\|}^{\infty}\frac{d^{n}v_{0}^{\alpha}}{\sqrt{2\pi}}e^{-\frac{1}{2}\sum_{\alpha,\beta}^{n}q_{\alpha\beta}^{-1}v_{0}^{\alpha}v_{0}^{\beta}-\frac{1}{2}\sum_{l}^{D}\sum_{\alpha,\beta}^{n}q_{\alpha\beta}^{-1}v_{l}^{\alpha}v_{l}^{\beta}+i\sum_{\alpha}^{n}c\left(s^{\alpha}\right)^{2}\hat{l}^{\alpha}}
\end{align}

Assuming replica symmetry:
\begin{align}
q_{\alpha\beta} & =q+(q_{0}-q)\delta_{\alpha\beta}\\
\hat{q}_{\alpha\beta} & =\hat{q}+(\hat{q}_{0}-\hat{q})\delta_{\alpha\beta}\\
\hat{l}^{\alpha} & =\hat{l}
\end{align}
we have:

\begin{align}
\frac{1}{n}\sum_{\alpha,\beta}^{n}q_{\alpha\beta}\hat{q}_{\alpha\beta} & \approx\hat{q}_{0}q_{0}-\hat{q}q\\
\log\det\left[q\right] & \approx n\log\left(q_{0}-q\right)+n\frac{q}{q_{0}-q}\\
\log\det|-2i\hat{q}-2i\hat{l}| & \approx n\log\left(-2i\hat{l}-2i\hat{q}_{0}+2i\hat{q}\right)+n\frac{-2i\hat{q}}{-2i\hat{l}-2i\hat{q}_{0}+2i\hat{q}}\\
q_{\alpha\beta}^{-1} & \approx\frac{1}{\left(q_{0}-q\right)}\delta_{\alpha\beta}-\frac{q}{\left(q_{0}-q\right)^{2}}
\end{align}

so that after changing $-i\hat{l}\to\hat{l}$,$-i\hat{q}\to\hat{q}$,$-i\hat{q}_{0}\to\hat{q}_{0}$ we get: 

\begin{align}
\left[V^{n}\right]_{x} & =\int dq\int dq_{0}\int d\hat{q}\int d\hat{q}_{0}\int\frac{d\hat{l}}{\sqrt{2\pi}}e^{-nNG_{0}-nNG_{1}}\\
G_{0} & =-\left[\hat{q}_{0}q_{0}-\hat{q}q\right]+\frac{1}{2}\left[\log\left(2\hat{l}+2\hat{q}_{0}-2\hat{q}\right)+\frac{2\hat{q}}{2\hat{l}+2\hat{q}_{0}-2\hat{q}}\right]-L\hat{l}\\
G_{1} & =\frac{\alpha\left(D+1\right)}{2}\left[\log\left(q_{0}-q\right)+\frac{q}{q_{0}-q}\right]\\
 & \cdots-\frac{\alpha}{n}\log\int\frac{d^{n}\vec{s}^{\alpha}}{\sqrt{2\pi}}\int d^{n\times D}\vec{v}_{l}^{\alpha}e^{-\frac{1}{2}\frac{1}{q_{0}-q}\sum_{l}^{D}\sum_{\alpha}^{n}\left(v_{l}^{\alpha}\right)^{2}+\frac{1}{2}\frac{q}{\left(q_{0}-q\right)^{2}}\sum_{l}^{D}\left(\sum_{\alpha}^{n}v_{l}^{\alpha}\right)^{2}-c\sum_{\alpha}^{n}\left(s^{\alpha}\right)^{2}\hat{l}}\times\\
 & \cdots\times\int_{1-s^{\alpha}+R\|\vec{v}^{\alpha}\|}^{\infty}\frac{d^{n}v_{0}^{\alpha}}{\sqrt{2\pi}}e^{-\frac{1}{2}\frac{1}{q_{0}-q}\sum_{\alpha}^{n}\left(v_{0}^{\alpha}\right)^{2}+\frac{1}{2}\frac{q}{\left(q_{0}-q\right)^{2}}\left(\sum_{\alpha}^{n}v_{0}^{\alpha}\right)^{2}}
\end{align}

Assuming the behavior in the thermodynamic limit $N\to\infty$ is dominated by the maximum of the integral, we calculate the derivatives of $G_0$:
\begin{align}
0 & =\frac{\partial G_{0}}{\partial\hat{q}}=q+\frac{1}{2}\left[-\frac{1}{\hat{l}+\hat{q}_{0}-\hat{q}}+\frac{\left(\hat{l}+\hat{q}_{0}-\hat{q}\right)+\hat{q}}{\left(\hat{l}+\hat{q}_{0}-\hat{q}\right)^{2}}\right]\\
0 & =\frac{\partial G_{0}}{\partial\hat{q}_{0}}=-q_{0}+\frac{1}{2}\left[\frac{1}{\hat{l}+\hat{q}_{0}-\hat{q}}-\frac{\hat{q}}{\left(\hat{l}+\hat{q}_{0}-\hat{q}\right)^{2}}\right]
\end{align}
so that:
\begin{align}
q & =-\frac{1}{2}\frac{\hat{q}}{\left(\hat{l}+\hat{q}_{0}-\hat{q}\right)^{2}}\\
q_{0} & =\frac{1}{2}\frac{1}{\hat{l}+\hat{q}_{0}-\hat{q}}+q\\
q_{0}-q & =\frac{1}{2}\frac{1}{\hat{l}+\hat{q}_{0}-\hat{q}}\\
\frac{q}{\left(q_{0}-q\right)^{2}} & =-\frac{1}{2}\frac{\hat{q}}{\left(\hat{l}+\hat{q}_{0}-\hat{q}\right)^{2}}/\frac{1}{4}\frac{1}{\left(\hat{l}+\hat{q}_{0}-\hat{q}\right)^{2}}=-2\hat{q}\\
q_{0}\hat{q}_{0}-q\hat{q} & =\frac{1}{2}\frac{\hat{q}_{0}}{\hat{l}+\hat{q}_{0}-\hat{q}}-\frac{1}{2}\frac{\hat{q}\hat{q}_{0}}{\left(\hat{l}+\hat{q}_{0}-\hat{q}\right)^{2}}+\frac{1}{2}\frac{\hat{q}\hat{q}}{\left(\hat{l}+\hat{q}_{0}-\hat{q}\right)^{2}}=\frac{1}{2}-\hat{l}q_{0}
\end{align}
and $G_{0}$ becomes:

\begin{align}
G_{0} & =-\left[\frac{1}{2}-\hat{l}\left(q_{0}-q\right)-q\hat{l}\right]+\frac{1}{2}\left[\log\left(\frac{1}{q_{0}-q}\right)-\left(q_{0}-q\right)\frac{q}{\left(q_{0}-q\right)^{2}}\right]-L\hat{l}\\
 & =-\frac{1}{2}+\left(q_{0}-L\right)\hat{l}-\frac{1}{2}\log\left(q_{0}-q\right)-\frac{1}{2}\frac{q}{q_{0}-q}
\end{align}

For $G_{1}$ we have:

\begin{align}
G_{1} & =\frac{\alpha\left(D+1\right)}{2}\log\left(q_{0}-q\right)+\frac{\alpha\left(D+1\right)}{2}\frac{q}{q_{0}-q}\\
 & \cdots-\frac{\alpha}{n}\log\int\frac{d^{n}\vec{s}^{\alpha}}{\sqrt{2\pi}}\int\frac{d^{n\times D}\vec{v}_{l}^{\alpha}}{\sqrt{2\pi}}e^{-\frac{1}{2}\frac{1}{q_{0}-q}\sum_{l}^{D}\sum_{\alpha}^{n}\left(v_{l}^{\alpha}\right)^{2}+\frac{1}{2}\frac{q}{\left(q_{0}-q\right)^{2}}\sum_{l}^{D}\left(\sum_{\alpha}^{n}v_{l}^{\alpha}\right)^{2}-c\sum_{\alpha}^{n}\left(s^{\alpha}\right)^{2}\hat{l}}\times\\
 & \cdots\times\int_{1-s^{\alpha}+R\|\vec{v}^{\alpha}\|}^{\infty}\frac{d^{n}v_{0}^{\alpha}}{\sqrt{2\pi}}e^{-\frac{1}{2}\frac{1}{q_{0}-q}\sum_{\alpha}^{n}\left(v_{0}^{\alpha}\right)^{2}+\frac{1}{2}\frac{q}{\left(q_{0}-q\right)^{2}}\left(\sum_{\alpha}^{n}v_{0}^{\alpha}\right)^{2}}
\end{align}
and using the Hubbard-Stratonovich transform $e^{ay^{2}/2}=\int\frac{dt}{\sqrt{2\pi}}e^{-t^{2}/2+\sqrt{a}yt}$
for $l=0..D$:
\begin{align}
e^{\frac{1}{2}\frac{q}{\left(q_{0}-q\right)^{2}}\left(\sum_{\alpha}^{n}v_{l}^{\alpha}\right)^{2}} & =\int Dt_{l}e^{t\frac{\sqrt{q}}{q_{0}-q}\sum_{\alpha}^{n}v_{l}^{\alpha}}
\end{align}
so that $G_{1}$ decouples into $n$ terms:

\begin{align}
G_{1} & =\frac{\alpha\left(D+1\right)}{2}\log\left(q_{0}-q\right)+\frac{\alpha\left(D+1\right)}{2}\frac{q}{q_{0}-q}\\
 & \cdots-\frac{\alpha}{n}\log\int\frac{d^{n}\vec{s}^{\alpha}}{\sqrt{2\pi}}\int\frac{d^{n\times D}\vec{v}_{l}^{\alpha}}{\sqrt{2\pi}}\int D^{D}\vec{t}e^{-\frac{1}{2}\frac{1}{q_{0}-q}\sum_{l}^{D}\sum_{\alpha}^{n}\left(v_{l}^{\alpha}\right)^{2}+t_{l}\frac{\sqrt{q}}{q_{0}-q}\sum_{\alpha}^{n}v_{l}^{\alpha}-c\sum_{\alpha}^{n}\left(s^{\alpha}\right)^{2}\hat{l}}\times\\
 & \cdots\times\int_{1-s^{\alpha}+R\|\vec{v}^{\alpha}\|}^{\infty}\frac{d^{n}v_{0}^{\alpha}}{\sqrt{2\pi}}\int Dt_{0}e^{-\frac{1}{2}\frac{1}{q_{0}-q}\sum_{\alpha}^{n}\left(v_{0}^{\alpha}\right)^{2}+t_{0}\frac{\sqrt{q}}{q_{0}-q}\sum_{\alpha}^{n}v_{0}^{\alpha}}\\
 & =\frac{\alpha\left(D+1\right)}{2}\log\left(q_{0}-q\right)+\frac{\alpha\left(D+1\right)}{2}\frac{q}{q_{0}-q}-\frac{\alpha}{n}\log\int D^{D}\vec{t}\int Dt_{0}\times\\
 & \cdots\times\left(\int\frac{ds}{\sqrt{2\pi}}\int\frac{d^{D}\vec{v}_{l}}{\sqrt{2\pi}}e^{-\frac{1}{2}\frac{1}{q_{0}-q}\sum_{l}^{D}\left(v_{l}\right)^{2}+t_{l}\frac{\sqrt{q}}{q_{0}-q}v_{l}-Cs^{2}\hat{l}}\int_{1-s+R\|\vec{v}\|}^{\infty}\frac{dv_{0}}{\sqrt{2\pi}}e^{-\frac{1}{2}\frac{1}{q_{0}-q}v_{0}^{2}+t_{0}\frac{\sqrt{q}}{q_{0}-q}v_{0}}\right)^{n}
\end{align}
and using the replica identity $\log\int Dt\,z(t)^{n}\approx\log\left[1+n\int Dt\log z(t)\right]\approx n\int Dt\log z(t)$
for $n\to0$: 

\begin{align}
G_{1} & =\frac{\alpha\left(D+1\right)}{2}\log\left(q_{0}-q\right)+\frac{\alpha\left(D+1\right)}{2}\frac{q}{q_{0}-q}\label{eq:si-checkpoint}\\
 & \cdots-\alpha\int D^{D}\vec{t}\int Dt_{0}\log\int\frac{d^{D}\vec{v}_{l}}{\sqrt{2\pi}}e^{-\frac{1}{2}\frac{1}{q_{0}-q}\sum_{l}^{D}\left(v_{l}\right)^{2}+t_{l}\frac{\sqrt{q}}{q_{0}-q}v_{l}}\int\frac{ds}{\sqrt{2\pi}}\int_{1-s+R\|\vec{v}\|}^{\infty}\frac{dv_{0}}{\sqrt{2\pi}}e^{-\frac{1}{2}\frac{1}{q_{0}-q}v_{0}^{2}+t_{0}\frac{\sqrt{q}}{q_{0}-q}v_{0}-Cs^{2}\hat{l}}
\end{align}
by changing $v_{0}\to v_{0}-s$ and integrating over $s$ using $\int\frac{dx}{\sqrt{2\pi}}e^{-x^{2}/2a+bx}=e^{\frac{1}{2}\log a+\frac{1}{2}b^{2}a}$:

\begin{align}
I & \doteq\int\frac{ds}{\sqrt{2\pi}}\int_{1-s+R\|\vec{v}\|}^{\infty}\frac{dv_{0}}{\sqrt{2\pi}}e^{-\frac{1}{2}\frac{1}{q_{0}-q}v_{0}^{2}+t_{0}\frac{\sqrt{q}}{q_{0}-q}v_{0}-Cs^{2}\hat{l}}\\
 & =\int\frac{ds}{\sqrt{2\pi}}\int_{1+R\|\vec{v}\|}^{\infty}\frac{dv_{0}}{\sqrt{2\pi}}e^{-\frac{1}{2}\frac{1}{q_{0}-q}\left(v_{0}-s\right)^{2}+t\frac{\sqrt{q}}{q_{0}-q}\left(v_{0}-s\right)-Cs^{2}\hat{l}}\\
 & =\int_{1+R\|\vec{v}\|}^{\infty}\frac{dv_{0}}{\sqrt{2\pi}}e^{-\frac{1}{2}\frac{1}{q_{0}-q}v_{0}^{2}+t\frac{\sqrt{q}}{q_{0}-q}v_{0}}\int\frac{ds}{\sqrt{2\pi}}e^{-\frac{1}{2}\left(\frac{1}{q_{0}-q}+2C\hat{l}\right)s^{2}+\frac{1}{q_{0}-q}\left(v_{0}-t\sqrt{q}\right)s}\\
 & =\int_{1+R\|\vec{v}\|}^{\infty}\frac{dv_{0}}{\sqrt{2\pi}}e^{-\frac{1}{2}\frac{1}{q_{0}-q}v_{0}^{2}+t\frac{\sqrt{q}}{q_{0}-q}v_{0}+\frac{1}{2}\log\frac{q_{0}-q}{1+2C\hat{l}\left(q_{0}-q\right)}+\frac{1}{2}\frac{q_{0}-q}{1+2C\hat{l}\left(q_{0}-q\right)}\left(\frac{v_{0}-t\sqrt{q}}{q_{0}-q}\right)^{2}}
\end{align}
and by completion to square:
\begin{align}
\int\frac{dv_{0}}{\sqrt{2\pi}}e^{-\frac{1}{2}\frac{1}{q_{0}-q}v_{0}^{2}+t_{0}\frac{\sqrt{q}}{q_{0}-q}v_{0}+\frac{1}{2}\log\frac{q_{0}-q}{1+2C\hat{l}\left(q_{0}-q\right)}+\frac{1}{2}\frac{\left(v_{0}-t_{0}\sqrt{q}\right)^{2}}{1+2C\hat{l}\left(q_{0}-q\right)}} & =\int\frac{dv_{0}}{\sqrt{2\pi}}e^{\frac{1}{2}\frac{qt_{0}^{2}}{q_{0}-q}-\frac{1}{2}\frac{\left(v_{0}-t_{0}\sqrt{q}\right)^{2}2C\hat{l}}{1+2C\hat{l}\left(q_{0}-q\right)}-\frac{1}{2}\log\frac{1+2C\hat{l}\left(q_{0}-q\right)}{q_{0}-q}}\\
\int\frac{d^{D}\vec{v}_{l}}{\sqrt{2\pi}}e^{-\frac{1}{2}\frac{1}{q_{0}-q}\sum_{l}^{D}v_{l}^{2}+t_{l}\frac{\sqrt{q}}{q_{0}-q}v_{l}} & =\int\frac{d^{D}\vec{v}_{l}}{\sqrt{2\pi}}e^{-\frac{1}{2}\frac{1}{q_{0}-q}\|\vec{v}-\sqrt{q}\vec{t}\|^{2}+\frac{1}{2}\frac{1}{q_{0}-q}q\|\vec{t}\|^{2}}
\end{align}
so that by inserting the $\log\left(q_{0}-q\right)$ into the integral
and taking out the term which depends only on $\vec{t},t_{0}$:

\begin{align}
G_{1} & =\frac{\alpha\left(D+1\right)}{2}\log\left(q_{0}-q\right)+\frac{\alpha\left(D+1\right)}{2}\frac{q}{q_{0}-q}\\
 & \cdots-\alpha\int D^{D}\vec{t}\int Dt_{0}\log\int\frac{d^{D}\vec{v}_{l}}{\sqrt{2\pi}}e^{-\frac{1}{2}\frac{1}{q_{0}-q}\|\vec{v}-\sqrt{q}\vec{t}\|^{2}+\frac{1}{2}\frac{1}{q_{0}-q}q\|\vec{t}\|^{2}}\times\\
 & \cdots\int_{1+R\|\vec{v}\|}^{\infty}\frac{dv_{0}}{\sqrt{2\pi}}e^{\frac{1}{2}\frac{qt_{0}^{2}}{q_{0}-q}-\frac{1}{2}\frac{\left(v_{0}-t_{0}\sqrt{q}\right)^{2}2C\hat{l}}{1+2C\hat{l}\left(q_{0}-q\right)}-\frac{1}{2}\log\frac{1+2C\hat{l}\left(q_{0}-q\right)}{q_{0}-q}}\\
 & =\frac{\alpha\left(D+1\right)}{2}\frac{q}{q_{0}-q}-\alpha\left[\frac{1}{2}\frac{q}{q_{0}-q}D+\frac{1}{2}\frac{q}{q_{0}-q}\right]-\alpha\int D^{D}\vec{t}\int Dt_{0}\log\int\frac{d^{D}\vec{v}_{l}}{\sqrt{2\pi\left(q_{0}-q\right)}}e^{-\frac{1}{2}\frac{1}{q_{0}-q}\|\vec{v}-\sqrt{q}\vec{t}\|^{2}}\times\\
 & \cdots\int_{1+R\|\vec{v}\|}^{\infty}\frac{dv_{0}}{\sqrt{2\pi}}e^{-\frac{1}{2}\frac{\left(v_{0}-t_{0}\sqrt{q}\right)^{2}2C\hat{l}}{1+2C\hat{l}\left(q_{0}-q\right)}-\frac{1}{2}\log\left(q_{0}-q\right)-\frac{1}{2}\log\frac{1+2C\hat{l}\left(q_{0}-q\right)}{q_{0}-q}}\\
 & =-\alpha\int D^{D}\vec{t}\int Dt_{0}\log\int\frac{d^{D}\vec{v}_{l}}{\sqrt{2\pi\left(q_{0}-q\right)}}\int_{1+R\|\vec{v}\|}^{\infty}\frac{dv_{0}}{\sqrt{2\pi}}e^{-\frac{1}{2}\frac{1}{q_{0}-q}\|\vec{v}-\sqrt{q}\vec{t}\|^{2}-\frac{1}{2}\frac{\left(v_{0}-t_{0}\sqrt{q}\right)^{2}2C\hat{l}}{1+2C\hat{l}\left(q_{0}-q\right)}-\frac{1}{2}\log\left(1+2C\hat{l}\left(q_{0}-q\right)\right)}
\end{align}
and by a change of variable $v_{l}=v_{l}\sqrt{q}$ for $l=0..D$:
\begin{align}
G_{1} & =-\alpha\int D^{D}\vec{t}\int Dt_{0}\log\int\frac{d^{D}\vec{v}_{l}\sqrt{q}}{\sqrt{2\pi\left(q_{0}-q\right)}}\int_{1/\sqrt{q}+R\|\vec{v}\|}^{\infty}\frac{dv_{0}\sqrt{q}}{\sqrt{2\pi}}e^{-\frac{1}{2}\frac{1}{q_{0}-q}\|\sqrt{q}\vec{v}-\sqrt{q}\vec{t}\|^{2}-\frac{1}{2}\frac{\left(\sqrt{q}v_{0}-\sqrt{q}t_{0}\right)^{2}2C\hat{l}}{1+2C\hat{l}\left(q_{0}-q\right)}-\frac{1}{2}\log\left(1+2C\hat{l}\left(q_{0}-q\right)\right)}\\
 & =-\alpha\int D^{D}\vec{t}\int Dt_{0}\log\int\frac{d^{D}\vec{v}_{l}}{\sqrt{2\pi}}\int_{1/\sqrt{q}+R\|\vec{v}\|}^{\infty}\frac{dv_{0}}{\sqrt{2\pi}}e^{-\frac{1}{2}\frac{q}{q_{0}-q}\|\vec{v}-\vec{t}\|^{2}-\frac{1}{2}\frac{\left(v_{0}-t_{0}\right)^{2}2C\hat{l}q}{1+2C\hat{l}\left(q_{0}-q\right)}-\frac{1}{2}\log\left(1+2C\hat{l}\left(q_{0}-q\right)\right)/q-\frac{D}{2}\log\left(q_{0}-q\right)/q}
\end{align}

Thus we conclude:

\begin{align}
\left[V^{n}\right]_{x} & =\int dq\int dq_{0}\int d\hat{q}\int d\hat{q}_{0}\int\frac{d\hat{l}}{\sqrt{2\pi}}e^{-nNG_{0}-nNG_{1}}\\
G_{0} & =-\frac{1}{2}+\left(q_{0}-L\right)\hat{l}-\frac{1}{2}\log\left(q_{0}-q\right)-\frac{1}{2}\frac{q}{q_{0}-q}\\
G_{1} & =-\alpha\int D^{D}\vec{t}\int Dt_{0}\log\int\frac{d^{D}\vec{v}_{l}}{\sqrt{2\pi}}\int_{1/\sqrt{q}+R\|\vec{v}\|}^{\infty}\frac{dv_{0}}{\sqrt{2\pi}}e^{-\frac{1}{2}\frac{q}{q_{0}-q}\|\vec{v}-\vec{t}\|^{2}-\frac{1}{2}\frac{\left(v_{0}-t_{0}\right)^{2}2C\hat{l}q}{1+2C\hat{l}\left(q_{0}-q\right)}-\frac{1}{2}\log\left(1+2C\hat{l}\left(q_{0}-q\right)\right)/q-\frac{D}{2}\log\left(q_{0}-q\right)}
\end{align}

Now we rename $\hat{l}=\hat{l}_{0}/\left(q_{0}-q\right)$ and take the limit $q\to q_{0}$: 
\begin{align}
\lim_{q\to q_{0}}\left(q_{0}-q\right)G_{0} & =\left(q_{0}-L\right)\hat{l}_{0}-\frac{1}{2}q_{0}\\
F &\doteq -\frac{1}{2}\frac{q}{q_{0}-q}\|\vec{v}-\vec{t}\|^{2}-\frac{1}{2}\frac{\left(v_{0}-t_{0}\right)^{2}2C\hat{l}_{0}q}{\left(1+2C\hat{l}_{0}\right)\left(q_{0}-q\right)}-\frac{1}{2}\log\left(1+2C\hat{l}_{0}\right)/q-\frac{D}{2}\log\left(q_{0}-q\right)\\
\lim_{q\to q_{0}}\left(q_{0}-q\right)G_{1} & =-\alpha\lim_{q\to q_{0}}\left(q_{0}-q\right)\int D^{D}\vec{t}\int Dt_{0}\log\int\frac{d^{D}\vec{v}_{l}}{\sqrt{2\pi}}\int_{1/\sqrt{q}+R\|\vec{v}\|}^{\infty}\frac{dv_{0}}{\sqrt{2\pi}} e^{F}\\
 & =-\alpha\int D^{D}\vec{t}\int Dt_{0}\lim_{q\to q_{0}}\left(q_{0}-q\right)\log\max_{v_{0}\ge1/\sqrt{q}+R\|\vec{v}\|}e^{F} \\
 & =-\alpha\int D^{D}\vec{t}\int Dt_{0}\lim_{q\to q_{0}}\left(q_{0}-q\right)\max_{v_{0}\ge1/\sqrt{q}+R\|\vec{v}\|}F\\
 & =\frac{\alpha q_{0}}{2}\int D^{D}\vec{t}\int Dt_{0}\min_{v_{0}\ge1/\sqrt{q}+R\|\vec{v}\|}\left\{ \|\vec{v}-\vec{t}\|^{2}+\frac{2C\hat{l}_{0}}{\left(1+2C\hat{l}_{0}\right)}\left(v_{0}-t_{0}\right)^{2}\right\} 
\end{align}

So combined into $G=G_1+G_2$:
\begin{equation}
\lim_{q\to q_{0}}\left(q_{0}-q\right)G=\left(q-L\right)\hat{l}_{0}-\frac{1}{2}q+\frac{\alpha q}{2}\int D^{D}\vec{t}\int Dt_{0}\min_{v_{0}\ge1/\sqrt{q}+R\|\vec{v}\|}\left\{ \|\vec{v}-\vec{t}\|^{2}+\frac{2C\hat{l}_{0}}{\left(1+2C\hat{l}_{0}\right)}\left(v_{0}-t_{0}\right)^{2}\right\} 
\end{equation}
Thus we have from $\lim_{q\to q_{0}}\left(q_{0}-q\right)G=0$:

\begin{align}
1 & =\left(1-L/q\right)2\hat{l}_{0}+\alpha\int D^{D}\vec{t}\int Dt_{0}\min_{v_{0}\ge1/\sqrt{q}+R\|\vec{v}\|}\left\{ \|\vec{v}-\vec{t}\|^{2}+\frac{2C\hat{l}_{0}}{\left(1+2C\hat{l}_{0}\right)}\left(v_{0}-t_{0}\right)^{2}\right\} 
\end{align}
and denoting $k=2\hat{l}_{0}$:

\begin{align}
L & =q+\frac{q\alpha}{k}\int D^{D}\vec{t}\int Dt_{0}\min_{v_{0}\ge1/\sqrt{q}+R\|\vec{v}\|}\left\{ \|\vec{v}-\vec{t}\|^{2}+\frac{ck}{1+ck}\left(v_{0}-t_{0}\right)^{2}\right\} -\frac{q}{k}
\end{align}

\bigskip

\subsection{\label{subsec:appendix-spheres-KKT}Solving the mean-field minimization problem for spheres}

Let us solve the following problem, so that we can write for it a closed form expression:
\begin{align}
F\left(\vec{t},t_{0};ck,q\right) & =\min_{v_{0}\ge1/\sqrt{q}+R\|\vec{v}\|}\left\{ \|\vec{v}-\vec{t}\|^{2}+\frac{ck}{1+ck}\left(v_{0}-t_{0}\right)^{2}\right\} \\
{\cal L} & =\frac{1}{2}\|\vec{v}-\vec{t}\|^{2}+\frac{1}{2}\frac{ck}{1+ck}\left(v_{0}-t_{0}\right)^{2}+\lambda\left(1/\sqrt{q}+R\|\vec{v}\|-v_{0}\right)
\end{align}
From KKT conditions by taking derivatives we have the equations:

\begin{align}
0=\frac{\partial L}{\partial v_{0}} & =\frac{ck}{1+ck}\left(v_{0}-t_{0}\right)-\lambda\\
0=\frac{\partial L}{\partial v_{l}} & =\left(v_{l}-t_{l}\right)+\frac{\lambda R2v_{l}}{2\|\vec{v}\|}\\
0 & =\lambda\left(1/\sqrt{q}+R\|\vec{v}\|-v_{0}\right)
\end{align}

so that denoting $v=\|\vec{v}\|\ge0$ and $t=\|\vec{t}\|$
the constraints are:
\begin{align}
\lambda & =\frac{ck}{1+ck}\left(v_{0}-t_{0}\right)\\
t_{l} & =\frac{v+\lambda R}{v}v_{l}\\
0 & =\lambda\left(1/\sqrt{q}+Rv-v_{0}\right)\\
v & \ge0
\end{align}
We solve for different regimes:
\begin{enumerate}
\item ``Interior'' regime defined by $\lambda=0$:
\begin{align}
v_{0} & =t_{0}\\
v_{l} & =t_{l}\\
v_{0} & \ge1/\sqrt{q}+Rv\\
t_{0} & \ge1/\sqrt{q}+Rt\\
F & =0
\end{align}
\item ``Embedded'' regime defined by $\lambda>0$ and $v=0$:
\begin{align}
v_{0} & =1/\sqrt{q}\\
F & =\frac{ck}{1+ck}\left(1/\sqrt{q}-t_{0}\right)^{2}+t^{2}
\end{align}
\item ``Touching'' regime defined by $\lambda>0$ and $v>0$:
\end{enumerate}
\begin{align}
v_{0} & =1/\sqrt{q}+Rv\\
v_{l} & =\frac{v}{v+\lambda R}t_{l}\\
\|\vec{v}-\vec{t}\|^{2} & =\left(\frac{-\lambda R}{v+\lambda R}\right)^{2}t^{2}\\
t-\lambda R=v & =\frac{v}{v+\lambda R}t\\
v_{0} & =1/\sqrt{q}+\frac{\left(t_{0}-1/\sqrt{q}\right)R^{2}+\frac{1+ck}{ck}tR}{\frac{1+ck}{ck}+R^{2}}\\
t_{0} & \le1/\sqrt{q}+Rt\\
t_{0} & \ge1/\sqrt{q}-\frac{1+ck}{ck}t/R\\
F & =\frac{\left(1/\sqrt{q}+Rt-t_{0}\right)^{2}}{\frac{1+ck}{ck}+R^{2}}
\end{align}
so that we got that the minimization problem depends only on $t_{0}$
and the norm $t=\|\vec{t}\|$:
\begin{equation}
F\left(t_{0},t\right)=\begin{cases}
\frac{ck}{1+ck}\left(1/\sqrt{q}-t_{0}\right)^{2}+t^{2} & -\infty<t_{0}\le1/\sqrt{q}-\frac{1+ck}{ck}t/R\\
\frac{ck}{1+ck\left(1+R^{2}\right)}\left(1/\sqrt{q}+Rt-t_{0}\right)^{2} & 1/\sqrt{q}-\frac{1+ck}{ck}t/R\le t_{0}\le1/\sqrt{q}+Rt\\
0 & 1/\sqrt{q}+Rt\le t_{0}\le\infty
\end{cases}\label{eq:si-spheres-F}
\end{equation}

As $t\sim\chi_{D}$ the Chi distribution with $D$ degrees of freedom, denoting $\chi_{D}\left(t\right)=\Gamma\left(D/2\right)^{-1}2^{1-D/2}t^{D-1}e^{-t^{2}/2}dt$:

\begin{align}
f\left(R,D,ck,q\right) & =\int\chi_{D}\left(t\right)\int_{-\infty}^{1/\sqrt{q}-\frac{1+ck}{ck}t/R}Dt_{0}\left[\frac{ck}{1+ck}\left(1/\sqrt{q}-t_{0}\right)^{2}+t^{2}\right]\cdots\\
 & +\int\chi_{D}\left(t\right)\int_{1/\sqrt{q}-\frac{1+ck}{ck}t/R}^{1/\sqrt{q}+Rt}Dt_{0}\frac{ck}{1+ck+ckR^{2}}\left(1/\sqrt{q}+Rt-t_{0}\right)^{2}
\end{align}

For $R=0$ we recover the result for soft classification of points $f =\frac{ck}{1+ck}\alpha_{0}^{-1}\left(1/\sqrt{q}\right)$
and for $c\to\infty$ we recover the expression for max-margin classifiers \cite{chung2016linear}, denoting $\kappa=1/\sqrt{q}$:

\begin{equation}
\alpha\left(\kappa\right)^{-1}=\int\chi_{D}(t)\int_{-\infty}^{\kappa-tR^{-1}}Dt_{0}\left[\left(\kappa-t_{0}\right)^{2}+t^{2}\right]+\int_{\kappa-tR^{-1}}^{\kappa+Rt}Dt_{0}\frac{\left(\kappa-t_{0}+Rt\right)^{2}}{1+R^{2}}
\end{equation}

\bigskip

\subsection{\label{subsec:appendix-spheres-self-consistent-equations}Self-consistent equations for spheres}

Assuming the optimal loss satisfies the saddle-point equations
$0=\frac{\partial L}{\partial k}=\frac{\partial L}{\partial q}$ we
have:

\begin{align}
1 & =\alpha f-\alpha k\frac{\partial}{\partial k}f\label{eq:si-spheres-mf1-abstract}\\
1-k & =\alpha f+\alpha q\frac{\partial}{\partial q}f\label{eq:si-spheres-mf2-abstract}
\end{align}

Taking the derivatives of $f$ with respect to $k,q$ we have that:
\begin{align}
\frac{\partial}{\partial k}f & =\frac{c}{\left(1+ck\right)^{2}}\int\chi_{D}\left(t\right)\int_{-\infty}^{1/\sqrt{q}-\frac{1+ck}{ck}t/R}Dt_{0}\left(1/\sqrt{q}-t_{0}\right)^{2}\\
 & +\frac{c}{\left(1+ck\left(1+R^{2}\right)\right)^{2}}\int\chi_{D}\left(t\right)\int_{1/\sqrt{q}-\frac{1+ck}{ck}t/R}^{1/\sqrt{q}+Rt}Dt_{0}\left(1/\sqrt{q}+Rt-t_{0}\right)^{2}\\
\frac{\partial}{\partial q}f & =-q^{-3/2}\frac{ck}{1+ck}\int\chi_{D}\left(t\right)\int_{-\infty}^{1/\sqrt{q}-\frac{1+ck}{ck}t/R}Dt_{0}\left(1/\sqrt{q}-t_{0}\right)\\
 & -q^{-3/2}\frac{ck}{1+ck\left(1+R^{2}\right)}\int\chi_{D}\left(t\right)\int_{1/\sqrt{q}-\frac{1+ck}{ck}t/R}^{1/\sqrt{q}+Rt}Dt_{0}\left(1/\sqrt{q}+Rt-t_{0}\right)
\end{align}
so the self-consistent equations:

\begin{align}
1 & =\alpha\frac{\left(ck\right)^{2}\left(1+R^{2}\right)}{\left(1+ck\left(1+R^{2}\right)\right)^{2}}\int\chi_{D}\left(t\right)\int_{1/\sqrt{q}-\frac{1+ck}{ck}t/R}^{1/\sqrt{q}+Rt}Dt_{0}\left(1/\sqrt{q}+Rt-t_{0}\right)^{2}\label{eq:si-spheres-mf1}\\
 & +\alpha\int\chi_{D}\left(t\right)\int_{-\infty}^{1/\sqrt{q}-\frac{1+ck}{ck}t/R}Dt_{0}\left[\frac{\left(ck\right)^{2}}{\left(1+ck\right)^{2}}\left(1/\sqrt{q}-t_{0}\right)^{2}+t^{2}\right]\\
1-k & =\alpha\int\chi_{D}\left(t\right)\int_{1/\sqrt{q}-\frac{1+ck}{ck}t/R}^{1/\sqrt{q}+Rt}Dt_{0}\frac{ck}{1+ck\left(1+R^{2}\right)}\left(1/\sqrt{q}+Rt-t_{0}\right)\left(Rt-t_{0}\right)\label{eq:si-spheres-mf2}\\
 & +\alpha\int\chi_{D}\left(t\right)\int_{-\infty}^{1/\sqrt{q}-\frac{1+ck}{ck}t/R}Dt_{0}\left[t^{2}-\frac{ck}{1+ck}\left(1/\sqrt{q}-t_{0}\right)t_{0}\right]
\end{align}

As for points, those equations can also be derived by combining the equations for the optimal loss, namely the loss definition, the mean-field
equation \ref{eq:spheres-L-mf}, and an optimality condition for the
loss (see details in section \ref{subsec:appendix-optimal-loss}):

\begin{align}
L & =q+\alpha c\left\langle s^{2}\right\rangle \\
L & =q+\frac{q}{k}\left(\alpha f-1\right)\\
L & =c\alpha\left\langle s\right\rangle \label{eq:si-spheres-L-opt}
\end{align}
where the slack moment equations \ref{eq:si-mean-slack-sqr-spheres}-\ref{eq:si-mean-slack-spheres} can be written as:

\begin{align}
\left\langle s^{2}\right\rangle  & =\frac{q}{c}\frac{\partial}{\partial k}f\label{eq:si-mean-slack-sqr-spheres-absract}\\
\left\langle s\right\rangle  & =-\frac{q}{ck}q\frac{\partial}{\partial q}f\label{eq:si-mean-slack-spheres-absract}
\end{align}
which leads to the self-consistent equations \ref{eq:si-spheres-mf1-abstract},\ref{eq:si-spheres-mf2-abstract}.

\bigskip

\subsection{\label{subsec:appendix-spheres-simplified-mf}Interesting regimes of the self-consistent equations for spheres}

The self-consistent equations \ref{eq:si-spheres-mf1}-\ref{eq:si-spheres-mf2}
can be simplified for several interesting cases.

In the limit $c\to\infty$ for $\alpha>\alpha_C^{Hard}$ the equations which can be derived by a replica theory from the Lagrangian:
\begin{equation}
L  = \|\vec{s}\|^{2}/N\ \ s.t.\ \forall\mu\ h_{min}^{\mu}\ge1-s^{\mu}
\end{equation}
the resulting equations are as follows, which are related to self-consistent equations from soft classification theory through $\lim_{c\to\infty}k=0$ while $q,K=\lim_{c\to\infty}ck$ are finite:
\begin{align}
1 & =\alpha\frac{\left(K\right)^{2}\left(1+R^{2}\right)}{\left(1+K\left(1+R^{2}\right)\right)^{2}}\int\chi_{D}\left(t\right)\int_{1/\sqrt{q}-\frac{1+K}{K}t/R}^{1/\sqrt{q}+Rt}Dt_{0}\left(1/\sqrt{q}+Rt-t_{0}\right)^{2}\label{eq:si-spheres-mf1-infC}\\
 & +\alpha\int\chi_{D}\left(t\right)\int_{-\infty}^{1/\sqrt{q}-\frac{1+K}{K}t/R}Dt_{0}\left[\frac{\left(K\right)^{2}}{\left(1+K\right)^{2}}\left(1/\sqrt{q}-t_{0}\right)^{2}+t^{2}\right]\\
1 & =\alpha\int\chi_{D}\left(t\right)\int_{1/\sqrt{q}-\frac{1+K}{K}t/R}^{1/\sqrt{q}+Rt}Dt_{0}\frac{K}{1+K\left(1+R^{2}\right)}\left(1/\sqrt{q}+Rt-t_{0}\right)\left(Rt-t_{0}\right)\label{eq:si-spheres-mf2-infC}\\
 & +\alpha\int\chi_{D}\left(t\right)\int_{-\infty}^{1/\sqrt{q}-\frac{1+K}{K}t/R}Dt_{0}\left[t^{2}-\frac{K}{1+K}\left(1/\sqrt{q}-t_{0}\right)t_{0}\right]
\end{align}
\clearpage

When $D\gg1$ the distribution of $\chi_{D}$ is narrow with a mode at $\sqrt{D-1}$ and a mean just below $\sqrt{D}$, so we will assume $t=\sqrt{D}$ and $\alpha_C\approx\frac{1+R^2}{R^2D}$ (equation \ref{eq:spheres-alpha-critical}). We consider several different cases:
\begin{enumerate}
\item For the case of ``large $D$, small $R$'' we assume $R\sqrt{D}=O\left(1\right)$
with $\sqrt{D}/R\gg1$. In this case $F$ is dominated by the contribution
of the ``touching'' regime and $\frac{1+ck}{ck}\sqrt{D}/R-1/\sqrt{q}\gg1$,
and the equations become:
\begin{align}
1 & =\frac{\left(ck\right)^{2}\left(1+R^{2}\right)}{\left(1+ck+ckR^{2}\right)^{2}}\alpha\alpha_{0}^{-1}\left(1/\sqrt{q}+R\sqrt{D}\right)\\
1-k & =\alpha\int_{-\infty}^{1/\sqrt{q}+R\sqrt{D}}Dt_{0}\frac{ck\left(1/\sqrt{q}+R\sqrt{D}-t_{0}\right)}{1+ck+ckR^{2}}\left(R\sqrt{D}-t_{0}\right)
\end{align}
In this case $q,k$ are non-trivial functions of $R,\sqrt{D},\alpha$.

\item For the case of ``large $D$, regular $R$'' we assume $R=O\left(1\right)$
with $R\sqrt{D}\gg1$. Again the leading contribution comes from the
``touching'' regime, but this time leading to the following equations:
\begin{align}
1 & =\alpha\frac{\left(ck\right)^{2}\left(1+R^{2}\right)}{\left(1+ck\left(1+R^{2}\right)\right)^{2}}\left(1/\sqrt{q}+R\sqrt{D}\right)^{2}\\
1-k & =\alpha\frac{ckR\sqrt{D}}{1+ck\left(1+R^{2}\right)}\left(1/\sqrt{q}+R\sqrt{D}\right)
\end{align}
which are combined as (for $\alpha_{C}$ given by equation \ref{eq:spheres-alpha-critical}):
\begin{align}
k & \approx1-\sqrt{\alpha/\alpha_{C}}\label{eq:si-spheres-k-largeD-R1}\\
\sqrt{q} & \approx\frac{1+R^{2}}{R\sqrt{D}}\frac{ck\sqrt{\alpha/\alpha_{C}}}{1+ck^{2}\left(1+R^{2}\right)\label{eq:si-spheres-q-largeD-R1}}
\end{align}

\item In the limit of $\alpha\to0$ we expect to have $q\to0$ and $k\to1$, as in the case of soft classification of points, so that $1/\sqrt{q}-\frac{1+ck}{ck}\sqrt{D}/R\gg1$
and $1/\sqrt{q}+R\sqrt{D}\gg1$, and the self-consistent equations are simplified to:
\begin{align}
1 & =\alpha\left[D+\frac{\left(ck\right)^{2}}{\left(1+ck\right)^{2}}\left(1+1/q\right)\right]\\
1-k & =\alpha\left[D+\frac{ck}{1+ck}\right]
\end{align}
and the resulting first-order approximations for small $\alpha$ are:
\begin{align}
k & \approx1-\alpha\left(1+D\right)\\
\sqrt{q} & \approx\sqrt{\alpha}\frac{ck}{1+ck}
\end{align}

\item On the other hand, for $\alpha\to\alpha_{C}^{Soft}$ we expect both
$q\to0$ and $k\to0$, so that we need to assume $1/\sqrt{q}+R\sqrt{D}\gg1$
and $\frac{1+ck}{ck}\sqrt{D}/R-1/\sqrt{q}\gg1$, leading to different
simplified equations:
\begin{align}
1 & =\frac{\alpha\left(ck\right)^{2}\left(1+R^{2}\right)}{\left(1+ck\left(1+R^{2}\right)\right)^{2}}\left(1+\left(1/\sqrt{q}+R\sqrt{D}\right)^{2}\right)\\
1-k & =\alpha\frac{ck}{1+ck\left(1+R^{2}\right)}\left(R\sqrt{D}/\sqrt{q}+R^{2}D+1\right)
\end{align}
so that the resulting order parameters $k$ and $q$ (for $\alpha_{C}$ given by equation \ref{eq:spheres-alpha-critical}):
\begin{align}
k & \approx\left(1-\sqrt{\alpha/\alpha_{C}}\right)/\left(\alpha c+1\right)\label{eq:si-spheres-alphac-k}\\
\sqrt{q} & \approx ck\frac{1+R^{2}}{R\sqrt{D}}\label{eq:si-spheres-alphac-q}
\end{align}

\end{enumerate}

\bigskip

\subsection{\label{subsec:appendix-spheres-capacity}Capacity in classification of spheres}

We first note that equations \ref{eq:si-spheres-mf1} , \ref{eq:si-spheres-mf2}
can be integrated over $t_{0}$ to yield the following self-consistent
equations:

\begin{align}
1 & =\alpha\int\chi_{D}\left(t\right)\left(t^{2}+\frac{\left(ck\right)^{2}}{\left(1+ck\right)^{2}}\left(1+1/q\right)\right)H(-1/\sqrt{q}+\frac{1+ck}{ck}t/R)\\
 & +\alpha\int\chi_{D}\left(t\right)\frac{ck}{1+ck}\left(\frac{ck}{1+ck}1/\sqrt{q}-\frac{R^{2}}{1+ck\left(1+R^{2}\right)}t\right)\frac{1}{\sqrt{2\pi}}e^{-\left(1/\sqrt{q}-\frac{1+ck}{ck}t/R\right)^{2}/2}\\
 & +\frac{\alpha\left(ck\right)^{2}\left(1+R^{2}\right)}{\left(1+ck\left(1+R^{2}\right)\right)^{2}}\int\chi_{D}\left(t\right)\left(1+\left(1/\sqrt{q}+Rt\right)^{2}\right)\left[H(-1/\sqrt{q}-Rt)-H(-1/\sqrt{q}+\frac{1+ck}{ck}t/R)\right]\\
 & +\frac{\alpha\left(ck\right)^{2}\left(1+R^{2}\right)}{\left(1+ck\left(1+R^{2}\right)\right)^{2}}\int\chi_{D}\left(t\right)\left(1/\sqrt{q}+Rt\right)\left[\frac{1}{\sqrt{2\pi}}e^{-\left(1/\sqrt{q}+Rt\right)^{2}/2}-\frac{1}{\sqrt{2\pi}}e^{-\left(1/\sqrt{q}-\frac{1+ck}{ck}t/R\right)^{2}/2}\right]\\
1-k & =\alpha\int\chi_{D}\left(t\right)\left[t^{2}+\frac{ck}{1+ck}\right]H(-1/\sqrt{q}+\frac{1+ck}{ck}t/R)\\
 & +\alpha\frac{ck}{1+ck\left(1+R^{2}\right)}\int\chi_{D}\left(t\right)Rt\left[\frac{1}{\sqrt{2\pi}}e^{-\left(1/\sqrt{q}+Rt\right)^{2}/2}-\frac{1}{\sqrt{2\pi}}e^{-\left(1/\sqrt{q}-\frac{1+ck}{ck}t/R\right)^{2}/2}\right]\\
 & +\alpha\frac{ck}{1+ck\left(1+R^{2}\right)}\int\chi_{D}\left(t\right)\left(Rt/\sqrt{q}+R^{2}t^{2}+1\right)\left[H(-1/\sqrt{q}-Rt)-H(-1/\sqrt{q}+\frac{1+ck}{ck}t/R)\right]
\end{align}

Now let us assume both $k,\sqrt{q}\ll1$ and further that $k=x\sqrt{q}$. 
For the first equation we have contributions only from the first term $\int_{0}^{xcR}\chi_{D}\left(t\right)\left(t^{2}+c^{2}x^{2}\right)$, and the third term $c^{2}x^{2}\left(1+R^{2}\right)\int_{xcR}^{\infty}\chi_{D}\left(t\right)$, leading to:
\begin{equation}
1=\alpha\int_{0}^{xcR}\chi_{D}\left(t\right)\left(t^{2}+c^{2}x^{2}\right)+\alpha c^{2}x^{2}\left(1+R^{2}\right)\int_{xcR}^{\infty}\chi_{D}\left(t\right)
\end{equation}
For the second equation we have contributions from the first term $\int_{0}^{xcR}\chi_{D}\left(t\right)t^{2}$ and the third term $xcR\int_{xcR}^{\infty}\chi_{D}\left(t\right)t$, leading to:
\begin{equation}
1=\alpha\int_{0}^{xcR}\chi_{D}\left(t\right)t^{2}+\alpha xcR\int_{xcR}^{\infty}\chi_{D}\left(t\right)t
\end{equation}

Combining those equations and replacing $xc\to x$ we have that $x=kc/\sqrt{q}$ and we get two equations which are independent of $c$, one for $x$ and another for $\alpha=\alpha_{C}$:
\begin{align}
x & =\frac{R\int_{xR}^{\infty}\chi_{D}\left(t\right)t}{\left(1+R^{2}\int_{xR}^{\infty}\chi_{D}\left(t\right)\right)}\\
\alpha_{C}^{-1} & =\int_{0}^{xR}\chi_{D}\left(t\right)t^{2}+xR\int_{xR}^{\infty}\chi_{D}\left(t\right)t
\end{align}

Now note that for $R\to0$ we have that $x=R\int_{xR}^{\infty}\chi_{D}\left(t\right)t=R\sqrt{2}\Gamma\left(\frac{D}{2}+\frac{1}{2}\right)\big/\Gamma\left(\frac{D}{2}\right)$ and $\alpha_{C}^{-1}=x^{2}$ (which converges to $R^{2}D$ for large $D$), whereas for $R\to\infty$ we have $x\approx0$ and $\alpha_{C}^{-1}=D$. 

When $D\gg1$ the distribution of $\chi_{D}$ is narrow around $\sqrt{D}$. If $\int_{0}^{xR}\chi_{D}\ll1$ we have a much simpler result; in this case $x\approx\frac{R\sqrt{D}}{1+R^{2}}$ and thus:

\begin{align}
\alpha_{C}^{-1} & =xR\sqrt{D}=\frac{R^{2}D}{1+R^{2}}
\end{align}
and from the above limits on $R$ we obtain that for large $D$ this approximation is valid for any $R$.

\bigskip

\subsection{\label{subsec:appendix-spheres-fields-and-slack-distribution}Field and slack distribution for spheres}

To derive the slack and field distribution we do not integrate away the slack variable in equation \ref{eq:si-checkpoint}, and instead use the notation $k=2l_{0}=2\hat{l}\left(q_{0}-q\right)$:

\begin{align}
G_{1} & =\frac{\alpha\left(D+1\right)}{2}\log\left(q_{0}-q\right)+\frac{\alpha\left(D+1\right)}{2}\frac{q}{q_{0}-q}-\alpha\int D^{D}\vec{t}\int Dt_{0}\log I\\
I & \doteq\int\frac{d^{D}\vec{v}_{l}}{\sqrt{2\pi}}e^{-\frac{1}{2}\frac{1}{q_{0}-q}\sum_{l}^{D}v_{l}^{2}+t_{l}\frac{\sqrt{q}}{q_{0}-q}v_{l}}\int\frac{ds}{\sqrt{2\pi}}\int_{1-s+R\|\vec{v}\|}^{\infty}\frac{dv_{0}}{\sqrt{2\pi}}e^{-\frac{1}{2}\frac{1}{q_{0}-q}v_{0}^{2}+t_{0}\frac{\sqrt{q}}{q_{0}-q}v_{0}-Cs^{2}\hat{l}}\\
 & =\int\frac{d^{D}\vec{v}_{l}}{\sqrt{2\pi}}e^{-\frac{1}{2}\frac{1}{q_{0}-q}\left(\vec{v}-\sqrt{q}\vec{t}\right)^{2}+\frac{1}{2}\frac{q\|\vec{t}\|^{2}}{q_{0}-q}}\int\frac{ds}{\sqrt{2\pi}}\int_{1-s+R\|\vec{v}\|}^{\infty}\frac{dv_{0}}{\sqrt{2\pi}}e^{-\frac{1}{2}\frac{1}{q_{0}-q}\left(v_{0}-\sqrt{q}t_{0}\right)^{2}+\frac{1}{2}\frac{qt_{0}^{2}}{q_{0}-q}-\frac{1}{2}\frac{1}{q_{0}-q}cks^{2}}
\end{align}
So that we can write:

\begin{align}
G_{1} & =\frac{\alpha\left(D+1\right)}{2}\frac{q}{q_{0}-q}+\frac{\alpha\left(D+1\right)}{2}\log\left(q_{0}-q\right)\\
 & \cdots-\alpha\int D^{D}\vec{t}\int Dt_{0}\log\int\frac{d^{D}\vec{v}_{l}}{\sqrt{2\pi}}e^{-\frac{1}{2}\frac{1}{q_{0}-q}\left(\vec{v}-\sqrt{q}\vec{t}\right)^{2}+\frac{1}{2}\frac{q\|\vec{t}\|^{2}}{q_{0}-q}}\int\frac{ds}{\sqrt{2\pi}}\int_{1-s+R\|\vec{v}\|}^{\infty}\frac{dv_{0}}{\sqrt{2\pi}}e^{-\frac{1}{2}\frac{1}{q_{0}-q}\left(v_{0}-\sqrt{q}t_{0}\right)^{2}+\frac{1}{2}\frac{qt_{0}^{2}}{q_{0}-q}-\frac{1}{2}\frac{1}{q_{0}-q}cks^{2}}
\end{align}
and denoting for brevity:
\begin{align*}
F & \doteq-\frac{D}{2}\log\left(q_{0}-q\right)-\frac{1}{2}\frac{1}{q_{0}-q}\left(\vec{v}-\sqrt{q}\vec{t}\right)^{2}-\frac{1}{2}\frac{1}{q_{0}-q}\left(v_{0}-\sqrt{q}t_{0}\right)^{2}-\frac{1}{2}\log\left(q_{0}-q\right)-\frac{1}{2}\frac{1}{q_{0}-q}cks^{2}\\
G_{1} & =-\alpha\int D^{D}\vec{t}\int Dt_{0}\log\int\frac{d^{D}\vec{v}_{l}}{\sqrt{2\pi}}\int\frac{ds}{\sqrt{2\pi}}\int_{1-s+R\|\vec{v}\|}^{\infty}\frac{dv_{0}}{\sqrt{2\pi}}e^{F}
\end{align*}
we have the limit:

\begin{align}
\lim_{q_{0}\to q}\left(q_{0}-q\right)G_{1} & =-\alpha\lim_{q_{0}\to q}\left(q_{0}-q\right)\int D^{D}\vec{t}\int Dt_{0}\log\int\frac{d^{D}\vec{v}_{l}}{\sqrt{2\pi}}\int\frac{ds}{\sqrt{2\pi}}\int_{1-s+R\|\vec{v}\|}^{\infty}\frac{dv_{0}}{\sqrt{2\pi}}e^{F}\\
 & =-\alpha\int D^{D}\vec{t}\int Dt_{0}\lim_{q_{0}\to q}\left(q_{0}-q\right)\log\max_{v_{0}\ge1-s+R\|\vec{v}\|}e^{F}\\
 & =-\alpha\int D^{D}\vec{t}\int Dt_{0}\lim_{q_{0}\to q}\left(q_{0}-q\right)\max_{v_{0}\ge1-s+R\|\vec{v}\|}F\\
 & =\alpha\int D^{D}\vec{t}\int Dt_{0}\min_{v_{0}\ge1-s+R\|\vec{v}\|}\left\{ \frac{1}{2}\left(\vec{v}-\sqrt{q}\vec{t}\right)^{2}+\frac{1}{2}\left(v_{0}-\sqrt{q}t_{0}\right)^{2}+\frac{1}{2}cks^{2}\right\} 
\end{align}

To solve the inner minimization problem we denote a Lagrangian:
\begin{align}
{\cal L} & =\frac{1}{2}\left(\vec{v}-\sqrt{q}\vec{t}\right)^{2}+\frac{1}{2}\left(v_{0}-\sqrt{q}t_{0}\right)^{2}+\frac{1}{2}cks^{2}+\lambda\left(1-s+R\|\vec{v}\|-v_{0}\right)
\end{align}
with derivative:

\begin{align}
\frac{\partial{\cal L}}{\partial v_{0}} & =v_{0}-\sqrt{q}t_{0}-\lambda=0\\
\frac{\partial{\cal L}}{\partial v_{l}} & =\left(v_{l}-\sqrt{q}t_{l}\right)+\frac{1}{2}\lambda R\frac{2v_{l}}{\|\vec{v}\|}=0\\
\frac{\partial{\cal L}}{\partial s} & =cks-\lambda=0
\end{align}

so from KKT conditions:

\begin{align}
0 & =\lambda\left(1-s+R\|\vec{v}\|-v_{0}\right)\\
\lambda & =v_{0}-\sqrt{q}t_{0}\\
\sqrt{q}t_{l} & =\frac{\|\vec{v}\|+\lambda R}{\|\vec{v}\|}v_{l}\\
\lambda & =cks
\end{align}
and denoting $v=\|\vec{v}\|$ and $t=\|\vec{t}\|$ we have three solution regimes:
\begin{enumerate}
\item ``Interior'' regime: assuming $\lambda=0$
\begin{align}
v_{0} & \ge1-s+Rv\\
v_{0} & =\sqrt{q}t_{0}\\
v_{l} & =\sqrt{q}t_{l}\\
s & =0\\
F & =0
\end{align}
which is valid for $t_{0} \ge1/\sqrt{q}+Rt$.
\item ``Touching'' regime: assuming $\lambda>0$, $v>0$
\begin{align}
v_{0} & =1-s+Rv\\
v_{l} & =\frac{v}{v+\lambda R}\sqrt{q}t_{l}\\
v & =\sqrt{q}t-\lambda R\\
\|\vec{v}-\sqrt{q}\vec{t}\|^{2} & =\lambda^{2}R^{2}\\
s & =\frac{1-\sqrt{q}t_{0}+R\sqrt{q}t}{1+ck\left(1+R^{2}\right)}\\
v_{0} & =\frac{\left(ckR^{2}+1\right)\sqrt{q}t_{0}+\left(1+R\sqrt{q}t\right)ck}{1+\left(1+R^{2}\right)ck}\\
0<v & =\frac{\left(1+ck\right)\sqrt{q}t-ckR+\sqrt{q}ckRt_{0}}{1+\left(1+R^{2}\right)ck}\\
v_{0}-\sqrt{q}t_{0} & =ck\frac{1+R\sqrt{q}t-\sqrt{q}t_{0}}{1+\left(1+R^{2}\right)ck}\\
F & =\frac{ck}{1+\left(1+R^{2}\right)ck}\left(1+R\sqrt{q}t-\sqrt{q}t_{0}\right)^{2}
\end{align}
which is valid for $t_{0} \ge1/\sqrt{q}-\frac{1+ck}{ck}tR^{-1}$.
\item ``Embedded'' regime: assuming $\lambda>0$, $v=0$ 
\begin{align}
v_{0} & =1-s\\
cks = \lambda & =v_{0}-\sqrt{q}t_{0}\\
0<s & =\frac{1}{1+ck}-\frac{\sqrt{q}}{1+ck}t_{0}\\
v_{0} & =\frac{ck}{1+ck}+\frac{\sqrt{q}}{1+ck}t_{0}\\
v_{0}-\sqrt{q}t_{0} & =\frac{ck-ck\sqrt{q}t_{0}}{1+ck}\\
F & =qt^{2}+\frac{ck}{1+ck}\left(1-\sqrt{q}t_{0}\right)^{2}
\end{align}
which is valid for $t_{0}<1/\sqrt{q}$.
\end{enumerate}
Using the conditions on $t_{0},t$ from each regime the following table summarizes the results:
\begin{equation}
\begin{array}{ccccccc}
\mathrm{Regime} &  & \mathrm{Embedded} &  & \mathrm{Touch} &  & \mathrm{Interior}\\
\mathrm{Range} &  & t_{0}\le1/\sqrt{q}-\frac{1+ck}{ck}t/R &  & 1/\sqrt{q}-\frac{1+ck}{ck}t/R\le t_{0}\le1/\sqrt{q}+Rt/ &  & 1/\sqrt{q}+Rt\le t_{0}\\
v_{0} &  & \frac{ck}{1+ck}+\frac{\sqrt{q}}{1+ck}t_{0} &  & \frac{\left(ckR^{2}+1\right)\sqrt{q}}{1+\left(1+R^{2}\right)ck}t_{0}+\frac{\left(1+R\sqrt{q}t\right)ck}{1+\left(1+R^{2}\right)ck} &  & \sqrt{q}t_{0}\\
v &  & 0 &  & \frac{\left(1+ck\right)\sqrt{q}t-ckR+\sqrt{q}ckRt_{0}}{1+\left(1+R^{2}\right)ck} &  & \sqrt{q}t\\
s &  & \frac{1}{1+ck}-\frac{\sqrt{q}}{1+ck}t_{0} &  & \frac{1+R\sqrt{q}t}{1+\left(1+R^{2}\right)ck}-\frac{\sqrt{q}}{1+\left(1+R^{2}\right)ck}t_{0} &  & 0\\
F &  & qt^{2}+\frac{ck}{1+ck}\left(1-\sqrt{q}t_{0}\right)^{2} &  & \frac{ck}{1+\left(1+R^{2}\right)ck}\left(1+R\sqrt{q}t-\sqrt{q}t_{0}\right)^{2} &  & 0
\end{array}\label{eq:si-spheres-fields-distribution}
\end{equation}
or written explicitly, the field and slack distribution conditioned on $t,t_{0}$:

\begin{align}
v_{0} & =\begin{cases}
\frac{ck}{1+ck}+\frac{1}{1+ck}\sqrt{q}t_{0} & t_{0}\le1/\sqrt{q}-\frac{1+ck}{ck}t/R\\
\frac{\left(1+R\sqrt{q}t\right)ck}{1+\left(1+R^{2}\right)ck}+\frac{\left(ckR^{2}+1\right)\sqrt{q}}{1+\left(1+R^{2}\right)ck}t_{0} & 1/\sqrt{q}-\frac{1+ck}{ck}t/R\le t_{0}\le1/\sqrt{q}+Rt\\
\sqrt{q}t_{0} & 1/\sqrt{q}+Rt\le t_{0}
\end{cases}\label{eq:si-spheres-v0}\\
v & =\begin{cases}
0 & t_{0}\le1/\sqrt{q}-\frac{1+ck}{ck}t/R\\
\frac{\left(1+ck\right)\sqrt{q}t-ckR}{1+\left(1+R^{2}\right)ck}+\frac{\sqrt{q}ckR}{1+\left(1+R^{2}\right)ck}t_{0} & 1/\sqrt{q}-\frac{1+ck}{ck}t/R\le t_{0}\le1/\sqrt{q}+Rt\\
\sqrt{q}t & 1/\sqrt{q}+Rt\le t_{0}
\end{cases}\label{eq:si-spheres-v}\\
s & =\begin{cases}
\frac{1}{1+ck}-\frac{1}{1+ck}\sqrt{q}t_{0} & t_{0}\le1/\sqrt{q}-\frac{1+ck}{ck}t/R\\
\frac{1+R\sqrt{q}t}{1+\left(1+R^{2}\right)ck}-\frac{\sqrt{q}}{1+\left(1+R^{2}\right)ck}t_{0} & 1/\sqrt{q}-\frac{1+ck}{ck}t/R\le t_{0}\le1/\sqrt{q}+Rt\\
0 & 1/\sqrt{q}+Rt\le t_{0}
\end{cases}\label{eq:si-spheres-s}
\end{align}
and the slack variable moments, used above for the self-consistent equations, are given by:

\begin{align}
\left\langle s^{2}\right\rangle  & =\frac{q}{\left(1+ck\right)^{2}}\int\chi_{D}\left(t\right)\int_{-\infty}^{1/\sqrt{q}-\frac{1+ck}{ck}t/R}Dt_{0}\left(1/\sqrt{q}-t_{0}\right)^{2}\label{eq:si-mean-slack-sqr-spheres}\\
 & +\frac{q}{\left(1+\left(1+R^{2}\right)ck\right)^{2}}\int\chi_{D}\left(t\right)\int_{1/\sqrt{q}-\frac{1+ck}{ck}t/R}^{1/\sqrt{q}+Rt}Dt_{0}\left(1/\sqrt{q}+Rt-t_{0}\right)^{2}\\
\left\langle s\right\rangle  & =\frac{\sqrt{q}}{1+ck}\int\chi_{D}\left(t\right)\int_{-\infty}^{1/\sqrt{q}-\frac{1+ck}{ck}t/R}Dt_{0}\left(1/\sqrt{q}-t_{0}\right)\label{eq:si-mean-slack-spheres}\\
 & +\frac{\sqrt{q}}{1+\left(1+R^{2}\right)ck}\int\chi_{D}\left(t\right)\int_{1/\sqrt{q}-\frac{1+ck}{ck}t/R}^{1/\sqrt{q}+Rt}Dt_{0}\left(1/\sqrt{q}+Rt-t_{0}\right)
\end{align}

\bigskip

\subsection{\label{subsec:appendix-spheres-classification-errors}Classification error for spheres}

Assuming $D\gg1$, $t\sim\chi_D$ is concentrated around $\sqrt{D}$ and the distribution of $v_0,v,s$ is a concatenation of truncated Gaussian (or $\delta$) distributions which correspond to the different regimes.
The slack distribution is then:
\begin{equation}
s	\sim\begin{cases}
{\cal N}\left(\frac{1}{1+ck},\frac{q}{\left(1+ck\right)^{2}}\right) & \frac{\sqrt{q}}{ck}\sqrt{D}/R<s\\
{\cal N}\left(\frac{1+R\sqrt{q}\sqrt{D}}{1+\left(1+R^{2}\right)ck},\frac{q}{\left(1+\left(1+R^{2}\right)ck\right)^{2}}\right) & 0<s\le\frac{\sqrt{q}}{ck}\sqrt{D}/R\\
\delta\left(0\right)H\left(1/\sqrt{q}+R\sqrt{D}\right) & s=0
\end{cases}
\end{equation}
From this distribution the probability of error anywhere on the manifold is $\varepsilon_{tr}^{manifold}=P\left(s>1\right)=H\left({\cal S}\right)$ for:
\begin{align}
{\cal S} & =\begin{cases}
ck/\sqrt{q} & \frac{\sqrt{q}}{ck}\sqrt{D}/R<1\\
\left(1+R^{2}\right)ck/\sqrt{q}-R\sqrt{D} & \frac{\sqrt{q}}{ck}\sqrt{D}/R\ge1
\end{cases}\label{eq:si-training-error-any}
\end{align}
\clearpage

Given any classifier $\boldsymbol{w}$, we assume the test error is calculated by sampling uniformly from the sphere, then adding noise.
When i.i.d Gaussian noise ${\cal N}\left(0,\sigma^{2}/N\right)$ is applied to each input component, as the weights are independent of this noise, the fields are affected
by Gaussian noise ${\cal N}\left(0,\sigma^{2}q\right)$. That is, the error is given by:
\begin{equation}
\varepsilon=P\left(h+\sigma\sqrt{q}\eta<0\right)=\left\langle H\left(\frac{h}{\sigma\sqrt{q}}\right)\right\rangle _{h}
\end{equation}
where $\eta$ is a standard Gaussian variable.

For a $D$-dimensional spheres of radius $R$, denote the fields $h(\vec{S})=y\boldsymbol{w}\cdot\boldsymbol{x}(\vec{S})=v_{0}+\vec{S}\cdot\vec{v}$.
For a given $\boldsymbol{w}$, we can always choose the coordinate system such as $u_{1}\propto w$ so that $v_{1}=w\cdot u_{1}$ and $v_{i}=0$ for $i>1$, so that $v=\|\vec{v}\|=v_{1}$. Denote $S_{1}=z$ we note that $\vec{S}\cdot\vec{v}=zv$ and thus $h=v_{0}+zv$.
As the joint distribution of $v_{0},v$ is given by theory (equations \ref{eq:si-spheres-v0},\ref{eq:si-spheres-v}) we shall now derive the distribution of $z$ under uniform sampling from the sphere. As $z\in\left[-R,R\right]$, we can denote $\boldsymbol{x}\in{\cal S}_{D-2}\left(\sqrt{R^{2}-z^{2}}\right)$ a sphere of all choices for the values of $S_{2..D}$, and thus wish to calculate the following integral:
\begin{equation}
\varepsilon\left(v,v_{0}\right)=\int_{-R}^{R}\mu z\int\mu^{D-1}\boldsymbol{x}\delta\left(\|x\|^{2}+z^{2}-R^{2}\right)H\left(\frac{v_{0}+zv}{\sigma\sqrt{q}}\right)
\end{equation}
where $\mu z$ and $\mu^{D-1}\boldsymbol{x}$ denote the corresponding measures on $z$ and $\boldsymbol{x}$.

Using the $n$-ball surface formula, $S_{n-1}\left(r\right)=\frac{2\pi^{\frac{n}{2}}}{\Gamma\left(\frac{n}{2}\right)}r^{n-1}$, the surface of the $D-1$ sphere with a radius $\|x\|$ is $S_{D-2}\left(\sqrt{R^{2}-z^{2}}\right)$, which needs to be normalized by the total surface, given by $S_{D-1}\left(R\right)$.
Furthermore, using polar coordinates the measure on $z$ is given by $\frac{R}{\sqrt{R^{2}-z^{2}}}$, yielding:

\begin{align}
\varepsilon\left(\vec{v},v_{0}\right) & =\frac{1}{S_{D-1}\left(R\right)}\int_{-R}^{R}\frac{Rdz}{\sqrt{R^{2}-z^{2}}}S_{D-2}\left(\sqrt{R^{2}-z^{2}}\right)H\left(\frac{v_{0}+z\|v\|}{\sigma\sqrt{q}}\right)\\
 & =\frac{\Gamma\left(\frac{D}{2}\right)}{2\pi^{\frac{D}{2}}R^{D-1}}\int_{-R}^{R}\frac{Rdz}{\sqrt{R^{2}-z^{2}}}\frac{2\pi^{\frac{D-1}{2}}}{\Gamma\left(\frac{D-1}{2}\right)}\left(\sqrt{R^{2}-z^{2}}\right)^{\left(D-2\right)}H\left(\frac{v_{0}+z\|v\|}{\sigma\sqrt{q}}\right)\\
 & =\frac{\Gamma\left(\frac{D}{2}\right)}{\Gamma\left(\frac{D-1}{2}\right)\sqrt{\pi}}\int_{-1}^{1}d\hat{z}\left(1^{2}-z^{2}\right)^{\left(D-3\right)/2}H\left(\frac{v_{0}+Rz\|v\|}{\sigma\sqrt{q}}\right)
\end{align}
by a change of variable $\hat{z}=z/R$. Thus we can write an expression for the full test error averaged on $v,v_{0}$:

\begin{align}
\varepsilon & =\frac{\Gamma\left(\frac{D}{2}\right)}{\Gamma\left(\frac{D-1}{2}\right)\sqrt{\pi}}\left\langle \int_{-1}^{1}d\hat{z}\left(1-\hat{z}^{2}\right)^{\left(D-3\right)/2}H\left(\frac{v_{0}+R\hat{z}v}{\sigma\sqrt{q}}\right)\right\rangle _{v_{0},v}
\end{align}

Using a change of variable $\hat{z}=\sqrt{t}$ and $d\hat{z}=\frac{1}{2}t^{-1/2}dt$ and using the Beta function:

\begin{align}
\int_{-1}^{1}d\hat{z}\left(1-\hat{z}^{2}\right)^{\left(D-3\right)/2} & =\int_{0}^{1}dtt^{-1/2}\left(1-t\right)^{\left(D-3\right)/2}\\
 & =B\left(x=1/2,y=\left(D-3\right)/2+1\right)\\
 & =\sqrt{\pi}\Gamma\left(\frac{D-1}{2}\right)\big/\Gamma\left(\frac{D}{2}\right)
\end{align}

Thus we define a bell-shaped distribution supported at $\hat{z}\in\left[-1,1\right]$:
\begin{align}
P\left(\hat{z}\right) & =\frac{\Gamma\left(\frac{D}{2}\right)}{\sqrt{\pi}\Gamma\left(\frac{D-1}{2}\right)}\left(1-\hat{z}^{2}\right)^{\left(D-3\right)/2}
\end{align}
with moments: $\left\langle \hat{z}\right\rangle _{P\left(\hat{z}\right)} =0$ and $\left\langle \hat{z}^{2}\right\rangle _{P\left(\hat{z}\right)}  =1/D$, and now write the error as an average with respect to $P\left(\hat{z}\right)$:
\begin{align}
\varepsilon & =\left\langle H\left(\frac{v_{0}+R\hat{z}v}{\sigma\sqrt{q}}\right)\right\rangle _{\hat{z},v_{0},v}
\end{align}

For $D\gg1$, by assuming that only the ``touching'' regime contributes to the error, we may evaluate the leading orders of $v_{0}+R\hat{z}v$ in order to write a simpler expression for the error. In the ``touching'' regime: 

\begin{align}
v_{0} & =\frac{\left(1+R\sqrt{q}t\right)ck+\left(ckR^{2}+1\right)\sqrt{q}t_{0}}{1+\left(1+R^{2}\right)ck}\\
v & =ckR\frac{\frac{1+ck}{ck}\sqrt{q}t/R+\sqrt{q}t_{0}-1}{1+\left(1+R^{2}\right)ck}
\end{align}
where $t_{0}\sim{\cal N}\left(0,1\right)$ and $t\sim\chi_{D}$. Noting that $\hat{z},t,t_{0}$ are pairwise independent and
also that $\hat{z}v$ is independent from $v_{0}$, the first two moments are:

\begin{align}
\left\langle v_{0}+R\hat{z}v\right\rangle _{\hat{z},t_{0}} & =\frac{\left(1+R\sqrt{q}t\right)ck}{1+\left(1+R^{2}\right)ck}\\
\left\langle \delta\left(v_{0}+R\hat{z}v\right)^{2}\right\rangle _{\hat{z},t_{0}} & =q\frac{\left(ckR^{2}+1\right)^{2}+\frac{R^{2}}{D+1}\left(ck\right)^{2}R^{2}\left[\left(\frac{1+ck}{ck}t/R-1/\sqrt{q}\right)^{2}+1\right]}{\left(1+\left(1+R^{2}\right)ck\right)^{2}}
\end{align}
and approximating $v_{0}+R\hat{z}v$ as Gaussian, we have using $\left\langle H\left(x/a\right)\right\rangle _{x\sim{\cal N}\left(\mu,s^{2}\right)} =H\left(\mu/\sqrt{s^{2}+a^{2}}\right)$ that:
\begin{align}
\varepsilon & \approx\left\langle H\left(\frac{\left(1+R\sqrt{q}t\right)ck}{1+\left(1+R^{2}\right)ck}\big/\sqrt{\sigma^{2}q+q\frac{\left(ckR^{2}+1\right)^{2}+\frac{R^{2}}{D}\left(ck\right)^{2}R^{2}\left[\left(\frac{1+ck}{ck}t/R-1/\sqrt{q}\right)^{2}+1\right]}{\left(1+\left(1+R^{2}\right)ck\right)^{2}}}\right)\right\rangle _{t}
\end{align}
so that denoting $\sigma_{0}^2$ the contribution of those terms to the variance we have the approximation:

\begin{align}
\sigma_{0}^2\left(t\right) & =\left(ckR^{2}+1\right)^{2}+\frac{R^{4}}{D}\left(ck\right)^{2}\left[\left(\frac{1+ck}{ck}t/R-1/\sqrt{q}\right)^{2}+1\right]\label{eq:si-spheres-snr-approx-sigma0}\\
\varepsilon & \approx\left\langle H\left(\left(1/\sqrt{q}+Rt\right)ck\big/\sqrt{\sigma_{0}^{2}+\left(1+\left(1+R^{2}\right)ck\right)^{2}\sigma^{2}}\right)\right\rangle _{t\sim\chi_{D}\label{eq:si-spheres-snr-approx}}
\end{align}
and the training error is given by setting $\sigma=0$.
Near $\alpha_{C}$ we have $k\to0$ such that $\sigma_{0}^{2}=1+1/\left(R^{-1}+R\right)^{2}\approx1$ and using equation \ref{eq:si-spheres-alphac-q}:
\begin{align}
\varepsilon & \approx H\left(\frac{R\sqrt{D}}{1+R^{2}}\frac{1}{\sqrt{1+\sigma^{2}}}\right)
\end{align}

\bigskip

\subsection{\label{subsec:appendix-general-manifold-iterative}Iterative algorithm for point-cloud manifolds}

From the mean-field equations of spheres we get that a theory of general-manifolds would imply:

\begin{align}
1 & =\left(1-L/q\right)k+\alpha\int D^{D}\vec{t}\int Dt_{0}F\left(\vec{t},t_{0}\right)\\
F\left(\vec{t},t_{0}\right) & =\min_{\min_{\vec{S}\in M}\left\{ v_{0}+\vec{v}\cdot\vec{S}\right\} \ge1/\sqrt{q}}\left\{ \|\vec{v}-\vec{t}\|^{2}+\frac{ck}{1+ck}\left(v_{0}-t_{0}\right)^{2}\right\} 
\end{align}

Denote a scalar function $g\left(\vec{v}\right) =\min_{\vec{S}\in M}\vec{v}\cdot\vec{S}$ and its subgradient:
\begin{align}
\tilde{S}\left(\vec{v}\right) & =\frac{\partial}{\partial v}g\left(\vec{v}\right)
\end{align}

Denote a Lagrangian:
\begin{align}
{\cal L} & =\frac{1}{2}\left(\vec{v}-\vec{t}\right)^{2}+\frac{1}{2}\frac{ck}{1+ck}\left(v_{0}-t_{0}\right)^{2}+\lambda\left(1/\sqrt{q}-v_{0}-g\left(\vec{v}\right)\right)
\end{align}
its derivations are

\begin{align}
\frac{\partial{\cal L}}{\partial v_{0}} & =\frac{ck}{1+ck}\left(v_{0}-t_{0}\right)-\lambda=0\\
\frac{\partial{\cal L}}{\partial v_{l}} & =\left(v_{l}-t_{l}\right)-\lambda\tilde{S}\left(\vec{v}\right)=0
\end{align}
so from KKT conditions:
\begin{align}
0 & =\lambda\left(1/\sqrt{q}-v_{0}-g\left(\vec{v}\right)\right)\\
\lambda & =\frac{ck}{1+ck}\left(v_{0}-t_{0}\right)\\
\vec{v} & =\vec{t}+\lambda\tilde{S}\left(\vec{v}\right)
\end{align}
so that we got equation \ref{eq:anchor-points} when $v_{0}\ne t_{0}$:
\begin{align}
\tilde{S}\left(\vec{v}\right) & =\frac{\vec{v}-\vec{t}}{\frac{ck}{1+ck}\left(v_{0}-t_{0}\right)}
\end{align}

Denoting $v=\|\vec{v}\|$ and $t=\|\vec{t}\|$ we have regimes:
\begin{enumerate}
\item ``Interior'' regime: assuming $v>0$ and $\lambda=0$ we have: 
\begin{align}
v_{0} & \ge1/\sqrt{q}-g\left(\vec{v}\right)\\
v_{0} & = t_{0}\\
v_{l} & = t_{l}\\
F & =0
\end{align}
\item ``Embedded'' regime: assuming $v=0$ and $\lambda>0$ we have:
\begin{align}
v_{0} & =1/\sqrt{q}-g\left(\vec{v}\right)=1/\sqrt{q}\\
\lambda & =\frac{ck}{1+ck}\left(1/\sqrt{q}-t_{0}\right)\\
\tilde{S}\left(\vec{v}\right) & =-\frac{\vec{t}}{\frac{ck}{1+ck}\left(1/\sqrt{q}-t_{0}\right)}\\
F & =t^{2}+\frac{ck}{1+ck}\left(1/\sqrt{q}-t_{0}\right)^{2}
\end{align}
\item ``Touching'' regime: assuming $v>0$ and $\lambda>0$ we have:
\begin{align}
v_{0} & =1/\sqrt{q}-g\left(\vec{v}\right)\\
\lambda & =\frac{ck}{1+ck}\left(1/\sqrt{q}-g\left(\vec{v}\right)-t_{0}\right)\\
\vec{v} & =\vec{t}+\frac{ck}{1+ck}\left(1/\sqrt{q}-\vec{v}\cdot\tilde{S}\left(\vec{v}\right)-t_{0}\right)\tilde{S}\left(\vec{v}\right)\\
\vec{v}\cdot\tilde{S} & =\frac{\left(1+ck\right)\vec{t}\cdot\tilde{S}\left(\vec{v}\right)+ck\left(1/\sqrt{q}-t_{0}\right)\tilde{S}^{2}}{1+ck\left(1+\tilde{S}^{2}\right)}\\
F & =\frac{ck}{1+ck\left(1+\tilde{S}^{2}\right)}\left(1/\sqrt{q}-t_{0}-\vec{t}\cdot\tilde{S}\right)^{2}
\end{align}
\end{enumerate}
so that for the ``touching'' regime we have equation \ref{eq:touching-regime-iteration}.

\end{document}